\documentclass[english]{iopart}

\usepackage{appendix}
\usepackage{graphicx}
\usepackage{url}
\usepackage{iopams}

\expandafter\let\csname equation*\endcsname\relax
\expandafter\let\csname endequation*\endcsname\relax

\usepackage[colorlinks=true,citecolor=blue]{hyperref}

\usepackage{amssymb}
\usepackage[latin1]{inputenc}
\usepackage{amsmath}
\usepackage{epsfig}
\newcounter{fig}
\usepackage{comment}

\newcommand{\Diag}{\textsf{Diag}}

\begin{document}

\title[Diagonals of rational functions ]
{\Large Ising $\,n$-fold integrals as diagonals of rational functions
and integrality of series expansions}

\vskip .3cm 

{\bf  March 17th, 2013} 

\author{A. Bostan$^\P$, S. Boukraa$||$, G. Christol$^\ddag$, S. Hassani$^\S$, 
J-M. Maillard$^\pounds$}
\address{$^\P$ \ INRIA, B\^atiment Alan Turing, 1 rue Honor\'e d'Estienne d'Orves,
 Campus de l'\'Ecole Polytechnique, 91120 Palaiseau, France}
\address{$||$  \ LPTHIRM and D\'epartement d'A{\'e}ronautique,
 Universit\'e de Blida, Algeria}
\address{$^\ddag$ \ Institut de Math\'ematiques de Jussieu, 
UPMC, Tour 25,
 4\`eme \'etage,   
 4 Place Jussieu, 75252 Paris Cedex 05, France}
\address{\S  \ Centre de Recherche Nucl\'eaire d'Alger, 
2 Bd. Frantz Fanon, B.P. 399, 16000 Alger, Algeria}
\address{$^\pounds$ \ LPTMC, UMR 7600 CNRS, 
Universit\'e de Paris 6, Tour 23,
 5\`eme \'etage, case 121, 
 4 Place Jussieu, 75252 Paris Cedex 05, France} 

\begin{abstract}
 We show that the $\, n$-fold integrals $\, \chi^{(n)}$ of the
magnetic susceptibility of the Ising model, as well as various other 
$\,n$-fold integrals of the ``Ising class'', or $\, n$-fold integrals from
enumerative combinatorics, like lattice Green functions, correspond to a
distinguished class of functions generalising algebraic functions:
they are actually \emph{diagonals of rational functions}. As a 
consequence, the power series
expansions of the, analytic at $\, x=0$, solutions 
of these linear differential equations ``Derived From
Geometry'' are {\em globally bounded}, which means that, after just one
rescaling of the expansion variable, they can be cast into 
series expansions with \emph{integer coefficients}. 
We also give several results showing that the unique analytical solution
 of  Calabi-Yau ODEs, and, more generally, Picard-Fuchs linear ODEs
with solutions of maximal weights, are 
always diagonal of rational functions. 
Besides, in a more enumerative combinatorics context,
generating functions whose coefficients are expressed in terms of nested sums
of products of binomial terms can also be shown to be {\em diagonals of
rational functions}. We finally address the question of the relations between
the notion of \emph{integrality\/} (series with integer coefficients, or, more
generally, globally bounded series) and the \emph{modularity\/} of ODEs.
\end{abstract}

\vskip .2cm
  
 {\em This paper is the short version of the larger (100 pages) 
 version~\cite{Big}, available on
  \url{http://arxiv.org/abs/1211.6031}, where all the detailed proofs are given 
 and where a  larger set of examples is displayed.}

\vskip .5cm


\noindent {\bf AMS Classification scheme numbers}: 34M55, 
47E05, 81Qxx, 32G34, 34Lxx, 34Mxx, 14Kxx 

\vskip .5cm


\maketitle

\section{Introduction}
\label{introduc}

The series expansions of many magnetic
susceptibilities (or many other quantities, like 
the spontaneous magnetisation)
of the Ising model on various 
lattices in arbitrary dimensions are actually series with 
{\em integer coefficients}~\cite{Butera,Vicari,Fujiwara}. This is 
a consequence of the fact that, in a van der Waerden type expansion 
of the susceptibility, all the contributing graphs are the ones
 with exactly two odd-degree vertices and the number of such graphs 
is an integer.
When series expansions in theoretical physics, or mathematical physics, do not
have such an obvious counting interpretation, the puzzling emergence of series
with {\em integer coefficients} is a strong indication that some fundamental
structure, symmetry, concept have been overlooked, and that a deeper
understanding of the problem remains to be 
discovered\footnote[2]{The emergence of
{\em positive integer} coefficients corresponds to the existence
of some underlying measure~\cite{Bessis}.}. 
Algebraic functions are known 
to produce series with {\em integer coefficients}. Eisenstein's 
theorem~\cite{Eisenstein} states that 
the Taylor series of a (branch of an) algebraic function can be recast
into a series with integer coefficients, up to a rescaling by a constant (Eisenstein
constant).
An intriguing result due to Fatou~\cite{Fatou} (see pp. 368--373)
states that a
power series with \emph{integer\/} coefficients and radius
 of convergence (at least)
one, is either rational, or transcendental. This result also appears in P\'olya
and Szeg\"o's famous Aufgaben book~\cite{Polya} (see Problem VIII-167).
P\'olya~\cite{Polya2} conjectured a stronger result, namely that a power
series with integer coefficients which converges in the open unit disk is
either rational, or  admits the {\em unit circle as a natural boundary}
(i.e. it has no analytic continuation beyond the unit disk). This was
eventually proved\footnote[5]{The P\'olya-Carlson result
 can be used to prove that some 
integer sequences, such as the sequence of prime 
numbers $\, (p_n)$~\cite{Flajolet},
 do not satisfy any linear recurrence relation with polynomial
 coefficients.} by Carlson~\cite{Carlson}.
Along this natural
boundary line, it is worth
recalling~\cite{ze-bo-ha-ma-04,Khi6,ze-bo-ha-ma-05c,bo-ha-ma-ze-07b,mccoy3}
that the series expansions of the full magnetic susceptibility of the 2D Ising
model~\cite{Importance} correspond to a power series 
with {\em integer} coefficients\footnote[1]{In
some variable $\,w$~\cite{ze-bo-ha-ma-04,Khi6,ze-bo-ha-ma-05c,bo-ha-ma-ze-07b}. 
In the modulus variable $\, k$, one needs to perform a simple rescaling by a
factor $\, 2$ or $\, 4$ according to the type of (high, or low temperature)
expansions.}. For them, the unit circle certainly arises as
 a natural boundary~\cite{bo-gu-ha-je-ma-ni-ze-08} (with
 respect to the modulus variable
$\, k$), but, unfortunately, this cannot be justified by Carlson's
theorem\footnote[9]{The radius of convergence is $\, 1$ with respect to the
modulus variable $\, k$, in which the series {\em does not have} integer
coefficients, being {\em globally bounded} only (this means that it can be
recast into a series with integer coefficients by one rescaling of the
variable $\, k$). If one considers the series expansion with respect to
another variable (such as~$\, w$) in which the series {\em does have} integer
coefficients, then the radius of convergence is not $\, 1$.}.

A series with natural boundaries {\em cannot be D-finite}\footnote[8]{D-finite
series are sometimes called \emph{holonomic}. A priori, for multivariate functions,  
these notions differ~\cite{Big}. The equivalence of these notions is proved by deep
results of Bern{\v{s}}te{\u\i}n~\cite{Bernstein} and
Kashiwara~\cite{Kashiwara}.}, i.e. solution of a linear differential equation
with polynomial coefficients~\cite{Stanley80,Lipshitz89}. For simplicity, let
us restrict to series with {\em integer coefficients} (or series that have
integer coefficients up to a variable rescaling), that are series expansions
of D-finite functions.
Wu, McCoy, Tracy and Barouch~\cite{wu-mc-tr-ba-76}  
have shown that the previous full magnetic susceptibility of the 2D Ising
model can be expressed (up to a normalisation
factor $\,(1-s)^{1/4}/s$, see~\cite{ze-bo-ha-ma-05c,ze-bo-ha-ma-05b}) as an
infinite sum of $n$-fold integrals, denoted by $\, \tilde{\chi}^{(n)}$, which
are {\em actually D-finite}\footnote[3]{For Ising models on higher dimensional 
lattices~\cite{Butera,Vicari,Fujiwara} no such decomposition
of susceptibilities, as an infinite sum of D-finite functions, 
should be expected at first sight.}.
We found out that the corresponding (minimal order)
differential operators are Fuchsian~\cite{ze-bo-ha-ma-04,ze-bo-ha-ma-05c},
and, in fact, ``special'' Fuchsian operators: the
 critical exponents for \emph{all\/} their
singularities are {\em rational numbers}, and their Wronskians are $\, N$-th
roots of {\em rational functions}~\cite{High}. Furthermore, it has been shown
later that these $\, \tilde{\chi}^{(n)}$'s are, in fact, solutions of 
{\em globally nilpotent} operators~\cite{bo-bo-ha-ma-we-ze-09}, or
 $\, G$-operators~\cite{Andre5,Andre6}. It is worth noting that the series
expansions, at the origin, of the $\, \tilde{\chi}^{(n)}$'s, 
in a well-suited variable~\cite{ze-bo-ha-ma-05c,ze-bo-ha-ma-05b} 
$\,w$, actually have {\em integer coefficients}, even if this  result
does not have an immediate proof\footnote[1]{We are interested in this paper
in the emergence of {\em integers} as coefficients of D-finite series. In general,
this emergence is {\em not obvious}: it cannot be simply explained at the level of
the linear recurrence satisfied by the coefficients, as illustrated by the
case of Ap\'ery's calculations (see also section (6.2) and \ref{proof}).} 
for all integers
$\, n$ (in contrast with the full susceptibility).
{}From the first truncated series expansions of $\, \tilde{\chi}^{(n)}$, the
coefficients for generic $\, n$ can be inferred~\cite{bo-bo-ha-ma-we-ze-09}
\begin{eqnarray} 
\label{closedn} 
\hspace{-0.9in}&&\tilde{\chi}^{(n)}(w) \, \, \,
= \, \, \, \, \, \, 2^n \cdot w^{n^2} \cdot 
\Bigl(1 \,\, + \, 4 \, n^2 \cdot w^2 \,\, + \,
2 \cdot (4\, n^4 \, +13\, n^2 \, +1)\cdot w^4 \, \, 
\nonumber \\
\hspace{-0.9in}&& \quad \quad + \, {{8} \over {3}} \cdot 
(n^2+4)\, (4\, n^4\,+23\, n^2+3) \cdot w^6 \\ 
\hspace{-0.9in}&& \quad \quad + \,{{1} \over {3}}
\cdot \, (32\,n^8\,+624\,n^6\,+4006\,n^4\,+8643\,n^2\,+1404)\cdot w^8
\nonumber \\ 
\hspace{-0.9in}&& \quad \quad + \,{{4} \over {15}} \cdot \,
(n^2+8)\cdot \, (32\, n^8 \, +784\, n^6 \, +6238\, n^4 \, +16271\, n^2 \,
+3180)\cdot w^{10} \,\, + \, \cdots \, \Bigr).
\nonumber 
\end{eqnarray} 
Note that the coefficients of $\, \tilde{\chi}^{(n)}(w)/2^n$, which depend
on $\, n^2$, are {\em integer coefficients} when $\, n$ is {\em any integer},
this integrality property of the coefficients for any integer $\, n$ being not
straightforward (see~\cite{Big}). These coefficients are valid up to $\, w^2$
for $\, n\, \ge \, 3$, $\, w^4$ for $\,n\, \ge \, 5$, $\, w^6$ for $\, n\, \ge
\, 7$, $w^8$ for $n \, \ge \, 9$, and $w^{10}$ for $n\ge 11$ (in particular it
should be noted that $\tilde{\chi}^{(n)}$ is an even function of $\, w$ only
for even $n$). Further studies on these $\,\tilde{\chi}^{(n)}$'s showed the
fundamental role played by the theory of elliptic functions\footnote[5]{Which
is not a surprise for Yang-Baxter integrability specialists.} (elliptic
integrals, {\em modular forms}) and, much more unexpectedly, 
{\em Calabi-Yau ODEs}~\cite{CalabiYauIsing1,CalabiYauIsing}. These 
recent structure results thus suggest to see the occurrence 
of series with {\em integer coefficients}
as a consequence of {\em modularity}~\cite{SP4} (modular forms, mirror
maps~\cite{CalabiYauIsing1,CalabiYauIsing,SP4,LianYau}, etc) in the Ising
model.

Along this line, many other examples of series with {\em integer coefficients}
emerged in mathematical physics (differential geometry, lattice statistical
physics, enumerative combinatorics, {\em replicable functions}\footnote[8]{The
concept of replicable functions is closely related to {\em modular
functions}~\cite{McKay}, (see the replicability of Hauptmoduls), Calabi-Yau
threefolds, and more generally the concept of {\em
modularity}~\cite{SP4,Livne,Schutt,Gouvea,Saito}. } \ldots). One must, of
course, also recall Ap\'ery's results~\cite{Apery}. \ref{modularapp} 
gives a list of {\em modular forms}, and their associated series with
integer coefficients, corresponding to various lattice Green
functions~\cite{GlasserGuttmann,Prell,Zucker,GoodGuttmann}, that are, often,
expressed in terms of HeunG functions\footnote[2]{Generically HeunG functions
are far from being modular forms.} which can be written as hypergeometric
functions with {\em two alternative pullbacks} (see also sections
(\ref{ex1}) and (\ref{ex1sec}) below). Let us underline, in \ref{modularapp}, 
the {\em Green function for the diamond
lattice}~\cite{GoodGuttmann}, the {\em Green function for the face-centred
cubic lattice} (see equation (19) in~\cite{GoodGuttmann}), and more examples
corresponding to the spanning tree generating functions~\cite{spanning} (and
Mahler measures). This {\em integrality} is also seen in the {\em nome } and in
other quantities like the {\em Yukawa coupling}~\cite{CalabiYauIsing1}.

 In this paper we restrict on series with {\em integer
coefficients}, or, more generally, {\em globally bounded}~\cite{Christol}
series of {\em one complex variable}, but it is clear that this integrality
property does also occur in physics with {\em several complex variables}: they
can, for instance, be seen for the previous (D-finite\footnote[3]{For several
 complex variables the ODEs of the paper are
replaced by Picard-Fuchs systems.}) $\, n$-fold integrals 
$\,\tilde{\chi}^{(n)}$ for the anisotropic Ising model~\cite{WuJPA} (or for the
Ising model on the checkerboard lattice), or on the example of the lattice
Ising models with a magnetic field\footnote[2]{Along this line, original
alternative representations of the partition function of the Ising model in a
magnetic field are also worth recalling~\cite{Barouch}.} (see for instance,
Bessis et al.~\cite{Bessis}).

We take, here, a learn-by-example approach: on such quite technical
questions one often gets a much deeper understanding from highly non-trivial
examples than from sometimes too general, or slightly obfuscated, 
mathematical demonstrations.

The main result of the paper will be to
 show that the $\,\tilde{\chi}^{(n)}$'s are {\em globally bounded} 
series, as a consequence of the fact that they are actually 
{\em diagonals of rational functions for any 
value of the integer} $\, n$. We will
generalise these ideas, and show that an extremely large class of problems of
mathematical physics can be interpreted in terms of {\em diagonals of rational
functions}: $\, n$-fold integrals with algebraic integrand of a certain type
that we will characterise, Calabi-Yau ODEs, MUM linear ODEs~\cite{Guttmann},
series whose coefficients are {\em  nested sums of products of binomials}, etc. 

Another purpose of this paper is to ``disentangle'' the notion of series with
integer coefficients ({\em integrality}) and the notion of 
{\em modularity}~\cite{SP4,Livne,Schutt,Gouvea,Saito,Roques1,Roques2}. In 
this ``down-to-earth'' paper we essentially restrict to Picard-Fuchs ODEs and
 to a ``Calabi-Yau'' framework, therefore 
modularity\footnote[8]{Modularity is a wider concept than this
``Calabi-Yau'' modularity (see modular up to a Tate twist,
 modular Galois representations~\cite{Hulek}).
Modular forms provide the simplest 
examples (see \ref{modularapp}) of
modularity (see also Serre's modularity conjecture, and 
the Taniyama-Shimura conjecture). For a first 
introduction to these ideas see~\cite{Yui}.}
 will just mean that the series 
solutions of Picard-Fuchs ODEs,
{\em as well} as  the corresponding nome series, and the Yukawa 
series, have {\em integer} coefficients. 

\vskip .1cm 

The paper is organised as follows. 
Section (\ref{Locally}) introduces the main concepts 
we need in this very paper, 
namely the concept of {\em diagonals of rational or algebraic functions}, 
and the concept of {\em globally bounded series},
recalling that diagonals of rational or algebraic functions 
are necessarily globally bounded series.
Section (\ref{nfoldas}) shows the main result of the paper, namely that 
the $\,n$-fold Ising integrals $\, \tilde{\chi}^{(n)}$ are {\em diagonals of 
rational functions} for any value of the integer $\, n$, the corresponding 
series being, thus, globally bounded.
Section (\ref{plan}) shows that series with (nested sums of products of) 
{\em binomials} coefficients are diagonals of 
rational functions.
Section (\ref{commspec}) discusses, in 
the most general framework, the conjecture 
that D-finite globally bounded
series could be necessarily diagonals of rational functions.
Section (\ref{versus}) provides a set of 
{\em modular forms} examples (in particular {\em lattice Green functions} 
see \ref{modularapp}).
Beyond modular forms, using new determinantal identities on the Yukawa 
couplings, and focusing on Hadamard products of modular forms,
section (\ref{learning}) analyses the difference between integrality
 and modularity, showing that the two concepts
 are actually quite different.
Section (\ref{hadamardmore}) addresses, more specifically, the Calabi-Yau
 modularity, and the difference between integrality 
and modularity, underlining that  the {\em integrality of the nome}
 series is crucial for modularity, the integrality of the Yukawa series
being {\em not} sufficient. The conclusion, section (\ref{concl}),
 emphasises the difference between the ``special properties'' 
of {\em geometrical} nature and the ones of {\em arithmetic} 
nature, emerging in theoretical physics.
Several large appendices provide detailed examples illustrating
pedagogically the previous sections. In particular \ref{modularapp} provides
 many modular forms examples
associated with {\em lattice Green functions}, and \ref{Yukawaratio} 
provides new representations 
of {\em Yukawa couplings as ratios of determinants}.

\vskip .2cm

\section{Series integrality, diagonal of rational functions}
\label{Locally}

Let us recall some concepts that will be fundamental in this paper, first the
notion of {\em globally bounded series}, and, then, the concept of {\em diagonal} of
a function\footnote[5]{The functions are
in fact defined by {\em series} of several complex variables: they have to be
Taylor, or Laurent, series (no Puiseux series).}, and some of its most important properties.
The main reason to introduce this concept of {\em diagonal} of function, not
very familiar to physicists, is that it enables to consider {\em diagonal of
rational functions}, this class of functions filling the gap between algebraic
functions and $\, G$-series: they can be seen as {\em generalisations of
algebraic functions}. Thus this class of functions can play a {\em key role to
decipher the complexity of functions occurring in theoretical physics}.

\subsection{Globally bounded series}
\label{locally}

Let us first recall the definition of being {\em globally
bounded}~\cite{Christol} for a series. Consider a series expansion with
rational coefficients, with non-zero radius of convergence\footnote[1]{A
series like the Euler-series $\, \sum_{n=0}^{\infty} \, n! \cdot \, x^n$ which
has integer coefficients is excluded.}. The series is said to be globally
bounded if there exists an integer $\, N$ such that the series can be recast
into a series with integer coefficients with just one rescaling $\, x \,
\rightarrow \, N\, x$.

A necessary condition for being globally bounded is that only a finite number
of primes occur as factors of the denominators of the rational number
series coefficients. There is also a condition on the growth of these
denominators, that must be bounded exponentially~\cite{Christol}, in such a
way that the series has a non-zero $p$-adic radius of convergence for all
primes~$p$. When this is the case, it is easy to see that these series can be
recast, with just one rescaling, into series with {\em integer
coefficients}\footnote[8]{For a first set of series with integer coefficients,
see \ref{modularapp}, where a set of such series with integer
coefficients corresponding to \emph{modular forms\/} is displayed. See also
(\ref{ex1}) and (\ref{ex1sec}) below.}.

\subsection{Definition of the diagonal of a rational function}
\label{def}

Assume that 
$\,{\cal F}(z_1, \ldots, z_n)\, = \,\, P(z_1, \ldots, z_n)/Q(z_1, \ldots, z_n)$ 
is a rational function, where $P$ and $Q$ are
 polynomials of $\, z_1, \, \cdots, \, z_n\,$ 
with {\em rational coefficients} 
such that $Q(0, \ldots, 0) \neq 0$. This assumption implies that ${\cal F}$ 
can be expanded at the origin as a Taylor series with rational number coefficients
\begin{eqnarray}
\label{defdiag}
\hspace{-0.90in}&&\quad \quad {\cal F}\Bigl(z_1, \, z_2, \, \ldots, \, z_n \Bigr)
\, \, \,\, = \, \,\, \,\sum_{m_1 \, = \, 0}^{\infty}
 \, \cdots \, \sum_{m_n\, = \, 0}^{\infty} 
 \,F_{m_1,  \, \ldots, \, m_n}
\cdot  \, z_1^{m_1} \,\,  \cdots \,\, z_n^{m_n}. 
\end{eqnarray}
The \emph{diagonal of ${\cal F}$\/} is defined as the series
of {\em one variable}
\begin{eqnarray}
\label{defdiag2}
\hspace{-0.6in}&&\Diag\Bigl({\cal F}\Bigl(z_1, \, z_2, \,
 \ldots, \, z_n \Bigr)\Bigr)
\, \, \, = \, \,  \quad \sum_{m \, = \, 0}^{\infty}
 \,F_{m, \, m, \, \ldots, \, m} \cdot \, z^{m}.
\end{eqnarray}

More generally, one can define, in a similar way, the diagonal of any
multivariate power series $ \, {\cal F}$, with rational number
coefficients, or with coefficients in a finite
 field\footnote[2]{The definition even extends to multivariate
Laurent power series, see e.g.~\cite{BA-JPB}.}.

\subsection{Main properties of diagonals} 
\label{propr-diag}

The concept of diagonal of a function has a lot of interesting properties (see
for instance~\cite{legacy2}). Let us recall, through examples, some of the
most important ones.

The study of diagonals goes back, at least, to P\'olya~\cite{Polya21}, in a
combinatorial context, and to Cameron and Martin~\cite{CaMa38} in an
analytical context {\em related to Hadamard products}~\cite{Hadamard}.  
P\'olya showed that the diagonal of a rational function
in {\em two variables\/} is always an {\em algebraic function}. The most
basic example is $\, {\cal F} =\,  1/(1\,-z_1\,-z_2)$, for which  
\begin{eqnarray}
\hspace{-0.3in}&&\Diag ({\cal F})\, \, \, = \, \, \, \,\,
\Diag \left(\sum_{m_1 = 0}^\infty \sum_{m_2 = 0}^\infty  \binom{m_1+m_2}{m_1} 
\cdot \, z_1 ^{m_1} z_2 ^{m_2}\right) 
\nonumber \\ 
\hspace{-0.3in}&& \qquad \quad \quad  \quad \,  \, =\, \,  \, \, \,
 \sum_{m = 0}^\infty \binom{2m}{m} \cdot \, z^m 
\, \,\, \, = \,  \, \, \, \, \,  \frac{1}{\sqrt{1-4z}}.
\end{eqnarray}

The proof of P\'olya's result is based on the simple observation that the
diagonal  $\textsf{Diag} ({\cal F})$ is equal to the coefficient of $\, z_1^0$
in the expansion of ${\cal F}(z_1,z/z_1)$. Therefore, by Cauchy's integral
theorem, $\, \textsf{Diag} ({\cal F})$ is given by the contour integral
\begin{eqnarray}
\hspace{-0.3in} \quad \textsf{Diag} ({\cal F})  \, \,\,  = \,  \, \, \,\,
\frac{1}{2\pi i}\, \oint_\gamma {\cal F}(z_1,z/z_1) \, \frac{dz_1}{z_1}, 
\end{eqnarray}
where the contour $\,\gamma$ is a small circle 
around the origin. Therefore, by
Cauchy's residue theorem, $\,\textsf{Diag} ({\cal F})$ is the sum of the
residues of the rational function ${\cal G} = \,{\cal F}(z_1,z/z_1)/z_1$ at all
its singularities $\,s(z)$ with zero limit at $z=\, 0$. Since the residues of a
rational function of two variables are algebraic functions,
 $\, \textsf{Diag}({\cal F})$ is itself an algebraic function.

For instance, when $\,{\cal F} = \, 1/(1-z_1-z_2)$, then 
${\cal G} =\, {\cal F}(z_1,z/z_1)/z_1$ has two poles at
 $\,s\, =\, \frac12 (1 \pm \sqrt{1-4z})$. The
only one approaching zero when $\, z\,\, \rightarrow\,\, 0$ is
 $\,s_0 = \,\frac12 (1 -\sqrt{1-4z})$. If $\, p(s)/q(s)$ has 
a simple pole at $s_0$, then its residue at
$\,s_0$ is $\, p(s_0)/q'(s_0)$. Therefore 
\begin{eqnarray}
\hspace{-0.4in}&&  \Diag ({\cal F})
\, \,\,  = \, \,\, \,\, \, \frac{1}{2\pi i}
\oint_\gamma \frac{dz_1}{z_1-z_1^2-z} 
\,\, \,  =\, \,\,\, \,
  \frac{1}{1-2s_0}
\, \,\,\,  = \,\,\, \, \, \frac{1}{\sqrt{1-4z}}.
\end{eqnarray}

\subsubsection{Diagonals of rational functions of more than two variables \newline} 
\label{fromtwotomore}

When passing from {\em two to more variables}, diagonalisation may still be
interpreted using contour integration of a multiple complex integral over a
so-called {\em vanishing cycle}~\cite{Deligne84}. However,
 the result {\em is not}
an algebraic function anymore. A simple example is 
$\,{\cal F}\, = \,1/(1-z_2-z_3-z_1 z_2-z_1 z_3)$, for which  
\begin{eqnarray}
\hspace{-0.7in}&& \Diag ({\cal F})\,\, \, = \, \,\,\,\,\, 
 1 \,\,\,\, +4 z\, \,+36 z^2 \,\, +400 z^3\, \,+4900 z^4 \,
 +63504 z^5 \,\,\,\,  + \, \cdots 
\end{eqnarray}
is equal to the complete elliptic integral of the first kind 
\begin{eqnarray}
\hspace{-0.7in}&& \qquad 
\Diag ({\cal F})\,\, \,  =\, \, \,\,\, \sum_{m \geq 0} \binom{2m}{m}^2 \cdot \, z^m
\, \,\, \, = \,\, \, \, \, 
_2F_1\Bigl([{{1} \over {2}}, \,{{1} \over {2}}], \,  [1]; \, 16 \, z \Bigr),
\end{eqnarray}
which is a \emph{transcendental\/} function. A less obvious 
example (see~\cite{Pech} for a related example with a combinatorial flavor) is
\begin{eqnarray}
\hspace{-0.4in}&& \Diag \left( \frac{1}{1-z_1-z_2-z_3-z_1z_2-z_2z_3-z_3z_1-z_1z_2z_3} \right)
\\
\hspace{-0.4in}&& \qquad  \qquad   \quad   \quad \,  = \,\,\, \,\,\, 
 {{1} \over { 1 \, -z}} \, \cdot \, _2F_1\Bigl(
[{{1} \over {3}}, \, {{2} \over {3}}], \, [1]; \, {{54 \, z } \over { (1\, -z)^3}}
\Bigr). \nonumber 
\end{eqnarray}

It was shown by Christol~\cite{Christol84,Christol85,Christol369} that the
diagonal $\Diag({\cal F})$ of \emph{any\/} rational function $\, {\cal F}$ is
\emph{D-finite}, in the sense that it satisfies a linear differential equation
with polynomial coefficients\footnote[5]{A more general result was proved by
Lipshitz~\cite{Lipshitz}: {\em the diagonal of any D-finite series is
D-finite}, see also~\cite{Purdue}.}. Moreover, the diagonal of any algebraic
power series with rational coefficients is a $G$-function {\em
coming from geometry}, i.e. it satisfies the Picard-Fuchs type
differential equation associated with some one-parameter family of algebraic
varieties. Diagonals of algebraic power series thus appear to be a {\em
distinguished class} of $G$-functions\footnote[8]{Such diagonals are solutions
of $G$-operators. They are functions that are {\em always algebraic modulo any prime}
 $\, p$. They fill the gap between algebraic functions and $\, G$-series: 
they can be seen as {\em generalisations of algebraic functions}.}.
It will be seen below (see (\ref{recall})) that algebraic functions
with $\, n$ variables can be seen as diagonals of rational functions with 
 $\, 2\, n$ variables. Thus diagonals of rational functions also 
 appear to be a {\em distinguished class} of $G$-functions. It is worth 
noting that this  distinguished class is stable by the Hadamard product: 
the {\em Hadamard product of two diagonals of rational functions
is the diagonal of rational function}.

An immediate, but important property of diagonals of rational functions,
 with rational number
coefficients, is that they are {\em globally bounded},
which means that they have {\em integer coefficients} up to a simple change
of variable $\,z \,\, \rightarrow \,\,N \, z$, where 
$ \,N \, \in \, \,\mathbb{Z}$. 

\subsubsection{Diagonals of rational functions modulo primes \newline} 
\label{diagprime}

Furstenberg~\cite{Fu} showed that 
 the diagonal of {\em any multivariate rational power series} 
with coefficients {\em in a field of positive characteristic} 
{\em is algebraic}. Deligne~\cite{Deligne84,BA-JPB} 
extended this result to
diagonals of algebraic functions.
For instance, when $\,{\cal F}\, =\, \,$
$  1/(1\,-z_2-z_3\,-z_1 z_2\,-z_1 z_3)$, one gets  modulo 7
\begin{eqnarray}
\hspace{-0.7in}&&  \Diag ({\cal F})  \quad   \bmod 7
 \, \,\, = \, \, \,\, \, \, \,
1 \,\, \, \, +4\,z \,\,  +z^2 \,  +z^3 \,  +4z^7 \,  +2z^8 \,  +4z^9
 \, \,\,  \, + \, \cdots  \, \, 
\nonumber \\ 
\hspace{-0.7in}&& \qquad \qquad \qquad 
=  \, \, \, \,\,  \frac{1}{\sqrt[6]{1\, +4z\,+z^2\, +z^3}} 
 \quad \,  \,\, \bmod 7.
\end{eqnarray}
More generally, in this example, for any prime $\, p$, one has
\begin{eqnarray}
\hspace{-0.7in}&&\qquad \quad \Diag ({\cal F}) \quad \,  
 \, \,  \,  \,  = \, \, \, \, \, \,    \,  \, P(z)^{1/(1-p)} 
\, \, \quad  \quad   \, \bmod p 
\end{eqnarray}
where the polynomial $\, P(z)$ is nothing, 
 but~\cite{Ihara,Honda,Koike99}
\begin{eqnarray}
\label{polP}
\hspace{-0.95in}&&P(z) \,   = \, \, 
_2F_1\Bigl([{{1} \over {2}}, \,{{1} \over {2}}], \,  [1];
 \, 16 \, z \Bigr)^{1-p}  \,\, \,  \,  \bmod \, p\, \,  = \, \, 
\sum_{n=0}^{(p-1)/2} \, {p \, -1/2 \choose n}^2 \cdot \, (16 \, z)^n. 
\end{eqnarray}

Note, however, that the Furstenberg-Deligne result~\cite{Fu,Deligne84},
 that we illustrate, here, 
with $\,{\cal F}\, =\, \,$ $  1/(1\,-z_2-z_3\,-z_1 z_2\,-z_1 z_3)$, 
goes {\em far beyond the case of hypergeometric functions}
 for which simple closed formulae can be displayed. 

\vskip .1cm 

\subsection{Hadamard product and other products of series}
\label{propr-diag}

Let us also recall the notion of {\em Hadamard product}~\cite{Hadamard,Necer}
 of two series, that 
we will denote by a star.
\begin{eqnarray}
\hspace{-0.7in}&&\hbox{If} \qquad \, \, 
f(x) \, = \, \, \, \sum_{n=0}^{\infty} \, a_n \cdot x^n,
 \qquad \, \,   
g(x) \, = \, \, \, \sum_{n=0}^{\infty} \, b_n \cdot x^n, 
\qquad \quad \, \, \hbox{then:} 
\nonumber \\
\hspace{-0.7in}&&\qquad  \qquad  f(x) \,\star\, g(x)
\, \, \,\, = \, \,\, \,  \,\, 
\sum_{n=0}^{\infty} \, a_n \cdot b_n \cdot x^n.
\end{eqnarray}

The notion of diagonal of a function and the notion 
of Hadamard product are obviously related:
\begin{eqnarray}
\hspace{-0.6in}\Diag\Bigl( f_1(x_1) \cdot
 f_2(x_2)\,\, \cdots \, \, f_n(x_n)\Bigr)
 \, \,  \,\, = \, \, \, \,  \, \,  
f_1(x) \, \star\, f_2(x) \, \star \, \cdots \, \star \,f_n(x).
\end{eqnarray}
In other words, the {\em diagonal of a product} of functions with separate variables is
equal to the {\em Hadamard product} of these functions in a common variable. In
particular, the Hadamard product of $n$ rational (or algebraic, or even
D-finite) power series is D-finite\footnote[3]{The Hadamard product of rational
 power series is still rational, but the Hadamard product 
of algebraic series is in general transcendental.}.

\vskip .1cm
 The Hadamard product of two series with integer coefficients is
straightforwardly a series with integer coefficients. Furthermore, the {\em
Hadamard product of two operators}, annihilating two series, defined as the
(minimal order, monic) linear differential operator annihilating the Hadamard
product of these two series, is a {\em product compatible with a large number
of structures and concepts}\footnote[2]{For instance, 
the Hadamard product of two globally
nilpotent~\cite{bo-bo-ha-ma-we-ze-09} operators is {\em also globally
nilpotent}.} that naturally occur in lattice statistical
mechanics.  We have a similar compatibility property between the 
diagonal and the Hurwitz product~\cite{Big,Hurwitz1899}.

\subsection{Furstenberg's result on algebraic functions }
\label{recall}

It was shown by Furstenberg~\cite{Fu} that \emph{any algebraic series\/} in one
variable can be written as the \emph{diagonal of a rational function of two
variables\/}.  The basis of Furstenberg's result is the fact that if 
$\, f(x)$ is a power series
without constant term, and is a root of a polynomial $ \, P(x,y)$ such that
$\, P_y(0,0)\,\neq\, 0$, then
\begin{eqnarray}
\label{Furstformula}
\hspace{-0.7in} f(x) \,\,\,\,  = \,\, \, \,\, \Diag \left( y^2 \cdot \, 
\frac{P_y(xy,\, y)}{P(xy,\, y)} \right)
 \qquad \, \hbox{where:} \quad \quad \quad 
P_y  \, \, = \, \, \, {{\partial P} \over {\partial y}}.
\end{eqnarray}
When $\, P_y(0,0) =\,  0$,  formula 
(\ref{Furstformula}) is not true anymore. However, Furstenberg's result
 still holds~\cite{Big}. 

Note that this representation as diagonal of a rational
function is, by no means unique, as can be 
seen on the algebraic function\footnote[1]{Here, $f$ 
is annihilated by $P(x,y) = (1-x)y^2 - x^2$, which is precisely
 such that  $\, P_y(0,0) =\,  0$.}
\begin{eqnarray}
\hspace{-0.9in}&&f\,\, =\,\,\,\,\, \frac{x}{\sqrt{1-x}}\,\,\, = \,\,\,
\,\,\, x\,\,\,\, +\frac12\,x^2\,\,
+\frac38\,x^3\,\, +\frac{5}{16}\,x^4\,\,
+\frac{35}{128}\,x^5\, \, +\frac{63}{256}x^6\,\,\,\,\, + \, \, \cdots 
\end{eqnarray}
which is the diagonal of $\,\,{(2\,x\,y - cx + cy)}/(x\,+y\,+2)\,$ 
for {\em any rational number} $c$.

Furstenberg's proof {\em does not necessarily produce
the simplest rational function} (see~\cite{Big}).

Furstenberg's result has been generalised to  power series 
expansions of algebraic functions
in an {\em arbitrary number of variables}~$n$: 
any  {\em algebraic power series}\footnote[5]{In the 
one-variable case, Puiseux 
 series could be considered 
but only after ramifying the variable.}  
with rational coefficients is 
the diagonal of a rational function with $\,2n$ variables 
(see Denef and Lipshitz~\cite{Denef}). 

\section{Selected $n$-fold integrals are
 diagonals of rational functions}
\label{nfoldas}

Among many multiple integrals that are important in various domains 
of mathematical physics, and before considering other $n$-fold integrals
of the ``Ising class\footnote[9]{Using the terminology introduced by 
Bailey et al.~\cite{Isingclass}, see also~\cite{bo-ha-ma-ze-07b}. 
}'', let us first consider the $n$-particle contribution 
to the magnetic susceptibility of the Ising model which we 
denote $\,\tilde{\chi}^{(n)}(w)$. They are given by $\,(n-1)$-dimensional 
integrals~\cite{ze-bo-ha-ma-04,bo-ha-ma-ze-07}:
\begin{eqnarray}
\label{chin}
\hspace{-0.6in}\tilde{\chi}^{(n)}(w) \,\, \,\, = \, \, \,\,\,\, 
\, \frac {(2w)^n}{n!}\ \Big(\prod_{j=1}^{n-1}\,
  \int_0^{2\pi}\, \frac{d\Phi_j}{2\pi} \Big)\ 
 \cdot  \,  Y\,\cdot  \, \frac{1+X}{1-X}\,\cdot\, X^{n-1} \, \cdot  \, G\ ,
\end{eqnarray}
where, defining  $\, \Phi_0$ by
 $ \, \, \sum_{i = 0}^{n-1}\,  \Phi_i \,=\, 0$, we set 
\begin{eqnarray}
\hspace{-0.9in}&&X \,=\, \,  \prod_{i = 0}^{n-1}\, x_i , \,\,  \,
\quad 
x_i\, = \,\, \frac{2w}{A_i  + \sqrt{A_i^2-4 w^2} },\,\, \, 
\quad \, Y \, = \,   \prod_{i = 0}^{n-1}\, y_i, \, 
\, \,\,  \quad \,
y_i\, =\,\, \frac{1}{\sqrt{A_i^2 \,-4 w^2} },
\nonumber \\
\hspace{-0.9in}&& G \, = \,\,
 \prod_{0\le i < j\le n-1}  \, \frac{2 \,-2 \cos{(\Phi_i-\Phi_j)}}{(1\,-x_i\, x_j)^2}, 
\quad \,  \hbox{where:} \qquad
A_i \,=\,\, \, 1 \, \, -2 \,w \cos(\Phi_i).
\end{eqnarray}
The integrality property (\ref{closedn}) had been checked~\cite{Khi6} 
for the first $\, \tilde{\chi}^{(n)}$'s  
 and inferred~\cite{bo-bo-ha-ma-we-ze-09} for generic $n$.
We are going to {\em prove it\footnote[2]{For the
 $\tilde{\chi}^{(n)}$, the rescaling 
factor (``Eisenstein constant'') is $\,2$ or $\,4$ according to the fact that
one considers high or low temperature series~\cite{Khi6,High}.}
 for any integer $n$, showing a much fundamental result, namely that
all the $(n-1)$-fold integrals  $\, \tilde{\chi}^{(n)}$'s are very special}:
they are actually {\em diagonals of rational functions}.

\vskip .1cm

\subsection{ $ \, \tilde{\chi}^{(3)}$ as a toy example}
\label{toy}

At first sight the $ \, \tilde{\chi}^{(n)}$'s are involved transcendental 
holonomic functions. Could it be possible that they correspond to the
 {\em distinguished class}~\cite{BA-JPB} of $G$-functions, 
generalising algebraic functions, which have an interpretation as
 diagonals of multivariate algebraic functions (and consequently diagonals 
of rational functions with twice more variables)? If this is the case, 
then the series of the $ \, \tilde{\chi}^{(n)}$'s
must {\em necessarily reduce modulo any prime to an algebraic function}
 (see (\ref{diagprime})). 
The $ \, \tilde{\chi}^{(1)}$ and $ \, \tilde{\chi}^{(2)}$ contributions
being too degenerate
(a rational function and a too simple elliptic function), 
let us consider the first non-trivial case, namely $ \, \tilde{\chi}^{(3)}$. Its 
series expansion has already been displayed in~\cite{ze-bo-ha-ma-04}. It reads
$\, \, {\tilde{\chi}}^{(3)}/8   \,\,=\,\,\, \, w^9 \cdot \, F(w)\, $ with: 
\begin{eqnarray}
\hspace{-0.9in}&& F(w)\,\, \,=\,\,\, \, 
1 \, \, + \, 36 \, w^2 \, + \, 4 \, w^3 \, 
+ \, 884 \, w^{13} \, + \, 196 \, w^5 \, + \, 18532 \, w^6 \,
 + \, 6084 \, w^7 \, \, \, \, + \, \, \cdots 
\nonumber 
\end{eqnarray}
Since we have obtained the exact ODE satisfied by $\, \tilde{\chi}^{(3)}$
we can produce as many coefficients as we want in its series expansion.
Let us consider this series modulo the prime $\, p \, = \, \, 2$. It now reads
the lacunary series
\begin{eqnarray}
\hspace{-0.9in}&&F(w) \quad  \, \bmod 2  \,\, \,\, = \,\,\, \,\,
1 \, \, + w^8\, +w^{24} \, +w^{56} \, +w^{120} \, +w^{248} \, +w^{504} \, \,
 +w^{1016} \,   \, + \cdots,   \nonumber
\end{eqnarray}
solution of the functional equations on $\, F(w)$ or, with
 $\, z \, = \, \, w^8$, on
 $\, G(z) \, = \, \, 1 \, + \, w^8 \cdot \,  F(w)$ 
\begin{eqnarray}
\hspace{-0.9in}&&\quad \quad \quad \quad  F(w)\,\,=\,\,\, 
1 \, \, + \, \, \, w^8 \cdot \, F(w^2), \qquad \quad 
G(z) \, \,  \, = \, \,  \, z \, \, + \, \, G(z^2),   
\end{eqnarray}
where one recognises, with equation $\, G(z) \,  = \,  \, z  \, + \,  G(z^2)$,
Furstenberg's example~\cite{Fu}
 of the simplest algebraic function in 
characteristic~2\footnote[3]{Modulo the prime $\, p\, = \, 2$,
 the previous functional equation becomes 
$ \, G(z) \, \,  \, = \, \,  \, z \, \, + \, \, G(z)^2$.}.
In fact $ \, H(w) \, = \, \, \, w^9 \, F(w)\, $ is solution of the 
 quadratic equation:
\begin{eqnarray}
\label{degree2}
\hspace{-0.9in}&&\qquad \qquad \quad 
H(w)^2\, \,  + w \cdot \, H(w) \,\,\,   + w^{10}
  \,   \, \, \,  \, = \, \, \, \, \,  \,   0 \quad \mod 2.  
\end{eqnarray}
The calculations are more involved modulo $\, p  = \, 3$.
Indeed, $H(w)=\, \, {\tilde{\chi}}^{(3)}(w)/8$ satisfies, modulo 3,
 the polynomial equation of degree nine
\begin{eqnarray}
\label{degree9}
\hspace{-0.95in}&&\quad \quad \quad \quad p_9 \cdot \, H(w)^9 \,\, \, 
 + \, \,  w^{6} \cdot \, p_3 \cdot \, H(w)^3  \, \, \, 
+ \, \,  w^{10} \cdot \, p_1 \cdot \, H(w) \,\, \,
 \nonumber   \\
\hspace{-0.95in}&& \qquad \qquad \qquad \quad 
 \qquad \qquad \qquad \qquad 
+ \, \, w^{19} \cdot \, p_0^{(1)}\cdot \, p_0^{(2)}\, 
 \, \,\,\, \,  = \, \,\, \,\,  \, 0, 
\end{eqnarray}
where:
\begin{eqnarray}
\hspace{-0.95in}&&\quad p_9 \, \, = \, \, \,\, 
  \, (w+1) ^{3} \, ({w}^{2}+1)^{18} \, (w-1)^{24}, 
\nonumber \\
\hspace{-0.95in}&&\quad p_3 \, \, = \, \, \,
 ({w}^{2}+1)^{18} \, (1-w)^{15} \, ({w}^{4}-{w}^{2}-1),
\quad \, \, p_1 \, \, = \, \, \,
 ({w}^{2}+1)^{20} \, (1-w)^{13}, \
\nonumber \\
\hspace{-0.95in}&&\quad p_0^{(1)} \, \, = \, \, \,
{w}^{6}+{w}^{5}+{w}^{4}-{w}^{2}-w+1,
  \\
\hspace{-0.95in}&&\quad p_0^{(2)} \,  = \,  \,{w}^{37} \, - {w}^{36} \, 
+{w}^{35} \, - {w}^{33} \, +{w}^{31} \, -{w}^{30} \, +{w}^{28} \,
 +{w}^{27} \, +{w}^{24} - {w}^{23} +{w}^{22}
 \nonumber  \\
\hspace{-0.95in}&&\qquad \quad  \quad   \, 
- {w}^{21} - {w}^{18} - {w}^{16}+{w}^{14} \, - {w}^{12} \, 
-{w}^{11} \, -{w}^{10}\, +{w}^{7} \, -{w}^{5} \, -{w}^{3}\, -1.
 \nonumber 
\end{eqnarray}

The calculations are even more involved modulo larger primes.
The (minimal order) linear differential
 operator annihilating
the  $ \, \tilde{\chi}^{(3)}$ series mod. $5$,
 reads\footnote[1]{This operator is of zero 
$\, 5$-curvature~\cite{bo-bo-ha-ma-we-ze-09}.
}:
\begin{eqnarray}
\hspace{-0.95in}&&\quad (x+1)  \,  ({x}^{2}+x+1)
  \,  (x+2) \cdot \,  {x}^{4} \cdot \,  D_x^{4}
\, \,\,  +2\,{x}^{3} \cdot \, ({x}^{3}+2\,{x}^{2}+4\,x+4) 
 \, (x+4) \cdot \,  D_x^{3} \, 
\nonumber \\
\hspace{-0.95in}&&\quad \quad \quad \quad \quad 
  +{x}^{2} \cdot \, ({x}^{4}+3\,{x}^{3}+4) \cdot \,   D_x^{2}
\,\,   +4  \cdot  \, ({x}^{4}+3) \cdot \,   x \cdot \,   D_x \,\,   +3
\end{eqnarray}
If one can easily get this  linear differential
 operator, finding the minimal polynomial of $\, \tilde{\chi}^{(3)}$ 
modulo $\,5$, generalising
 (\ref{degree2}) or (\ref{degree9}),
or rather, the polynomial $\, \tilde{P}(\kappa, \, w)$,
where  $\, \kappa \, = \, \, \tilde{\chi}^{(3)}(w)/w^9$, 
such that $\, \tilde{P}(\kappa, \, w)\, = \, \, 0 \, \,$ mod. $5$, 
requires a {\em very large} 
number of coefficients. The 
polynomial~\cite{Big} $\, \tilde{P}(\kappa, \, w)$
is of degree 50 in $\, \kappa$, 
of degree 832 in $\, x$ and is the sum of 4058 monomials. 

One can imagine, in a first step, that the
 $ \, \tilde{\chi}^{(3)}$ series mod. {\em any prime} $\, p$
are {\em also algebraic functions}, and, in a second step,
 that $ \, \tilde{\chi}^{(3)}$ 
may be the diagonal of a rational function. In fact we are going to show, 
in the next section, a stronger result: 
the $ \, \tilde{\chi}^{(n)}$'s are 
{\em actually diagonals of rational functions, for any integer $\, n$}. 

\subsection{The $ \, \tilde{\chi}^{(n)}$'s are diagonals of rational functions}
\label{calcula}
Let us, now, consider the general case where $\, n$ is an arbitrary 
integer.
With the change of variable $\,z_k\,=\,\exp(i\, \Phi_k)$ (where $\, i^2 =\, -1$),
 one clearly gets
\begin{eqnarray}
\hspace{-0.9in}&&\qquad \qquad \qquad \quad  \prod_{k = 0}^{n-1} z_k \,\,=\,\,\, 1\ , 
\quad \quad \quad  \, \, \frac{dz_j}{z_j}\,  = \, \, i \,d \Phi_j\ ,
\\ 
\hspace{-0.9in}&& \qquad \qquad 2 \cos(\Phi_k)\,\,\, =\,\,\, \, z_k\, +\frac1{z_k}\ , 
\quad \quad \quad \,\,\,
  2 \cos( \Phi_k - \Phi_j )\,\, =\,\, \,  \frac{z_k}{ z_j} + \frac{z_j}{ z_k} \ ,
\nonumber 
\end{eqnarray}
and \eqref{chin} becomes
\begin{eqnarray}
\label{prop1}
\hspace{-0.4in}&&\tilde{\chi}^{(n)}(w) \,\, \, = \, \,\,\, \, \frac {(2w)^n}{n!}\ 
 \Big( \prod_{j=1}^{n-1} \, \frac{1}{2\, i \pi}\, 
 \oint_{{\cal C}} \,  \frac{dz_j}{z_j} \Big) \cdot 
\, F(w,\, z_1,\,\dots,\, z_{n-1} )\ ,
\end{eqnarray}
where $\, {\cal C}$ is the path
 ``turning once counterclockwise around the unit circle''
and where $\, F$ is algebraic over
 $\, \mathbb{Q} (w, z_1,\,\dots,\, z_{n-1})$ and reads:
\begin{eqnarray} 
\label{integrantF}
 F(w,\, z_1,\dots,\,  z_{n-1}) \, \,\,\, = \, \,\, \,\,\,  \, 
 Y \cdot \, X^{n-1}  \cdot \,{{1 \, + \, X} \over {1 \, - \, X}}  \cdot \, G. 
\end{eqnarray}

\vskip .1cm

Now, let us suppose that $ \,F\, $  {\em is analytic\footnote[1]{One could  
consider Laurent, instead of Taylor, expansions, but this is a slight 
generalisation~\cite{BA-JPB,Sathaye,Kauers}.}
at the origin}, namely that it has a Taylor expansion \eqref{defdiag}.
Then applying $(n-1)$ times the residue formula, one finds 
\begin{eqnarray}
\label{prop2}
\hspace{-0.6in}\tilde{\chi}^{(n)}(w) 
\, \,\,\,  = \, \, \, \, \,\,
 \Diag\Big(\frac {(2\,z_0\cdots z_{n-1})^n}{n!}\ 
 \cdot 
\, F(z_0 \, \cdots\,  z_{n-1},\,\,  z_1,\, \dots,\,  z_{n-1})\Big).
\end{eqnarray}
To check that this is actually true,
 we introduce an auxiliary set, namely
 ${\cal T}_n$  
 the subset of Laurent series
  $\mathbb{Q} [ z_1,\, \dots,\,  z_{n-1}, \,  z_1^{-1}, \, \dots , \,  z_{n-1}^{-1} ] [[w]]$, 
consisting of series 
\begin{eqnarray}
\hspace{-0.2in}f(w,\, z_1,\, \dots,\,  z_{n-1} )
\,\,\, \,\, \,  = \, \, \,\, \, \,\, 
 \sum_{m=0}^\infty\,  P_m \cdot \,  w^m,
\nonumber
\end{eqnarray}
where 
$\, P_m$ belongs to
 $\,\,  \mathbb{Q} [ z_1, \, \dots,\,  z_{n-1},\,  z_1^{-1},\,  \dots ,\,  z_{n-1}^{-1} ]$
and is such that the degree of $\, P_m$, in each of the $\, z_k^ {-1}$,
is at most~$\, m$.

Then to prove that~$F(z_0 z_1 \dots z_{n-1}, z_1, \dots ,z_{n-1})$
 has a Taylor expansion, we only have
 to  verify that $\, F(w,z_1,\dots, z_{n-1} )$ belongs
 to this auxiliary set $\, {\cal T}_n$.
Checking this is a straightforward step-by-step computation on auxiliary functions:
\begin{eqnarray}
\label{G}
\hspace{-0.8in}&& \quad   A_k\,\,  =\, \,\,\, 
 \,   1 \,\, \,\,    -w \cdot \left(z_k \, +\frac{1}{z_k}\right), 
\quad \quad \quad   \quad
 \quad \quad \quad \quad    \,  
\hbox{for:} \quad \quad
 1\, \le\, i \,\le \, n-1,
 \nonumber \\
\hspace{-0.8in}&&  \quad  
 A_k\,\,  =\, \,\, \,   \, \,   1 \,\, \, \,  
 -w \cdot \, \left(\frac{1}{z_1 \cdots \, z_{n-1}} \, + z_1\cdots \, z_{n-1}\right),  
\quad \quad \quad \quad \hbox{for:} \quad \quad  i\,= \,\, 0.
\nonumber 
\end{eqnarray}

Hence $\, A_k$ belongs to this auxiliary set $\, {\cal T}_n$.
So to be sure that the inverse or the square root of some function
 in this auxiliary set $\, {\cal T}_n$ is also in this auxiliary set 
$\, {\cal T}_n$  we only have to check that its first Taylor coefficient 
is actually $\, 1$, or $\, w$, $\, w^2$, or $\, w^n$, $\, n$ integer. It 
is straightforward to see that:
\begin{eqnarray}
\hspace{-0.9in}&&  A_k^2\,\, -4\,w^2\,\,\,\, =
\,\,\,\,\, \, \,\, \,  1\,\, \,  \,\, \,
-2\,w \cdot \,\left(z_k\,+\frac{1}{z_k}\right)\,\,  \,
 + w^2 \cdot \,\left(z_k\,-\frac{1}{z_k}\right)^2, 
 \\
\hspace{-0.6in}&&\quad \quad \quad \hbox{hence:}
 \quad  \quad  \quad  \quad \quad
 \sqrt{A_k^2\,-4 w^2}\,\,\,  =\,\,\,\, \, \, \,
1\, \, \,\, +\,\,  \cdots, 
\nonumber
\end{eqnarray}
\begin{eqnarray}
\hspace{-0.9in}&&\quad \quad   y_k\, \,  =\, \, \,\, 
  \frac{1}{\sqrt{A_k^2\, -4w^2}}
 \,\,  =\,\, \,\, \,  \,
  1\, \,\,   \, +\, \cdots,  
\qquad \quad Y\,\,  =\,\,  \,  1\,\,  \, +\, \cdots,
 \nonumber \\
\hspace{-0.9in}&&\quad \quad   x_k\,\,  =\, \,\,   
\frac{2\,w}{A_k\,+\sqrt{A_k^2\,-4 w^2}}
 \, =\, \,\,  \, w\,\,\,  \, + \,\, \cdots, 
\quad  \quad \, \,  \, \,  \,   x_k \, x_j \,\, =\,\,\,
 w^2\, \, +\,  \, \cdots,
 \nonumber 
\end{eqnarray}
\begin{eqnarray}
\hspace{-0.8in}&&\quad  X\,= \,\,\, \, 
w^n\,  \,\, +\,\,  \cdots, 
\quad \quad \quad \quad \quad 
\frac{1+X}{1-X}\,\,\,  = \, \,\, \,  1\,\,\,   +\, \cdots, 
\nonumber 
\end{eqnarray}
\begin{eqnarray}
\hspace{-0.8in}&&\quad 
\, G\,\,=\,\,\,
 \prod_{0\le k<j\le n-1} \Bigl(\frac{ - \,(z_k\,-z_j)^2}{(1\,-x_k x_j)^2 \cdot \,z_k\,z_j }\Bigr) 
\,\, \,  \,\,= \,\,\,\,
\,\prod_{0\le k<j\le n-1} \Bigl( {{- \, (z_k-z_j)^2} \over {z_k\,z_j }}\Bigr) 
\,\,\, \, \, 
 +\,\, \cdots 
\nonumber
\end{eqnarray}

Thus, this shows that $\, F$ belongs to the auxiliary set $\,{\cal T}_n$:
 \begin{eqnarray}
\hspace{-0.7in}&&F(w,z_1,\dots, z_{n-1} )\,\, \,\,  = \\
\hspace{-0.7in}&& \quad \quad \quad  \quad\sum_{m=0}^\infty 
\sum_{m_1=-m}^{M_1(m)} \dots 
\sum_{m_{n-1}=-m}^{M_{n-1}(m)}\, a_{m,m_1,\dots,m_{n-1}} \cdot\,  
w^m \, \cdot\,  z_1^{m_1+m} \, \,\dots \, \,
z_{n-1}^{m_{n-1}+m}.  \nonumber
\end{eqnarray}

Consequently, it makes sense to take its
diagonal. The residue theorem requires searching 
the terms not containing $z_j$ i.e.
such as $\,\, m_1=\, \dots\, =\, m_{n-1} \, = \, 0$. One therefore
gets:
\begin{eqnarray}
\label{m000}
\tilde{\chi}^{(n)}\,\,=\,\,\,\,\,\, \sum_{m=0}^\infty \, a_{m,0,\dots,0} \cdot \, w^m.
\end{eqnarray}

In particular, we find
\begin{eqnarray}
\label{diagm000}
\hspace{-0.7in}&&\tilde{\chi}^{(n)}\, =\, \,\,\, 
\Diag\Bigl( 
\sum_{m=0}^\infty \sum_{m_1=-m}^{M_1(m)} \dots \,
\sum_{m_{n-1}=-m}^{M_{n-1}(m)} a_{m,m_1,\dots,m_{n-1}} 
\cdot \, z_0^m z_1^{m_1+m} \, \dots \, z_{n-1}^{m_n-1+m}\Bigr) 
\nonumber \\
\hspace{-0.7in}&&\qquad =\,\,\,\,  \Diag \Bigl(  
 F(z_0\, z_1\, \dots \,z_{n-1},\, z_1,\, \dots,\, z_{n-1})\Bigr). 
\end{eqnarray}

As~$\, F$ is algebraic, $\, \tilde{\chi}^{(n)}$ {\em is the diagonal
of an algebraic function} of $\, n$ variables and, consequently, {\em the
diagonal of a rational function} of $ \, 2\, n$ variables.

We thus see that we can actually {\em find explicitly}
the algebraic function such that its diagonal is 
the $\, n$-fold integrals $\, \widetilde{\chi}^{(n)}$: it is
 {\em nothing but the integrand} of the $\, n$-fold integral, 
up to simple transformations (namely
 $ \, F(w,\, z_1,\, \dots ,\,z_{n-1})\, \rightarrow$
$ \, F(z_0 \,z_1 \,\dots\, z_{n-1}, \,z_1, \,\dots ,\, z_{n-1})$).

\vskip .1cm

{\bf Remark}: $\tilde{\chi}^{(n)}$ is a solution of a linear 
differential equation, and has a radius of convergence equal to $\, 1/4$ 
in  $w$. Among the other solutions 
of this equation, there is the function obtained by changing the square root 
appearing in $\, x_k$ into its opposite. A priori there are $\, 2^n$
 ways to do this, hence
$\, 2^n$ new solutions but, not all distinct. At first sight, for
these new solutions, the $\, x_k$'s are no longer in $\, {\cal T}_n$. 

In fact, we find some quite interesting structure. Let us consider, for
instance, the case of $\, \tilde{\chi}^{(3)}$. If one considers other choices of
sign in front of the nested square roots in the integrand,
the series expansions of the corresponding $\, n$-fold integrals read:
\begin{eqnarray}
\fl \qquad w+6\,{w}^{2}+28\,{w}^{3}+124\,{w}^{4}+536\,{w}^{5}+2280\,{w}^{6}
+9604\,{w}^{7}+40164\,{w}^{8}
 \nonumber \\
\fl \qquad \qquad \quad  + 167066\,{w}^{9}+692060\,{w}^{10}+2857148\,{w}^{11}
 \,  \, + \cdots 
\nonumber \\
\fl \qquad {w}^{2}+6\,{w}^{3}+30\,{w}^{4}+140\,{w}^{5}+628\,{w}^{6}+2754\,{w}^{7}
+11890\,{w}^{8}+50765\,{w}^{9}
 \nonumber \\
\fl \qquad \qquad \quad +214958\,{w}^{10}+904286\,{w}^{11} \,  \, + \cdots 
\nonumber
\end{eqnarray}
One does remark that {\em all these alternative series} are,
as $\, \tilde{\chi}^{(3)}$, series with
 {\em integer coefficients}~\cite{Big}.

\vskip .1cm

\subsection{More $\, n$-fold integrals of the Ising class 
and a simple integral  of the Ising class}
\label{Isingclass}

It is clear that the demonstration we have performed on the $\, \chi^{(n)}$'s
can also be performed straightforwardly, mutatis mutandis, with other $\, n$-fold 
integrals of the ``Ising class\footnote[2]{For the purpose
of this section $n$-fold integrals of the ``Ising class'' will mean 
$n$-fold integrals that are known to arise in the study 
of the two-dimensional Ising model susceptibility.}~\cite{Isingclass,bo-ha-ma-ze-07b}''
 like the $\, n$-fold integrals 
$\, \Phi_H$ in~\cite{bo-ha-ma-ze-07}, which amounts to getting
 rid of the fermionic term 
$\, G$ (see (\ref{integrantF})), the  $\, \chi_d^{(n)}$'s corresponding to  $\, n$-fold 
integrals associated with the diagonal\footnote[1]{Of course this 
 ``diagonal~\cite{mccoy3,CalabiYauIsing} wording'' should not be confused
 with the notion of diagonal
of a function.} susceptibility~\cite{mccoy3,CalabiYauIsing}
(the magnetic field is located on a diagonal of the square lattice), 
 the $\, \Phi_D^{(n)}$'s
 in~\cite{bo-ha-ma-ze-07b} which are simple integrals, and also 
for all the lattice Green functions displayed 
in~\cite{GoodGuttmann,Guttmann}, and the list is
far from being exhaustive. For instance, the simple integral 
$\, \Phi_D^{(n)}$ is the
 diagonal of the algebraic function:

\vskip .1cm 

\begin{eqnarray}
\label{good}
\hspace{-0.6in}\frac{2}{n!} \cdot (1-t^2)^{-1/2} \cdot { G_n \, {F_n^{n-1} } \over {
G_n \, F_n^{n-1}  \, -\, (2 \, w \,t)^n}} \, \, \, \,\,   - \, \frac{1}{n!} \,,
 \qquad   \quad  \hbox{where:} 
\end{eqnarray}
\begin{eqnarray}
\hspace{-0.6in}&&F_n \, \, = \, \, \, \,\, \,
1\,\,\,  \, \,-2\, w \,\, \, \, 
+(1-4\,w\,+4\,w^2\,- 4\,w^2\,t^2)^{1/2},  \, \\
\hspace{-0.6in}&&G_n \, \, = \, \, \, \,\, \,
1 \, \,\,\,   \,-2\,w \, t\cdot \,T_{n-1}\Bigl({{1} \over {t}}\Bigr)\, \,\, \,
+\Bigl(\big(1 \, \,\,  \,-2\,w \, t\cdot \,T_{n-1}\Bigl({{1} \over {t}}\Bigr) \big)^2
 \, - 4\,w^2 \cdot \,t^2\Bigr)^{1/2},
 \nonumber 
\end{eqnarray}
and where $\, T_{n-1}(t)$ is the $\, (n-1)$-th Chebyshev polynomial of the first kind.
The way we have obtained these Chebyshev results
 (\ref{good}) is displayed in~\cite{Big}.

\vskip .1cm 

As opposed to the $\, \chi^{(n)}$'s, the integral $\,
\Phi_D^{(n)}(w) $ is the diagonal of an algebraic function of {\em two}
variables (and, thus, the diagonal of a rational function of \emph{four\/}
variables) {\em independently of the actual value of $n$}.

\vskip .1cm 

If the $\, \chi^{(n)}$'s  are fundamental to understand the
 Ising model~\cite{Importance,wu-mc-tr-ba-76}, or 
the $\, \chi_d^{(n)}$'s have a physical meaning associated with the diagonal
susceptibility~\cite{mccoy3,CalabiYauIsing} for the Ising model, most of
the $\, n$-fold integrals of the ``Ising class~\cite{Isingclass}''
do not have that importance, or even that physical meaning (even if they 
have played a crucial role to understand the singularities of the 
Ising model~\cite{bo-ha-ma-ze-07b}). What we see here with, for instance, 
the $\, \Phi_H$'s~\cite{bo-ha-ma-ze-07b,bo-ha-ma-ze-07}, is that the 
demonstration they are  diagonal of rational functions 
is exactly the same as for the $\, \chi^{(n)}$'s (see section \ref{calcula}),
because of a key analyticity assumption of the integrand is
also fulfilled.

\vskip .2cm 

\subsection{More general $\, n$-fold integrals as diagonals}
\label{gener}

More generally   the demonstration we have performed
 on the  $\, \tilde{\chi}^{(n)}$'s
can be performed for {\em any} $\, n$-fold integral that can be recast
in the following form:
\begin{eqnarray}
\label{form}
\hspace{-0.5in}\quad \int_{C}  \, \int_{C} \,  \cdots \, \int_{C} \,
 {{dz_1} \over {z_1}}  \, {{dz_2} \over {z_2}} 
\,\, \, \cdots \,\,\,  {{dz_n} \over {z_n}}
\,\cdot \,  {\cal A}\Bigl(x, \, z_1, \, z_2, \, \cdots, \, z_n \Bigr), 
\end{eqnarray}
where the subscript $\, C$ denotes the unit circle, and where $\,{\cal A}$ denotes 
an algebraic function of the $\,n$ variables, which (this is the crucial ingredient),
as a function of several variables $x$ and the $\, z_k$'s, has 
an {\em analytical} expansion at $(x , \, z_1, \, z_2, \, \cdots,  \, z_n) \, = \, \,$
$ (0 , \, 0, \, 0, \, \cdots,  \, 0)$:
\begin{eqnarray}
\label{form2}
\hspace{-0.6in}&&{\cal A}\Bigl(x, \, z_1, \, z_2, \, \cdots, \, z_n \Bigr)
\, \,\,  \,\, = \\
\hspace{-0.6in}&& \quad \quad 
\sum_{m \, = \, 0}^{\infty} \, \sum_{m_1 \, = \, 0}^{\infty} \, \sum_{m_2\, = \, 0}^{\infty} 
 \,\, \cdots \, \sum_{m_n\, = \, 0}^{\infty} 
 \,A_{m, \, m_1, \, m_2, \, \cdots, \, m_n}
\cdot  \, z_1^{m_1} \, z_2^{m_2} \,\,  \cdots \, \,z_n^{m_n} \, \cdot x^m.
\nonumber
\end{eqnarray}

Consequently, an extremely large set of $\, n$-fold integrals occurring in
theoretical physics (lattice statistical mechanics, enumerative combinatorics,
number theory, differential geometry, ...) can actually be seen to be
\emph{diagonals of rational functions}. These $\, n$-fold
integrals correspond to series expansions (in the variation parameter $\,x$)
that are {\em globally bounded} (can be written after one rescaling into
series with integer coefficients), and are solutions of {\em globally
nilpotent}~\cite{bo-bo-ha-ma-we-ze-09} linear differential operators.

Such a general $\, n$-fold integral is, thus, the diagonal of an algebraic
function (or of a rational function with twice more variables~\cite{Denef})
which is essentially the {\em integrand} of such $\, n$-fold integral. Furthermore,
such a general $\, n$-fold integral is solution of a 
(globally nilpotent~\cite{bo-bo-ha-ma-we-ze-09})
linear differential operator, that can be obtained exactly from the integrand,
using the creative telescoping method~\cite{Big}. \vskip .1cm

Finally, in the case of Calabi-Yau ODEs (see below), these functions
can be interpreted as periods of Calabi-Yau varieties, these 
algebraic varieties being essentially the
 integrand of such $\, n$-fold integrals.
The {\em integrand} is thus the {\em key ingredient} to wrap, in the same bag,
the {\em algebraic geometry viewpoint, the differential geometry viewpoint, and 
the analytic and arithmetic approaches} (series with integer coefficients).

\section{Calabi-Yau ODEs solutions and series
 with  binomials seen as diagonals}
\label{plan}

\subsection{Recalling Calabi-Yau ODEs}
\label{planb}

Calabi-Yau ODEs have been defined in~\cite{Almkvist} as order-four linear
differential ODEs that satisfy the following conditions: they are maximal
unipotent monodromy~\cite{Morrison,Batyrev} (MUM), they satisfy a ``Calabi-Yau
condition'' which amounts to imposing that the exterior squares of these
order-four operators are of order {\em five} (instead of the order six one
expects in the generic case), the series solution, analytic at $\, x=\, 0$, is
globally bounded (can be reduced to integer coefficients), the series of their
nome and Yukawa coupling are globally bounded.
In the literature, one finds 
also a cyclotomic condition on the monodromy at the point at
 $\, \infty$, $\, x \, = \, \,  \infty$, and/or
 the conifold\footnote[1]{The local exponents
are $0, 1, 1, 2$. For the cyclotomic condition on the
monodromy at $\, \infty$, see Proposition 3 in~\cite{TablesCalabi}.} 
character of one of the singularities~\cite{chen-yang-yui-08}.

Let us recall that a linear ODE has MUM ({\em maximal unipotent
monodromy}~\cite{CalabiYauIsing1,TablesCalabi}) if all the
exponents at (for instance) $x= \, 0$ are zero. In a hypergeometric framework
 the MUM condition amounts to restricting to
hypergeometric functions of the type
$\, _{n+1}F_n([a_1,\,a_2,\, \cdots \, a_n], \, [1, \, 1, \, \cdots \, 1], \, x)$,
since the indicial exponents at $\, x \, = 0$  are the solutions of
 $\rho \,(\rho+b_1-1) \cdots (\rho+b_n-1) \, = \, \,$
$\rho^{n+1} \, = \, \, 0$,
where the $b_j$ are the lower parameters which 
are here all equal to $\, 1$.

Let us consider a MUM order-four linear differential operator.
The four solutions $y_0, \,y_1, \,y_2, \,y_3$ of this order-four 
linear differential operator read:
\begin{eqnarray}
\hspace{-0.8in}&&y_0, \quad \quad \, \, \,\,\,\,
y_1 \, = \,\,\, \,\, y_0 \cdot \ln(x)\,\,   \,  +\,   \tilde{y}_1,
 \quad \quad  \, \, \,\,\,\,
y_2 \, = \,\,\, \, y_0 \cdot  {{\ln(x)^2} \over {2}} \,
  \,  +\, \tilde{y}_1\cdot \ln(x)
 \,\,  \,  + \,\,\, \,   \tilde{y}_2,
\nonumber \\
\hspace{-0.8in}&&y_3 \, \, = \,\,\, \,\, \, 
 y_0 \cdot  {{\ln(x)^3} \over {6}} \, \,\, \, 
 + \, \, \tilde{y}_1 \cdot  {{\ln(x)^2} \over {2}} \, \, \, \, 
+ \, \, \tilde{y}_2 \cdot  \ln(x) \, \, \, \, 
+ \,  \, \tilde{y}_3, 
 \nonumber
\end{eqnarray}
where $\, y_0$, $\,\tilde{y}_1$,
 $\,\tilde{y}_2$, $\,\tilde{y}_3$
 are  analytical at $\, x \, = \, \, 0$
(with also 
$\,  \tilde{y}_1(0) \, = \,   \tilde{y}_2(0) \, = \, \tilde{y}_3(0) \, = \,  0$).

The nome of this linear differential operator
reads:
\begin{eqnarray}
\label{nome}
q(x) \, \, \, = \,  \,\, \, \, 
\exp\Bigl({{y_1 } \over {y_0 }}  \Bigr)
 \, \, \, \,  = \, \,  \, \, \, \,
x \cdot \, \exp\Bigl({{\tilde{y}_1 } \over {y_0 }}  \Bigr). 
\end{eqnarray}

\vskip .1cm

Calabi-Yau ODEs have been defined as being MUM, thus having 
one solution analytical at $\, x \, = \, \, 0$. 
As far as Calabi-Yau ODEs are concerned, the fact that 
this solution analytical at $\, x \, = \, \, 0$ has an integral representation,
and, furthermore, an integral representation of the form (\ref{form})
together with (\ref{form2}),
is far from clear, even if one may have a ``Geometry-prejudice'' that 
this solution, analytical at $\, x \, = \, \, 0$, can be interpreted 
as a ``Period'' and ``Derived From Geometry''~\cite{Andre5,Andre6,Andre7}. 

Large tables of Calabi-Yau ODEs have been obtained by 
Almkvist et al.~\cite{TablesCalabi,Almkvist1,Almkvist2}. 
It is worth noting that the coefficients $\, A_n$ of the 
series corresponding to the solution analytical at $\, x \, = \, \, 0$,
are, most of the time, {\em nested sums of product of binomials},
 less frequently 
 nested sums of product of binomials and of harmonic
 numbers\footnote[5]{The generating function of Harmonic numbers is 
$\, H(x) \, = \, \,  \sum \, H_n \cdot x^n$
$ \, \, = \, \, \, -\, \ln(1-x)/(1-x)$.} $\, H_n$,
and, in rare cases, no ``closed formula'' is known for these coefficients.

Let us show, in the case of  $\, A_n$
coefficients being {\em nested sums of product of binomials}, that the 
solution of the Calabi-Yau ODE, analytical at $\, x \, = \, \, 0$,
which is by construction a series with integer coefficients, 
is {\em actually a diagonal of rational function}, and furthermore,
that this rational function can actually be easily built. 

\vskip .1cm 

\subsection{Calculating the rational function
 for nested product of binomials}
\label{theorem}

For pedagogical reasons we will just consider, here, 
a very simple 
example\footnote[3]{See Proposition 7.3.2 in~\cite{Batyrev}.} 
 of a series $\, {\cal S}(x)$, with integer coefficients, 
given by a sum of product of binomials
\begin{eqnarray}
\label{112}
\hspace{-0.7in}&&{\cal S}(x) 
\,\, \,\, =\, \,\, \, \,\,
\sum_{n=0}^{\infty} \, \sum_{k=0}^{n}
 \, {n\choose k}^3 \, \cdot x^n  \,\,  = \, \,\,  \,\,\,
 HeunG(-1/8, 1/4, 1, 1, 1, 1; \, -x)
 \\
\hspace{-0.7in}&&\quad \,\, \, = \, \,\,  \,\,\,
1\,\, \,\, +2\,x \,\,\, +10\,x^2 \,\, +56\,x^3 \, +346 \,x^4\,\,
 +2252\, x^5 \,\, +15184\, x^6 \,\,\,  + \, \,\, \cdots 
 \nonumber 
\end{eqnarray}
This is the generating function of sequence {\bf A} in Zagier's tables
 of binomial coefficients sums (see p.~354 in~\cite{Zagier}).

The calculations of this section can
straightforwardly (sometimes tediously) be generalised
to more complicated~\cite{Egorychev} nested sums of product of
 binomials\footnote[1]{Not necessarily corresponding to modular forms
as can be seen on (\ref{sol1}), (\ref{modcurve}).}. 
 
Finding that a series is a diagonal of a rational function
amounts to framing it into a residue form like (\ref{form}). In order to
achieve this, we write the binomial  $\, {n\choose k}$ as the residue
\begin{eqnarray}
{n\choose k} \,\, \,  = \, \, \, \,\, {{1} \over {2 \, i \, \pi}} \cdot \, 
\int_{C} {{(1\, +z)^n} \over {z^k}} \cdot {{dz } \over {z}},
\end{eqnarray}
and, thus, we can rewrite $\, {\cal S}(x)$ as
\begin{eqnarray}
\label{112bis}
\hspace{-0.95in}&& (2\, i\, \pi)^3 \cdot \, {\cal S}(x)
\, \, \,    =  \,\, \, \, 
\nonumber \\
\hspace{-0.95in}&& \, \,   =  \,\, \,  \, \,
\sum_{n=0}^{\infty} \int \int  \int  \,
\sum_{k=0}^{n} \, {{1} \over {(z_1\, z_2 \, z_3)^k}}
 \,  \cdot \, \Bigl(
 (1+z_1)\,  (1+z_2)\,  (1+z_3)\cdot \, x\Bigr)^n 
\,  \cdot \, {{dz_1\, dz_2 \, dz_3 } \over { z_1\, z_2 \, z_3}} 
\nonumber 
\end{eqnarray}
\begin{eqnarray}
\hspace{-0.95in}&& 
=  \,  \int \int \int  
\sum_{n=0}^{\infty} \, {{1 \, -\Bigl(1/(z_1\, z_2 \, z_3)\Bigr)^{(n+1)}} \over
  {1 \, -\Bigl(1/(z_1\, z_2 \, z_3)\Bigr)}}
\cdot \Bigl(
 (1+z_1)\,  (1+z_2)\,  (1+z_3)\cdot \, x\Bigr)^n 
\,  \cdot \, {{dz_1\, dz_2 \, dz_3 } \over { z_1\, z_2 \, z_3}}
\nonumber 
\end{eqnarray}
\begin{eqnarray}
\hspace{-0.95in}&& \quad 
 =  \,\, \,\,  \,  -\, \int \int \int  \,  
\sum_{n=0}^{\infty} \, {{z_1\, z_2 \, z_3 } \over
  {1 \, -\, z_1\, z_2 \, z_3}}
\cdot \Bigl(
 (1+z_1)\,  (1+z_2)\,  (1+z_3)\cdot \, x\Bigr)^n 
\,  \cdot \, {{dz_1\, dz_2 \, dz_3 } \over { z_1\, z_2 \, z_3}}
\nonumber \\
\hspace{-0.95in}&& \quad \quad 
\, \,\, \,\,  \,  + \, \, \int \int \int  
\sum_{n=0}^{\infty} \, {{1 } \over
  { 1 \, -\, z_1\, z_2 \, z_3}}
\cdot \Bigl(
 {{(1+z_1)\,  (1+z_2)\,  (1+z_3)\cdot  x} \over {z_1\, z_2 \, z_3 }}\Bigr)^n 
\,  \cdot \, {{dz_1\, dz_2 \, dz_3 } \over { z_1\, z_2 \, z_3}}
\nonumber 
\end{eqnarray}
\begin{eqnarray}
\hspace{-0.95in}&& \quad 
=  \,\, \,\,  \,  \, \int \int \int  \,  R(x; \, z_1, \, z_2,  \, z_3) \,
\cdot \, {{dz_1\, dz_2 \, dz_3 } \over { z_1\, z_2 \, z_3}},
 \qquad  \qquad  \qquad
\end{eqnarray}
where $\, R(x; \, z_1, \, z_2,  \, z_3)$ reads: 
\begin{eqnarray}
\hspace{-0.65in}{{z_1\, z_2 \, z_3 } \over {
 \big(1\, - \, x \cdot \,(1+z_1)(1+z_2)(1+z_3)\big)
\, \big( z_1\, z_2 \, z_3  \, - \, x \cdot \,(1+z_1)(1+z_2)(1+z_3)\big)}}.
 \nonumber 
\end{eqnarray}
{}From this last result one deduces immediately that (\ref{112})
is actually the diagonal of:
\begin{eqnarray}
\label{gooddiag}
\hspace{-0.9in}&&\quad {{ 1} \over {
 \big(1\,-z_0 \cdot \,(1+z_1)(1+z_2)(1+z_3)\big) \cdot 
\,\big(1\,-z_0\,z_1\, z_2\,z_3\,(1+z_1)(1+z_2)(1+z_3)\big)
}}.
\nonumber 
\end{eqnarray}

Note that, as a consequence of a combinatorial identity due 
to Strehl and Schmidt~\cite{Strehl,Schmidt,Zudilin},
  $\, {\cal S}(x)$ can also be written as 
\begin{eqnarray}
\label{112sec}
\hspace{-0.95in}&&\quad  {\cal S}(x) 
\,\, \, =\, \,\, \, \,
\sum_{n=0}^{\infty} \, \sum_{k=0}^{n}
 \, {n\choose k}^2 \, {2\, k\choose n} \, \cdot x^n 
\,\, \, =\, \,\, \, 
\sum_{n=0}^{\infty} \, \sum_{k=[n/2]}^{n}
 \, {n\choose k}^2 \, {2\, k\choose n} \, \cdot x^n.  
\end{eqnarray}
Calculations similar to  (\ref{112bis}) on this alternative binomial
representation (\ref{112sec}), enable to express
 (\ref{112}) as the diagonal of 
an alternative rational function:
\begin{eqnarray}
\hspace{-0.9in}&&\quad {{ 1} \over {
 \big(1\,\,  -z_0 \cdot \,(1+z_1)(1+z_2)(1+z_3)^2\big) \cdot
 \,\big(1\, \, -z_0\,z_1\, z_2 \cdot \,(1+z_1)(1+z_2) \big)
}}.
\end{eqnarray}

\vskip .1cm 

We thus see that we can actually {\em get explicitly}, from straightforward
calculations, the rational function (\ref{gooddiag}) for the Calabi-Yau-like
ODEs (occurring from {\em differential geometry} or {\em enumerative
combinatorics}) when series with nested sums of binomials take place, and,
more generally, for enumerative combinatorics problems (related or not to
Calabi-Yau manifolds) where series with {\em nested sums of binomials} take
place.

These effective calculations are actually algorithmic, and guarantee to
obtain an {\em explicit expression} for the rational function
(\ref{gooddiag}). However the rational function is far from being unique, and
worse, the number of variables, the rational function depends on, is far from
being the smallest possible number. Finding the ``minimal'' rational function
(whatever the meaning of ``minimal'' may be) is a very difficult problem.
\ref{minimalrational} provides a non-trivial illustration of this fact with
explicit calculations on the well-known Ap\'ery series and its rewriting due
to Strehl and Schmidt~\cite{Strehl,Schmidt,Zudilin}. We see in a crystal clear
way in \ref{minimalrational} that, when a given function is a diagonal of a
rational function, the rational function is far from being unique, the
``simplest'' representation (minimal number of variables, lowest degree
polynomials, ...) being hard to find. Similar
 computations\footnote[1]{These results are given 
in section (5.1) of~\cite{Big}.} 
show that the
generating function of sequence {\bf B} and {\bf E} in Zagier's
list~\cite{Zagier} are both diagonals of rationals function in four variables.

\vskip .1cm 

All these calculations can systematically be performed on any series defined
by {\em nested sums of product of binomials}. We have performed such
calculations on a large number of the series corresponding to the list of
Almkvist et al~\cite{TablesCalabi}, that are given by such {\em nested sums of
product of binomials}.

\section{Comments and speculations}
\label{commspec}

\subsection{A theorem of~\cite{Christol}}
\label{subtheorem}

In~\cite{Christol} (page 61 Theorem 12, see also Proposition 7 
in page 50 of~\cite{Christol369}) it is proved 
that any power series with an {\em integral representation} 
 and of {\em maximal weight} for  the corresponding
 {\em Picard-Fuchs linear differential equation}
is the {\em diagonal of a rational function}
 and, in particular, is {\em globally bounded}.

The technical nature of the original papers is such that
the result itself is difficult to find. This paragraph is devoted to explain,
 in down-to-earth terms, the somewhat esoteric expressions used
 in its wording, and to explain what it means on explicit examples.
As the original proof is very obfuscated its principle is 
 sketched  in~\cite{Big}.

Disappointingly, when applied to a hypergeometric~$\, _{n+1}F_n$,
this result becomes somewhat trivial. More precisely, the hypergeometric
function is of maximal weight if and only if $\, b_j \, = \, 1$ for all $j$
(there is only $n!$'s in the denominator of coefficients). In that case it is
obviously the Hadamard product of algebraic functions,
 therefore diagonal of a rational function:
 \begin{eqnarray}
\label{hyperdiag} \hspace{-0.6in}&&\quad _{n}F_{n-1}([\alpha_1, \,\alpha_2, \,
\cdots, \, \alpha_n], \,[1, \, 1, \, \cdots \, 1], \, x) \\
 \hspace{-0.6in}&&
\quad \quad \quad \quad \quad \quad \, \, = \, \, \, \, \, \, (1\,
-x)^{-\alpha_1} \, \star \, (1\, -x)^{-\alpha_2} \, \, \cdots \, \, \star \,
(1\, -x)^{-\alpha_n}. \nonumber 
\end{eqnarray}

Therefore, we now have (at least) three sets
 of problems yielding diagonal of rational 
functions: the $\, n$-fold integrals of the form 
(\ref{form}) with (\ref{form2}),
the Picard-Fuchs linear ODEs with solution of maximal monodromy weight
 and, finally, the problems of
enumerative combinatorics where nested sums
 of products of binomials take place. 

{\em Diagonal of rational functions, thus, occur in a
 quite large set of problems
of theoretical physics}. At first sight, one can see the frequent 
appearance of diagonals 
of rational functions in physics just as a mathematical 
curiousity\footnote[1]{In 1944 the occurrence
of elliptic functions in Onsager's solution of the Ising model was also 
seen as a mathematical curiosity ...}, and be surprised that, for instance,
so many series in physics are, modulo a prime, algebraic functions.
Being diagonal of rational functions is not just as a mathematical 
curiousity: it corresponds (see next section) to G-operators, and 
their {\em rational number exponents}, and can be seen as a first step 
to modularity properties (see sections below) 
in some work-in-progress integrability.

\subsection{A conjecture of~\cite{Christol}}
\label{conjec}

The diagonal of a rational function is 
 globally bounded  (i.e. it has non zero radius
 of convergence and
 integer coefficients up to one rescaling) and  {\em D-finite}
 (i.e. solution of a linear differential equation with polynomial 
coefficients)\footnote[2]{The series expansion of the susceptibility
of the isotropic 2-D Ising model can be recast into a series with {\em integer
coefficients} (see~\cite{Khi6,bo-gu-ha-je-ma-ni-ze-08,High,bernie2010}),
 but it {\em cannot be the diagonal of rational functions}
since the full susceptibility is
 {\em not a D-finite function}~\cite{bernie2010}.}.

The converse statement is the conjecture in~\cite{Christol} saying
that {\em any D-finite, globally bounded series is necessarily the diagonal of
a rational function}.

A remarkable result of Chudnovski's (\cite{DGS} Chapter VIII) 
asserts that the minimal linear differential operator of a $G$-function 
(and in particular of a
D-finite globally bounded series) is a $\, G$-operator (i.e. at least, 
a globally nilpotent operator)~\cite{bo-bo-ha-ma-we-ze-09,Andre5,Andre6}. The
 conjecture in~\cite{Christol} amounts to saying something more: 
if the solution of this globally nilpotent
linear differential operator is, not only a \, $G$-series, but a {\em globally
bounded series, then it is the diagonal of a rational function}.

Conversely the solution, analytical at $\,0$, of a globally nilpotent linear
differential operator is necessarily a $\, G$-function~\cite{Andre5,Andre6}.
Moreover, a ``classical'' conjecture, with numerous avatars, claims that any
$G$-function comes from geometry i.e. roughly speaking, it has an integral
representation\footnote[3]{Bombieri-Dwork conjecture
 see for instance~\cite{Andre6}.}.

To test the validity of the
conjecture of~\cite{Christol} we look for counter-examples not
contradicting classical conjectures. For instance,
 we search D-finite power series with
{\em integer coefficients} which are {\em not algebraic} but have an integral
representation and are not of maximal weight for the corresponding
Picard-Fuchs linear ODE.

As a first step let us limit ourselves to hypergeometric functions 
$\,_{n+1}F_n$. The monodromy weight $\, W$ is exactly 
the number of $\,1$ among the $\, b_i$.

When $\, _{n+1} F_n\, $ is globally bounded and has no integer parameters 
$ \,b_i$ ($W=\, 0$), its minimal ODE has a $p$-curvature zero for almost 
all primes $\, p$. However, a Grothendieck conjecture, 
proved for $\, _3F_2\, $
in~\cite{Hodge}, and generalised to $ \, _{n+1} F_n\, $ in~\cite{BeHe89},
asserts that, under these circumstances, the hypergeometric function is 
{\em algebraic}. We display in \ref{ratiooffac} a set of 
$ \, _{n+1} F_n\, $ hypergeometric functions which yield, naturally, 
series with {\em integer coefficients}, many of them corresponding to 
such {\em algebraic hypergeometric functions}. Even if such examples 
are quite non-trivial, the purpose of our paper is to focus on 
{\em transcendental} (non algebraic) functions.

So we are looking for  {\em  globally bounded} 
hypergeometric functions satisfying $\quad$
$ \, 1 \, \leq  \, W \, \leq \,  n-1$.
In general such hypergeometric functions are $\, G$-series but are very far 
from being globally bounded. The hypergeometric world extends largely 
outside the  world of diagonal of rational functions.

Such an example in the first case $\, n=\, 2$, $\, W=\, 1 \, $ 
was given in~\cite{Christol}:
\begin{eqnarray}
\label{contre1}
\hspace{-0.8in}&&\quad 
_3F_2\left(\left[{{1} \over {9}},\, {{4} \over {9}},\,{{5} \over {9}}\right],
\, \left[{{1} \over {3}},\, 1\right]; \,  3^6\, x\right)
\, \,\, \,  \, = \,\,  \, \, \, \,\, 
1\,\,\,\, +60\,x\, \, +20475\,x^2\, \, +9373650\,x^3 
\nonumber \\
\hspace{-0.8in}&&\,\quad \, \, \,  +4881796920\,x^4\,\, 
+2734407111744\,x^5\,\, 
+1605040007778900\,x^6 \, \,  \,\,+ \,\, \,  \cdots  
\end{eqnarray}
The integer coefficients read with the rising factorial
 (or Pochhammer) symbol
\begin{eqnarray}
{{ (1/9)_n \cdot (4/9)_n \cdot (5/9)_n } \over {
(1/3)_n \cdot (1)_n \cdot n!}} \cdot \,  3^{6n}
\, \,  \,= \, \, \,\, \, {{\rho(n)} \over {\rho(0)}}, 
\end{eqnarray}
where:
\begin{eqnarray}
\label{far1}
\rho(n)\,\,\, = \, \, \,  \, \, 
 {{ \Gamma(1/9+n)\,\Gamma(4/9+n) \, \Gamma(5/9+n)} \over {
\Gamma(1/3+n)\, \Gamma(1+n) \, \Gamma(1+n)}} \cdot \,  3^{6n}.
\end{eqnarray}
Note that, at first sight, it is 
{\em far from clear}\footnote[3]{In contrast with cases 
where binomial (and thus integers) 
expressions take place.} on (\ref{far1}), or 
 on the simple recursion
on the $\, \rho(n)$ coefficients (with the initial 
value $\, \rho(0)\, = \, \, 1$)
\begin{eqnarray}
\label{ratio1}
{{\rho(n+1)} \over {\rho(n)}}\, \, \,= \, \, \, \,  \,
3 \cdot \, 
{{(1 + 9 n) \, (4 + 9 n) \, (5 + 9 n)  } \over {(1 + 3 n) (1 + n)^2 }}, 
\end{eqnarray}
to see that the $\, \rho(n)$'s are actually integers. A 
sketch of the (quite arithmetic) proof that the  $\, \rho(n)$'s 
are actually integers, is given in \ref{proof}.

Because of the $\, 1/3$ in the right (lower) parameters of (\ref{contre1}),
the hypergeometric function (\ref{contre1}) is not an obvious
 Hadamard product 
of algebraic functions (and thus a diagonal of a rational function), and 
one can see that it is not an algebraic hypergeometric function
either by calculating its $\, p$-curvature and finding that
 it is not zero~\cite{Andre7} (see also~\cite{BeHe89,JAW}). 
 Proving that an algebraic function  is
the diagonal of a rational function and proving that a solution 
of maximal weight for a Picard-Fuchs equation is
the diagonal of a rational function use two entirely 
distinct ways. The hope is to  combine 
both techniques to conclude in the intermediate situation.

This example remained for twenty years, the only ``blind spot''
of the conjecture in~\cite{Christol}. We have recently found many other 
 $\, _3F_2 \, $ examples\footnote[2]{$_2F_1$ cases 
are straightforward, and cannot provide counterexamples 
to  conjecture in~\cite{Christol}.}, 
such that their series expansions have {\em integer coefficients} 
but are not obviously diagonals of rational functions.
Some of these new hypergeometric examples\footnote[1]{See also~\cite{Big}.}
read for instance:
\begin{eqnarray}
\hspace{-0.9in}&& \quad  _3F_2\Bigl([{{1} \over {9}},
 \,{{2} \over {9}}, \, {{7} \over {9}}, ], 
\, [{{2} \over {3}}, \, 1], \, \, 3^6 \, x\Bigr) 
\,\,  \,\, = \,\, \,\, \,  \, \, 1\,\,\, \,  
+21\, x\, \,\, +5544\, x^2\, +2194500\, x^3\,
\nonumber \\
\hspace{-0.9in}&&  \quad  \quad  \quad \quad  \quad
  +1032711750\, x^4\, \, 
 +535163031270\, x^5\,\,  + 294927297193620\, x^6
\nonumber \\
\hspace{-0.9in}&&   \quad \quad\,\quad \quad  \quad  \quad  \quad 
 +169625328357359160\, x^7 \,  +100668944872954458000\, x^8\, 
\,\, \, \,\, \,+ \, \cdots \nonumber 
\end{eqnarray}
or:
\begin{eqnarray}
\hspace{-0.9in}&&  \, \, \,  _3F_2\Bigl([{{1} \over {7}},
 \,{{2} \over {7}}, \, {{4} \over {7}}, ], 
\, [{{1} \over {2}}, \, 1], \, \, 7^4 \, x\Bigr), \qquad 
 _3F_2\Bigl([{{1} \over {11}},
 \,{{2} \over {11}}, \, {{6} \over {11}}, ], 
\, [{{1} \over {2}}, \, 1], \, \, 11^4 \, x\Bigr).
\end{eqnarray}

Unfortunately these hypergeometric 
examples are on the same ``frustrating footing'' as Christol's 
example (\ref{contre1}): we are not able to show that 
one of them is actually a diagonal of a rational function, or, 
conversely, to show that one of them cannot be the 
diagonal of a rational function.

\section{Integrality versus modularity: learning by examples}
\label{versus}

A large number of examples of integrality of  series-solutions
comes from modular forms. Let us just display two such modular 
forms associated with HeunG functions of the form 
$\, HeunG(a, q, 1, 1, 1, 1; x)$.  Many more similar examples 
can be found in~\cite{Big}.

\subsection{First modular form example}
\label{ex1}

 One can, for instance, rewrite the example (\ref{112}) 
of subsection (\ref{theorem}),  
namely $\, HeunG(-1/8, 1/4, 1, 1, 1, 1; -x)$, 
as a hypergeometric function
 with {\em two rational pullbacks}:
\begin{eqnarray}
\label{sol1}
\hspace{-0.6in}&&HeunG(-1/8, 1/4, 1, 1, 1, 1; -x)\
\, \, \, \, \, = \, \, \, \,  \,
\, \,  \sum_{n=0}^{\infty} \, \sum_{k=0}^{n} \, {n\choose k}^3 \, x^n
  \, \, \, \, \,
 \\
\hspace{-0.6in}&&\quad \quad \,  
 = \,\, \,  \,\Bigl((1\, +4\, x) \cdot
 (1\, +228\, x \,+48\, x^2\, +\, 64\, x^3 ) \Bigr)^{-1/4} 
\nonumber \\
\hspace{-0.6in}&&\quad \quad \quad \quad \quad \quad \times \, 
 _2F_1\Bigl([{{1} \over {12}}, \, {{5} \over {12}}], \, [1]; \, {{
1728 \cdot  (1\, -8 \, x)^6 \cdot (1\, +x)^3 \cdot x } \over {
(1\, +228\, x \,+48\, x^2\, +\, 64\, x^3)^3 \cdot 
(1\, +4\, x)^3 }}   \Bigr)
 \nonumber 
\end{eqnarray}
\begin{eqnarray}
\hspace{-0.6in}&&\quad \quad 
\,  = \, \, \,  \, 
\Bigl((1\, -2 \, x) \cdot (1\, -6 \,x\,+228 \,x^2\,-8\,x^3 ) \Bigr)^{-1/4} 
\nonumber \\
\hspace{-0.6in}&&\quad \quad \quad \quad \quad \quad   \times \, 
 _2F_1\Bigl([{{1} \over {12}}, \, {{5} \over {12}}], \, [1]; \, {{
1728 \cdot  (1\, -8 \, x)^3 \cdot (1\, +x)^6 \cdot x^2 } \over {
 (1\, -6 \,x\,+228 \,x^2\,-8\,x^3)^3  \cdot  (1\, -2 \, x)^3}}   \Bigr).
\nonumber 
\end{eqnarray}
The relation between the two pullbacks,
that are related by the ``Atkin'' involution\footnote[5]{In previous
 papers~\cite{Renorm,CalabiYauIsing1}, with some abuse of language, we
called such an involution an \emph{Atkin-Lehner involution}. In fact
 this terminology is commonly used in the mathematical community 
for an involution $ \, \tau \, \rightarrow \, -N/\tau$,
 on~$\tau$, the ratio of periods,
 and {\em not} for our $\, x$-involution. 
This is why we switch to the wording 
\emph{``Atkin" involution}.}
 $\, x \, \leftrightarrow \, -1/8/x$,
 gives the modular curve:
\begin{eqnarray}
\label{modcurve}
\hspace{-0.8in}&&1953125\, y^3 \, z^3 \, \,  \,  \, 
 -187500 \, y^2 \, z^2 \cdot (y+z) \,
  \,  +375 \, y \, z \cdot (16\, z^2\, -4027\, y\, z\, +16\, y^2)
\nonumber \\
\hspace{-0.8in}&& \qquad \quad \quad    -64 \cdot \, (z+y) \cdot
 (y^2+z^2\, +1487 \, y\, z)
\,   \, \,  \,  +110592 \cdot \, y\, z 
\, \, \,\, \, \,   = \, \,\, \, \,   \, \, 0. 
\end{eqnarray}

Series (\ref{sol1}) is solution of the  (exactly)
 {\em self adjoint} linear differential operator
$\, \Omega$ where ($\theta \, = \, \, x \cdot D_x$):
\begin{eqnarray}
\label{aux1}
\hspace{-0.2in}&&x \cdot \Omega\,\, \,\,\, = \, \, \, \, \, \, \, 
\theta^2 \,\, \,  \,  \,  -\, x \cdot 
(7\, \theta^2 \, +7 \,\theta \, +2) 
\,  \, \, \,  -8 \, x^2 \cdot   (\theta \, +\, 1)^2.
\end{eqnarray}

\vskip .1cm 

\subsection{Second modular form example}
\label{ex1sec}

The integrality of  series-solutions
 can be quite non-trivial like 
the solution of the 
 Ap\'ery-like operator
\begin{eqnarray}
\hspace{-0.7in}&&\Omega \,\,  \, = \, \, \, \,\, \, 
x \cdot (1 \, -11 \, x \, -x^2) \cdot D_x^2 \,\,  \,\,\,
 + (1 \, -22 \, x \, -3 \, x^2) \cdot D_x\,\,
\, \, -(x+3), \\
\hspace{-0.7in}&& \hbox{or:} \qquad \quad x \cdot \, \Omega
 \,\,  \, = \, \, \,\, \,\,  \,  
\theta ^2 \,\,\,  \,  -x \cdot \, (11 \, \theta^2 \, +11 \, \theta\, +\, 3)
\,\, \,    -x^2 \cdot \, (\theta\, +1)^2, \nonumber 
\end{eqnarray}
which can be written  as a HeunG
function. 
This (at first sight involved) HeunG
function reads: 
\begin{eqnarray}
\label{11253}
\hspace{-0.9in}&&HeunG\Bigl(-\, {{123} \over {2}} \,
 +{{55} \over {2}} \cdot 5^{1/2},\,
 -\, {{33} \over {2}} \, +{{15} \over {2}} \cdot 5^{1/2},1,1,1,1; \,\,
 \Bigl({{11-5^{3/2}} \over {2}} \Bigr)\, \cdot \, x\Bigr) 
\nonumber 
\end{eqnarray}
\begin{eqnarray}
\hspace{-0.9in}&& \quad \, \, = \, \, \,  \,  \, \, 
\sum_{n\, = \, 0}^{\infty} \,
 \sum_{k\, = \, 0}^{n} \, {n\choose k}^2 \, {n+k\choose k} \cdot \, x^n 
 \, \, = \, \, \, \, \, \, \,\, 
1 \,\,\,\,  \, +3 \cdot x \, \,+19 \cdot x^2 \,\, +147 \cdot x^3 
 \,\, \,  + \,\,  \cdots 
 \nonumber 
\end{eqnarray}
but  {\em actually corresponds  to a modular form}, which
 can be written in
two different ways using  {\em two pullbacks}:
\begin{eqnarray}
\label{cinqun}
\hspace{-0.6in}&&(x^4\,+12\,x^3+14\, x^2\,-12\,x\,+1)^{-1/4} \,
\nonumber \\
\hspace{-0.6in}&&\qquad \quad \quad \, \,\times  \,
   _2F_1\Bigl([{{1} \over {12}}, \,{{5} \over {12}}], \, [1]; \,
 {{1728 \cdot x^5 \cdot (1 \, -11\, x \, -x^2) } \over {
(x^4\,+12\, x^3\,+14\,x^2\,-12\,x\,+1)^3 }}\Bigr)
\nonumber \\
\hspace{-0.6in}&&\,\quad  = \, \, \, \, \,\,\,  
(1 \, + \, 228 \, x \, +\, 494 \, x^2 \, -228 \, x^3 \,
 + \, x^4)^{-1/4} \,  \\
\hspace{-0.6in}&&\qquad \quad \quad \quad  \, 
\times  \,  _2F_1\Bigl([{{1} \over {12}}, \,{{5} \over {12}}], \, [1]; \,
 {{1728 \cdot x \cdot (1 \, -11\, x \, -x^2)^5 } \over {
(1 \, + \, 228 \, x \, +\, 494 \, x^2 \, -228 \, x^3 \, + \, x^4)^3 }}\Bigr).
 \nonumber 
\end{eqnarray}

\vskip .1cm

Modular form examples of series with integer coefficients 
displayed in \ref{modularapp}, 
 correspond to lattice Green functions~\cite{Guttmann}.
Therefore, they have {\em $\, n$-fold integral 
representations}\footnote[1]{In contrast the modular form examples displayed 
in Appendix H of~\cite{Big} correspond to 
differential geometry examples discovered by 
Golyshev and Stienstra~\cite{Golyshev}, 
where no $\, n$-fold integral representation is available at first sight.},
and, {\em after} section (\ref{gener}), can be seen 
to {\em be diagonals of rational functions}.

\vskip .1cm

\section{Integrality versus modularity}
\label{learning}
\vskip .1cm

\subsection{Diffeomorphisms of unity pullbacks}
\label{diffeofunit}

Let us consider a first simple example of a hypergeometric function
which is solution of a Calabi-Yau ODE, and which occurred, at least
two times in the study of the Ising 
susceptibility $\, n$-fold integrals~\cite{CalabiYauIsing1,CalabiYauIsing}
$\, \chi^{(n)}$ and $\, \chi_d^{(n)}$, namely 
$_4F_3([1/2,\,1/2,\,1/2,\,1/2 ], \, [1, \, 1, \, 1]; 256 \,x)$, where 
we perform a (diffeomorphism of unity) pullback:
\begin{eqnarray}
\label{diffeo}
\hspace{-0.9in}&&_4F_3\Bigl([{{1} \over {2}},\,{{1} \over {2}},
\,{{1} \over {2}},\,{{1} \over {2}}], 
\, [1, \, 1, \, 1];  \, \,  \,{{256 \,\, x} \over {
1 \,+c_1\,x \, +c_2\, x^2+ \, \cdots }}\Bigr)
\, \, \, \,= \,\,  \,\,\,\, \,   1 \,\,\,\,  +16\cdot \, x \,
    \\
\hspace{-0.9in}&&\quad \quad  \quad \,  \, \, \, +16 \cdot (81-\,c_1)\cdot x^2 
\,\, \,  +16 \cdot (10000 \,+ c_1^2 \, -\,c_2 \, -162\, c_1)\cdot x^3
\,\,\,\,  \,  + \,\,  \cdots 
\nonumber 
\end{eqnarray}
If the pull-back in (\ref{diffeo}) is such that the
coefficients $\, c_n$, at its denominator, are integers, one finds that the 
series expansion is actually a series with integer coefficients, for {\em every
such pullback} (i.e. for every integer coefficients $\, c_n$).
Furthermore, a straightforward calculation of the corresponding nome $\, q(x)$ 
and its compositional inverse (mirror map) $\, x(q)$, 
{\em also yields series with integer coefficients}:
\begin{eqnarray}
\label{q}
\hspace{-0.7in}&& q(x) \, \,= \,\,\, \,  \,\,\,x\, \,  \,  \,
 +(64-c_1)\cdot x^2 \, \, 
+(c_1^2\,+7072 \, -c_2\,-128\,c_1)\cdot x^3 \, \, \, \, + \, \cdots, \\
\hspace{-0.7in}&&x(q) \, \,  = \,\, \,  \, \,\,   q \,\, \,\, \,  
 +(c_1 \, -64)\cdot q^2 \,\, 
 +(c_1^2 \, +1120 \, +c_2 \, -128\, c_1)\cdot q^3 \, \, \, \,  + \, \cdots, 
\end{eqnarray}
when its Yukawa coupling~\cite{CalabiYauIsing1}, seen
 as a function of the nome $\, q$,
 $\, K(q)$ is also a series with integer coefficients
and is {\em independent of the pullback}:
\begin{eqnarray}
\label{Yuku}
\hspace{-0.4in}K(q) \, \,= \,\, \, \,\, \,1 \,\, \,\, \, + \, 32 \cdot \,q\,  \,
 +  4896 \cdot \, q^2 \,\, +702464 \cdot \, q^3 \, \,\,\, \,   + \, \cdots 
\end{eqnarray}

This independence of the Yukawa coupling with regards to  pullbacks,
is a known property, and has been proven in~\cite{Almkvist}, for
any pullback of the diffeomorphism of unity 
form $\,\, p(x) \, = \, \, \, x \, \, + \, \cdots$

\vskip .1cm

Seeking for Calabi-Yau ODEs, Almkvist et al. 
have obtained~\cite{TablesCalabi} a quite large list
 of fourth order ODEs, which are MUM
by definition and have, by construction, the {\em integrality} for 
the solution-series  analytic at $\, x \, = \, \, 0$. Looking
 at the Yukawa coupling of these ODEs is a way to define 
{\em equivalence classes up to pullbacks} of ODEs sharing the 
same Yukawa coupling. This ``wraps in the same bag'' all the linear ODEs that 
are the same {\em up to pullbacks}. Let us recall how difficult it is to see
if a given Calabi-Yau ODE has, up to operator equivalence, and 
up to pullback, a hypergeometric function 
solution~\cite{CalabiYauIsing1,CalabiYauIsing}, because
finding the pullback is extremely
 difficult~\cite{CalabiYauIsing1,CalabiYauIsing}. We may
 have, for the Ising model,
some $\, _{n+1}F_n$ hypergeometric function
 prejudice~\cite{CalabiYauIsing1,CalabiYauIsing}:
it is, then, important to have an invariant that is independent of 
this pullback that we cannot find most of the time.

\vskip .1cm 

Finally, let us remark that the Yukawa coupling is {\em not preserved by 
the operator equivalence}. Two linear differential operators, that are 
 homomorphic, {\em do not necessarily have the same Yukawa
 coupling} (see \ref{Yukawaratio}).

\vskip .1cm

\subsection{Yukawa couplings in terms of determinants}
\label{Yukdet}

Another way to understand this fundamental 
{\em pullback invariance}, amounts to rewriting
 the Yukawa coupling~\cite{Almkvist,mirror},
not from the definition usually given 
in the literature (second derivative with
respect to the ratio of periods), but in terms 
of determinants of solutions (Wronskians, ...) that naturally 
present nice covariance properties with respect 
to pullback transformations (see \ref{Yukawaratio}).

We have the alternative definition for the 
{\em Yukawa coupling} given in \ref{Yukawaratio}: 
\begin{eqnarray}
\label{Yukawa}
K(q) \,\,\, = \, \, \,\,\, 
 \Bigl( q \cdot {{d} \over {dq }} \Bigr)^2
 \Bigl(  {{y_2} \over {y_0}}\Bigr)
 \, \,\,\, = \, \,\,\,\, \, 
 {{W_1^3 \cdot W_3 } \over {W_2^3 }}, 
\end{eqnarray}
where the determinantal variables $\, W_m$'s 
are  determinants built from the four solutions 
of the MUM differential operator. 
This alternative definition, in terms of these $\, W_m$'s, 
enables to understand the {\em remarkable invariance 
of the Yukawa coupling by pullback 
transformations}~\cite{CalabiYauIsing}. 
These determinantal variables $\, W_m$ 
quite naturally, and canonically, yield to introduce another 
 ``Yukawa coupling'' (which, in fact, 
{\em corresponds to the Yukawa coupling
of the adjoint operator} (see \ref{Kstar})). This
 ``adjoint Yukawa coupling''
 is {\em also invariant by pullbacks}. It 
has, for the previous example, the following series 
expansion with integer coefficients:
\begin{eqnarray}
\label{Yukstar}
\hspace{-0.4in}K^{\star}(q) \,\,\, = \, \,\, \, \,\, \,
1 \,\,\,\,  + \, 32 \cdot \,q\,  
\, +  4896 \cdot \, q^2 \,\, +702464 \cdot \, q^3
 \, \,\,\, \,\, + \,\, \cdots 
\end{eqnarray}
which actually identifies with (\ref{Yuku}). 
The equality of the Yukawa coupling for this order-four
 operator, and for its 
(formal) adjoint  operator, is a straightforward 
consequence of the fact
that the order-four operator annihilating 
$ \, _4F_3([{{1} \over {2}},\,{{1} \over {2}},
\,{{1} \over {2}},\,{{1} \over {2}}], 
\, [1, \, 1, \, 1];  \, \, 256 \,x)$
is exactly {\em self-adjoint}, and, more generally, 
 of the fact that the order-four operator, 
annihilating (\ref{diffeo}),
is conjugated to its adjoint by a simple function.

\vskip .1cm

\subsection{Modularity}
\label{modu}

This example, with its corresponding relations 
(\ref{diffeo}), (\ref{q}), (\ref{Yuku}), (\ref{Yukstar})
 may suggest a quite wrong prejudice 
that the {\em integrality of the solution} of an 
order-four linear differential operator
automatically yields to the integrality of the
 nome, mirror map and Yukawa
coupling, that we will call,
 for short, ``{\em modularity}''. This is {\em far from being
the case}, as can be seen, for instance, in the following interesting
example, where the nome and Yukawa coupling $\, K(q)$ 
{\em do not correspond to globally bounded series}, when the $\, _4F_3$
solution of the order-four operator as well as the 
Yukawa coupling {\em seen as a function of $\, x$},  $\, K(x)$,
are, actually, both {\em series with integer coefficients}. 

\vskip .1cm

Let us consider the following $\, _4F_3$ hypergeometric function
which is clearly a Hadamard product of algebraic functions
 and, thus, the diagonal of a rational function:
\begin{eqnarray}
\label{saoud}
\hspace{-0.8in}&&\quad _4F_3\Bigl([{{1} \over {2}},\, {{1} \over {3}},
 \,{{1} \over {4}},\, {{3} \over {4}}],
\, [1,\,1,\,1 ]; \, x\Bigr) 
\,\, \,  \, 
\nonumber   \\
\hspace{-0.8in}&&\quad \qquad \quad \,  \,= \, \, \,\,  \, 
(1\,-x)^{-1/3} \,\star \,  (1 \,-x)^{-1/2} \,
 \star \, (1 \,-x)^{-1/4} \,\star \,  (1 \,-x)^{-3/4}
\nonumber  \\
\hspace{-0.8in}&&\quad \qquad \quad  \,  \,= \, \, \,  \,\, \, 
\Diag\Bigl( 
(1-z_1)^{-1/3} \, (1-z_2)^{-1/2} \, (1-z_3)^{-1/4} \, (1-z_4)^{-3/4}
\Bigr) , \nonumber
\end{eqnarray}
It is therefore globally bounded:
\begin{eqnarray}
\label{saoud}
\hspace{-0.9in}&&_4F_3([{{1} \over {2}},\, {{1} \over {3}}, \,
{{1} \over {4}},\, {{3} \over {4}}],
\, [1,\,1,\,1 ]; \, 2304\, x) 
\,   \, \,  \,  \,= \,\, \, \,\, \,  \,1 \, \,\,  \, +72\, x\,\, 
+45360\, x^2\,\, +46569600\, x^3\,
 \nonumber \\
\hspace{-0.9in}&&\qquad \quad \qquad  +59594535000\, x^4 \,\,
 +86482063571904\, x^5\,\,\,\, \,  +  \,\cdots 
\end{eqnarray}

Its Yukawa coupling, seen as a function of $\, x$,  is actually a  
{\em series with integer coefficients} in $\, x$:
\begin{eqnarray}
\label{Ksaoudx}
\hspace{-0.9in}&&K(x) \,\, = \, \,\, \,\,\,\, \,  
1\,\, \,\, +480\,x\,\, +872496\,{x}^{2}\, \,
+1728211968 \,{x}^{3}\,\, +3566216754432\,{x}^{4}\, 
\nonumber  \\
\hspace{-0.9in}&& \quad \quad  \quad \,\qquad  
 +7536580798814208 \,{x}^{5}\,\,
 +16177041308360579328 \,{x}^{6}\,\, \,\,   + \, \, \cdots 
\end{eqnarray}
However, do note that the series, in term of the nome, is 
{\em not globally bounded}:
\begin{eqnarray}
\label{Ksaoudq}
\hspace{-0.9in}&&K(q) \,\,\, = \,\, \,\,\,\,\, 
 \, 1\,\,\, \, +480\,\, q\,\,\, +653616\,  \,{q}^{2}\,\,
 +942915456\,  \,{q}^{3}\,\, +1408019875200 \, \,{q}^{4}
\,\,\,  + \,\,  \, \cdots \nonumber \\
\hspace{-0.9in}&&  \quad \quad \quad \quad \qquad  
+ \, 571436303929319146711343817202689132288 
\,  \,{{ \, \, q^{12}} \over {11}}
\, \,\, \,+\, \, \,   \cdots 
\end{eqnarray}

In fact, the nome  $ \, q(x)$, and the mirror map $\, x(q)$, 
 are {\em also not globally bounded}.
Note that in this example, the non integrality appears at order twelve 
(for $ \, x(q)$, $ \, q(x)$ and $ \, K(q)$).
If the prime 11 in the denominator in (\ref{Ksaoudq})
was the only one, one could recast the series into 
a series with integer coefficients introducing
 another rescaling
 $2304 \,  x \, \rightarrow \,11 \times 2304 \, \,x$.
But, in fact, we do see the appearance of an {\em infinite number of 
other primes} at higher 
orders  denominators in $\, x(q)$, $ \, q(x)$ and $\, K(q)$.

\vskip .1cm
We do not have modularity because we do not have (up to rescaling) the 
nome integrality: the nome series is not globally bounded.

\vskip .1cm

\subsection{Order-two differential operators $\, \omega_n$ 
associated with modular forms}
\label{Hadomegan}
After Maier~\cite{Maier1} let us underline that
modular forms can be written as hypergeometric 
functions with {\em two different pullbacks}, and, consequently,
one can associate order-two differential operators  
to these modular forms.

Let us consider the two order-two operators 
\begin{eqnarray}
\label{tau2}
\hspace{-0.3in}&& \omega_2 \,\, = \,\, \,\, \,  \,\,\,
D_x^{2} \, \,\,\, \,
 +{\frac { (96\,x+1) }{(64\,x+1)  \cdot \,x }} \, \cdot D_x \, \,\,
+ \,{\frac {4}{(64\,x+1)\, x }},
\\
\label{tau3}
\hspace{-0.3in}&&\omega_3 \,\, = \,\, \, \,\, \,\,\,
D_x^{2} \, \,\, \, \,
+{\frac { (45\,x+1) }{ (27\,x+1) \cdot \, x}} \cdot D_x
\, \,\,\, +\,{\frac {3}{ \left( 27\,x+1 \right) x}},
\end{eqnarray}
which are associated with two modular forms corresponding, on their associated
nomes $\,q$, to the transformations $\, q \, \rightarrow \, q^2$ and 
$\, q \, \rightarrow \, q^3$ respectively (multiplication 
of $\, \tau$, the ratio of their periods by $\, 2$ and $\, 3$),
as can be seen on their respective solutions:
\begin{eqnarray}
\label{stau2}
\hspace{-0.9in}&&_2F_1\Bigl([{{1} \over {4}}, \,{{1} \over {4}} ],
 \, [1]; \, -64 \, x)
\, \,\,  = \, \, \,\,  \,
(1 \, +256\, x)^{-1/4} 
\cdot \, _2F_1\Bigl([{{1} \over {12}}, \,{{5} \over {12}} ],
 \, [1]; \, {{1728 \, x } \over {(1 \, +256\, x)^3 }}   \Bigr) 
 \nonumber \\
\hspace{-0.9in}&&\, \,  \qquad \, \, \,\,  \, \,
\, \,\,  = \, \, \,\,  \,
(1 \, +16\, x)^{-1/4} 
\cdot \, _2F_1\Bigl([{{1} \over {12}}, \,{{5} \over {12}} ],
 \, [1]; \, {{1728 \, x^2} \over {(1 \, +16\, x)^3 }}   \Bigr)
 \\
\hspace{-0.9in}&&\, \, \,\,    = \, \, \,\,  \, \, \,
1 \,\,\,  \,  \, -4\,x\,\,  +100\,{x}^{2}\,\,
  -3600\,{x}^{3}\,\,  +152100\,{x}^{4}\,\, 
-7033104\,{x}^{5}\,\,  +344622096\,{x}^{6} 
\nonumber \\
\hspace{-0.9in}&&\quad \quad  \quad  \quad  \quad \, \, 
\,-17582760000\,{x}^{7} \,+924193822500\,{x}^{8}\,\, 
-49701090010000\,{x}^{9}\, \, \, \, \, \,  + \, \,  \,\cdots 
\nonumber  
\end{eqnarray}
and:
\begin{eqnarray}
\label{stau3}
\hspace{-0.9in}&&\Bigl((1 \, +27\, x) \, (1 \, +243\, x)^3\Bigr)^{-1/12} 
\cdot \, _2F_1\Bigl([{{1} \over {12}}, \,{{5} \over {12}} ],
 \, [1]; \, {{1728 \, x } \over {
(1 \, +243\, x)^3\, (1\, +27 \, x) }}  \Bigr) 
\nonumber \\
\hspace{-0.9in}&&\,    = \, \, 
\Bigl((1 \, +27\, x) \, (1 \, +3\, x)^3\Bigr)^{-1/12} 
\cdot \, _2F_1\Bigl([{{1} \over {12}}, \,{{5} \over {12}} ],
 \, [1]; \, {{1728 \, x^3 } \over {
(1 \, +3\, x)^3\, (1\, +27 \, x) }}  \Bigr) 
 \\
\hspace{-0.9in}&&\,    = \, \, 
\,  \,_2F_1\Bigl([{{1} \over {3}}, \,{{1} \over {3}} ],
 \, [1], \, -27 \, x) \, \,\,  = \, \, \, \,\,\, 
1 \,\,\,   -3\,x\, +36\,{x}^{2}\, 
 -588\,{x}^{3}\,  +11025\,{x}^{4}\,
-223587\,{x}^{5}\, 
\nonumber   \\
\hspace{-0.9in}&&\quad \quad \,\, \, \,  \, \,    \, \,\,  
 +4769856\,{x}^{6}\,\, -105423552\,{x}^{7}\, +2391796836\,{x}^{8}\,\, 
-55365667500\,{x}^{9} \,\,\,  \,\, \,  + \,\, \, \cdots
 \nonumber
\end{eqnarray}

The relation between the two Hauptmodul pullbacks in (\ref{stau2}) 
\begin{eqnarray}
\label{pu1}
\hspace{-0.9in}&&\quad \quad \quad \,\, \,   u \,\, \,   = \, \, \,\,
 {{1728 \, x } \over {(1 \, +256\, x)^3 }}, \quad \quad  \, \, \, 
v \,\, \,   = \, \, \,\, {{1728 \, x^2} \over {(1 \, +16\, x)^3 }} 
\, \,   = \, \,\,  u\Bigl( {{1} \over { 2^{12} \,  \, x}} \Bigr), 
\end{eqnarray}
corresponds to the (genus-zero) fundamental modular curve:
\begin{eqnarray}
\hspace{-0.9in}&&\quad  \, \,   5^9\, \cdot \,  {u}^{3}{v}^{3} \, \, \,
-12\cdot \, 5^6 \,\cdot \,  {u}^{2}{v}^{2} \cdot \, (u+v)\,\,\,  
 +375\,uv \cdot \, (16\,{u}^{2}+16\,{v}^{2}\,  -4027\,uv)\,
\nonumber \\ 
\hspace{-0.9in}&&\quad \quad \quad \quad  \quad  -64\, \, (u\,  +v) \cdot \, 
 \, ({v}^{2} +1487\,uv +{u}^{2})\,\, 
 + 2^{12} \,   3^3 \cdot  \,uv \,\,\,   \, = \, \,\, \,  \, 0. 
\end{eqnarray}
The relation between the two Hauptmodul pullbacks in (\ref{stau3}) 
\begin{eqnarray}
\label{pu2}
\hspace{-0.95in}&& \,\, u \,\, \,   = \, \, \,\,
{{1728 \, x} \over {(1 \, +243\, x)^3 \, (1 \, +27\, x) }},
 \quad \,
v \,\, \,   = \, \, \,\,
 {{1728 \, x^3 } \over {(1 \, +3\, x)^3 \, (1 \, +27\, x)  }}\, \,   = \, \, \, 
u\Bigl({{1} \over {3^6 \, x }} \Bigr) , 
\end{eqnarray}
corresponds to the (genus-zero) modular curve:
\begin{eqnarray}
\hspace{-0.95in}&&\,  2^{27} \, 5^9 \,\cdot \, 
{u}^{3}{v}^{3}  \cdot \, (u \, +v) \, 
 \, +2^{18} \, 5^6 \,\, {u}^{2}{v}^{2} \cdot \, 
(27\,{v}^{2}+27\,{u}^{2}-45946\,uv) \, 
\nonumber \\ 
\hspace{-0.95in}&&\quad  \,
 +2^{9} \, 3^5 \, 5^3 \, \,uv \cdot \,  (u \,  + v) \cdot \, 
 \, ({v}^{2}+241433\,uv+{u}^{2}) \,  \, 
 \\ 
\hspace{-0.95in}&&\quad \quad  \, \, \,  
+729\,({u}^{4}\, +{v}^{4}) \,\,\,\,
 - 39628 \cdot \,3^9 \, \cdot \, ({u}^{2} \,+\,{v}^{2}) \cdot \, u \, v\,\,\,
 +   15974803 \cdot  \, 2 \, \cdot  \,  3^{10}  \, \cdot \,{u}^{2}{v}^{2}
\nonumber \\ 
\hspace{-0.95in}&&\quad \qquad \quad  \quad \quad \, 
+\, 31 \cdot \, 2^9\, 3^{11} \, \,uv \cdot \, (u\, + v)
 \, \,\, \, -\,  2^{12}\, 3^{12}\,uv \, \,\,\,  = \,\, \,\,  \, 0.
 \nonumber 
\end{eqnarray}

Similarly, one can consider the order-two operators
$\,\omega_n$ associated with other modular forms corresponding to 
$\, \tau \, \rightarrow \, n \cdot \tau$.
The $\, \omega_{n}$'s can be simply deduced from Maier~\cite{Maier1},
 for modular forms corresponding to {\em genus-zero} curves i.e. for 
$\, n \, = \, 2, \, 3,  $
$4, \, 5, \, $ $6, \, 7, \, 8, \, $ $9, \, 10, \, 12, \, $
 $ 13, \, 16, \, 18, \, 25$. Since the solutions 
can be written as $\, _2F_1$ hypergeometric up to 
{\em rational pullbacks}, these genus-zero $\, \omega_{n}$'s
are obviously {\em order-two} operators. After a simple rescaling,
 the solutions analytic at $\, x \, = \, \, 0$, 
can be rewritten as series with {\em integer coefficients}. 

One can also consider the other $\, \omega_{n}$'s corresponding
 to {\em higher-genus} modular curves. In 
these cases, one does not have a
rational parametrisation like (\ref{pu1}) or (\ref{pu2}), but
 {\em one still has an identity 
of the same hypergeometric function with two different pullbacks}, 
these two pullbacks being {\em algebraic functions and not rational 
functions} (see (\ref{pu1}) or (\ref{pu2})). These algebraic functions
correspond to the so-called {\em modular polynomials}~\cite{Big}.

For instance for $\, \tau \, \rightarrow \, 11 \cdot \tau$, 
one has a {\em genus-one} modular curve, the modular polynomial reads:
\begin{eqnarray}
\label{Phibis}
\hspace{-0.9in}&&P^{*}_{11}(x, \, H)
\, \,\,\, = \, \, \,\, \,\,\,
 (1\, +228\, x \, +486\, x^2 \, -540\, x^3 +225\, x^4)^3 \cdot \, H^2
\\
\hspace{-0.9in}&& \qquad \qquad \quad  \quad  \quad \quad 
\,  -1728 \cdot \, {\cal Q}_1(x) \cdot x \, \cdot \, 
H \,\,\,\, \,\, +1728^2 \, x^{12}, 
\qquad \quad \quad 
\hbox{with:} 
\nonumber
\end{eqnarray}
\begin{eqnarray}
\hspace{-0.9in}&&{\cal Q}_1(x) \,  \, \,  =  \, \,   \, \,  \,  \, \,\,
1 \, \, \,\,\,  -55\, x\,\,\, +1188\, x^2 \,\,
 -12716\, x^3 \,\, +69630\, x^4\, \, -177408\, x^5\,
+133056\, x^6 \nonumber \\
\hspace{-0.9in}&& \quad \quad \quad \quad \quad \quad \quad \quad 
+132066\, x^7 \, -187407\, x^8\,
+40095\, x^9\, +24300\, x^{10}\, -6750\, x^{11}.
\nonumber
\end{eqnarray}
One has the identity
\begin{eqnarray}
\label{defining}
\hspace{-0.6in}\quad  _2F_1\Bigl([{{ 1} \over {12}}, 
\, {{5 } \over {12 }}], \,[1];
   \, H_1  \Bigr)\, \,   \,\,\,
 = \,  \,  \,  \, \, \, A(x) \cdot \,  _2F_1\Bigl([{{ 1} \over {12}}, 
\, {{5 } \over {12 }}], \,[1];
   \, H_2  \Bigr),
\end{eqnarray}
where the two Hauptmoduls $\, H_1$ and $\, H_2$ are the two solutions
of $\, P^{*}_{11}(x, \, H) \, = \, \, \, 0$, and where
 $\, A(x)$ is the algebraic function such that 
\begin{eqnarray}
\hspace{-0.9in}\quad {{A(x)^4} \over {11^2}} \, + \, \,  {{11^2} \over {A(x)^4}} 
 \,\,  \,\,  = \, \,\, \,  \,\,  {{2} \over {11^2}} \cdot \, \, 
{{7321-87612\,x+73206\,x^2+21060\,x^3-23175\,x^4 } \over {
1+228\,x+486\,x^2-540\,x^3+225\,x^4 }}, 
\nonumber 
\end{eqnarray}
this last relation between $\, A(x)^4$ and $\, x$ corresponding to 
a genus-one curve with the {\em same  $\, j$-invariant} as 
the genus-one curve  $P^{*}_{11}(x, \, H) \, = \, \, \, 0$, namely 
$\, j \,\, = \, \, \,-\, 496^3/11^5$.
The two hypergeometric functions in (\ref{defining}) are 
{\em actually series with integer 
coefficients}:
\begin{eqnarray}
\hspace{-0.9in}&&_2F_1([1/12, \, 5/12], \,[1]; \, H_1) \, = \,\,   \, \, \,
1\,\,\,\, +60\,x\, -4560\,{x}^{2}
\,+614400\,{x}^{3} \, -95660400\,{x}^{4} 
\nonumber \\
\hspace{-0.9in}&&\quad \quad \quad \qquad \, \, +16231863060\,{x}^{5}\, 
 -2905028387700\,{x}^{6}\, \,  \,
+ \, \, \,  \cdots \, \, \,   \,
\end{eqnarray}
\begin{eqnarray}
\label{twoseries}
\hspace{-0.9in}&&_2F_1([1/12, \, 5/12], \,[1];\, H_2)
\, \,\,\,\, = \, \, \,\,\,\,
 1\,\,\,  \, +60\, x^{11}\,\, +3300\, x^{12} \,
 +110220\, x^{13}\,\, +2904660\, x^{14}
\nonumber \\
\hspace{-0.9in}&& \qquad \quad \quad\, \,
+66599940\, x^{15} \, \, +1394683620 \, x^{16} \, \, +27425371380\, x^{17}
\,\, \,\, \,  \,  + \, \,  \cdots 
\end{eqnarray}
More details are given for this 
$\, \tau \, \rightarrow \, 11 \cdot \tau$ case
in Appendix I of~\cite{Big}. 
Note that although $\, _2F_1([1/12, \, 5/12], \,[1]; \, H_1)$
 and  $\, _2F_1([1/12, \, 5/12], \,[1]; \, H_2)$ are solutions 
of the {\em same  order-four} operator~\cite{Big}, one
 can find an appropriate
algebraic function $\, {\cal A}(x)$, such that 
$\, {\cal A}(x) \cdot \, \, _2F_1([1/12, \, 5/12], \,[1],\, H_1)$
is solution of an  {\em order-two}
 operator $\, \omega_{11}$ (see~\cite{Big} for more details). 

The other $\, \omega_{n}$'s, corresponding 
to higher genus modular curves~\cite{Hibino},
are actually {\em also order-two operators}. 
The explicit expressions of  $\, \omega_{n}$'s 
for the elliptic  values
 $\, n\, = \, \, 17, \, 19$, and the
 {\em hyperelliptic} values~\cite{Hibino}
$ n\, = \,  23, \, 29, \,  31, \,  \,41, \, 47, \, 59, \, 71 $
 are given in~\cite{Big}. The  genus of the associated 
modular curves~\cite{Hibino}, is
respectively~\cite{Big}  {\em genus-one}
 for  $\,{\tilde \omega_{17}}(x)$, 
$\,{\tilde \omega_{19}}(x)$,
{\em genus-two} for $\,{\tilde \omega_{23}}(x)$, 
$\,{\tilde \omega_{29}}(x)$,
 $\,{\tilde \omega_{31}}(x)$,
{\em genus-three} for $\,{\tilde \omega_{41}}(x)$,  {\em genus-four }
for $\,{\tilde \omega_{47}}(x)$,
{\em genus-five} for  $\,{\tilde \omega_{59}}(x)$, and {\em genus-six} 
 for $\,{\tilde \omega_{71}}(x)$. 

\subsection{Hadamard products of $\, \omega_n$'s}
\label{Hadomegan}

The two operators  $\, \omega_{2}$ and $\, \omega_{3}$ 
have a ``modularity'' property: their series expansions 
analytic at $\, x=\, 0$, 
(\ref{stau2}) and (\ref{stau3}), {\em as well as}
 the corresponding nomes, mirror maps
are series with integer coefficients. The Hadamard product 
is a quite natural operation to introduce 
because {\em it preserves the global nilpotence
of the operators},  {\em it preserves the integrality of series-solutions}, 
and it is a {\em natural operation to introduce when seeking for
diagonals of rational functions}\footnote[5]{And,
 consequently, has been heavily used
 to build Calabi-Yau-like ODEs 
(see Almkvist et al.~\cite{Almkvist}).}.  Let us perform the 
 Hadamard product of these two operators.
With some abuse of language~\cite{CalabiYauIsing}, the Hadamard product 
of the two order-two operators
 (\ref{tau2}) and  (\ref{tau3}) 
\begin{eqnarray}
\label{H23}
\hspace{-0.9in}&&\quad H_{2,3}\,  \, \, = \, \, \, \,
\omega_2 \, \star \, \omega_3  \, \, = \, \, \, \, \, 
 D_x^{4} \, \, \, \,\,
+6\,{\frac { (2064\,x-1)}{ (1728\,x-1) \cdot \, x }} \cdot D_x^3 \, \, \,
\,+{\frac { (19020\,x-7) }{ (1728\,x-1) \cdot \, x^2 }} 
 \cdot D_x^2 \, \,
\nonumber \\
\hspace{-0.9in}&& \quad \quad \qquad  \quad \quad  \quad \quad 
\, \,+{\frac { (4788\,x-1) }{(1728\,x-1) \cdot \, x^3 }}
\cdot D_x \,\,\,
\,+ \,{\frac {12}{(1728\,x-1) \cdot \, x^3  }}, 
\end{eqnarray}
is defined as the  (minimal order) linear differential 
operator having, as a solution,
the Hadamard product of the solution-series (\ref{stau2}) and (\ref{stau3}),
which is, by construction, a series with integer coefficients.  
This series is,  of course, nothing but the expansion of
the hypergeometric function:
\begin{eqnarray}
\label{1728}
\hspace{-0.3in}&& _4F_3([{{1} \over {4}}, \, {{1} \over {4}}, 
\, {{1} \over {3}}, \, {{1} \over {3}}],
 \, [1, \, 1, \, 1]; 1728 \, x) \,  \\
\hspace{-0.3in}&& \qquad \qquad \, \, = \, \, \, \,\,\, 
 _2F_1([{{1} \over {4}}, \, {{1} \over {4}}], \, [1]; \, -64 \, x) \, \star \,
 _2F_1([{{1} \over {3}}, \, {{1} \over {3}}], \, [1]; \, -27 \, x). 
\nonumber    
\end{eqnarray}
\vskip .1cm
In a similar way one can consider (see~\cite{Big}) 
 $\,H_{2,2}\, \,  \, = \, \,\, \, \omega_2 \, \star \,\omega_2$
(resp.  $\, H_{3,3}\, \,  \, = \, \,\, \, \omega_3 \, \star \,\omega_3$)
the Hadamard product of the order-two operator (\ref{tau2}) 
(resp. (\ref{tau3})) with itself (Hadamard square). These two operators 
have respectively the hypergeometric solutions
\begin{eqnarray}
\hspace{-0.9in}&&\quad _4F_3([{{1} \over {4}}, \, {{1} \over {4}}, \,
 {{1} \over {4}}, \, {{1} \over {4}}],
 \, [1, \, 1, \, 1];  \, 4096 \, x), \quad \quad
 _4F_3([{{1} \over {3}}, \, 
 {{1} \over {3}}, \, {{1} \over {3}}, \, {{1} \over {3}}],
  \, [1, \, 1, \, 1]; 729 \, x), 
\end{eqnarray}
corresponding to series expansions with {\em integer coefficients}.
These operators $\, H_{2,2}$, $\, H_{3,3}$ are MUM operators.  We can, therefore,
 define, without any ambiguity,
the nome (and mirror map) and Yukawa coupling 
of this order-four operator~\cite{CalabiYauIsing}.
One finds out that the nome\footnote[1]{The nome of the Hadamard product
of two operators has no simple relation with the nome
of these two linear differential operators.}, and
 the mirror map (and the Yukawa coupling 
as a function of the $\, x$ variable), are {\em not globally bounded}:
they {\em cannot} be reduced, by one rescaling, 
to series with integer coefficients.

\vskip .1cm 

The three linear differential operators  $\, H_{2,3}$,
 $\,H_{2,2}$ and $\,H_{3,3}$ , are MUM and of order four, however, 
they {\em are not of the Calabi-Yau type}.

\subsection{Hadamard products versus Calabi-Yau ODEs}
\label{hadamard}

The occurrence of Calabi-Yau type operators, that we could
imagine, at first sight, to be extremely rare, 
 is in fact quite frequent among such Hadamard products,
 as can be seen  with other values of $\, n$ and $\, m$.
 For instance,
 one can introduce\footnote[2]{To get the Hadamard product
of two linear differential operators use, for instance,
 Maple's command \textsf{gfun[hadamardproduct]}.} 
 $\, H_{4,4} \, = \, \,\omega_4 \, \star \, \omega_4$,
the Hadamard square of $\, \omega_4$, which is an irreducible
order-four linear differential operator, and has 
the hypergeometric solution already encountered 
for some $\, n$-fold integrals of the decomposition of the full
 magnetic susceptibility of
the Ising model~\cite{CalabiYauIsing1,CalabiYauIsing}
 (see also subsections (\ref{diffeofunit}) 
and (\ref{Yukdet})): 
\begin{eqnarray}
\label{256}
\hspace{-0.5in}&& _4F_3([{{1} \over {2}}, \, {{1} \over {2}},
 \, {{1} \over {2}}, \, {{1} \over {2}}],
 \, [1, \, 1, \, 1]; \, 256 \, x) \, 
 \\
\hspace{-0.5in}&& \qquad \qquad \qquad \, \, = \, \,  \, \, \,\, 
 _2F_1([{{1} \over {2}}, \, {{1} \over {2}}], \, [1]; \, -16 \, x) \, \star \,
 _2F_1([{{1} \over {2}}, \, {{1} \over {2}}], \, [1]; \, -16 \, x). 
\nonumber 
\end{eqnarray}

The associated operator having (\ref{256}) as a solution, 
obeys the ``Calabi-Yau condition'' that its exterior square
is of {\em order five}.

Let us give in a table the orders (which go from $\, 4$ to $\, 20$)
 of the various $\, H_{m,n}\, = \,\, H_{n,m}\,  $ Hadamard products
of the order-two operators associated with the (genus-zero) modular forms 
 operators $\, \omega_n$ and $\, \omega_m$: 

\vskip .1cm 

\vskip .2cm 
\hspace{-0.2in}
\begin{tabular}{|l|p{.4cm}|p{.4cm}|p{.4cm}|p{.4cm}|p{.4cm}|p{.4cm}|p{.4cm}|p{.4cm}|p{.4cm}|p{.4cm}|p{.4cm}|p{.4cm}|p{.4cm}|p{.4cm}|}
\hline
n\textbackslash m   & $2$ & $3$ & $4$ & $5$ & $6$ & $7$ & $8$ & $9$ & $10$ & $12$ & $13$ & $16$ & $18$  & $25$  \\ \hline
2   & $4$ & $4$ & $4$ & $6$ & $4$ & $6$ & $4$ & $4$ & $10$ & $8$  & $10$  & $8$ & $12$ & $14$  \\ \hline
3   &     & $4$ & $4$ & $6$ & $4$ & $6$ & $4$ & $4$ & $10$ & $8$  & $10$  & $8$ & $12$ & $14$  \\ \hline
4   &     &     & $4\,\, *$ & $6$ & $4\,\, *$ & $6$ & $4\,\, *$ & $4\,\, *$ & $10$ & $8$  & $10$  & $8$ & $12$ & $14$  \\ \hline
5   &     &     &     & $6$ & $6$ & $8$ & $6$ & $6$ & $12$ & $10$ & $12$  & $10$ & $14$ & $16$  \\ \hline
6   &     &     &     &     & $4\,\, *$ & $6$ & $4\,\, *$ & $4\,\, *$ & $10$ & $8$  & $10$  & $8$ & $12$ & $14$  \\ \hline
7   &     &     &     &     &     & $6$ & $6$ & $6$ & $12$ & $10$ & $12$   & $10$ & $14$ & $16$ \\ \hline
8   &     &     &     &     &     &     & $4\,\, *$ & $4\,\, *$ & $10$ & $8$  & $10$   & $8$ & $12$ & $14$ \\ \hline
9   &     &     &     &     &     &     &     & $4\,\, *$ & $10$ & $8$  & $10$   & $8$ & $12$ & $14$ \\ \hline
10  &     &     &     &     &     &     &     &     & $10$ & $14$ & $16$   & $14$ & $18$ & $20$\\ \hline
12  &     &     &     &     &     &     &     &     &      & $8$  & $14$   & $12$ & $16$ & $18$ \\ \hline
13  &     &     &     &     &     &     &     &     &      &      & $10$   & $14$ & $18$ & $20$ \\ \hline
16  &     &     &     &     &     &     &     &     &      &      &        & $8$ & $16$ & $18$ \\ \hline
18  &     &     &     &     &     &     &     &     &      &      &        &     & $12$ & $20$ \\ \hline
25  &     &     &     &     &     &     &     &     &      &      &        &     &      & $14$ \\ \hline
\hline
\end{tabular}

\vskip .1cm 

\vskip .1cm 

\hskip -.7cm 
where the star $\, *$ denotes Calabi-Yau 
ODEs\footnote[8]{Recall that Calabi-Yau ODEs are defined by a list of 
constraints~\cite{Almkvist}, the most important ones being, besides
being MUM, that their {\em exterior square
 are of order five}. There are more exotic conditions like 
the cyclotomic condition on the monodromy at $\, \infty$,
 see Proposition 3 in~\cite{TablesCalabi}.}. 

\vskip .1cm 

The following operators are of order four:
$\, H_{2,2}$,  $\, H_{2,3}$,   $\, H_{2,4}$, 
 $\, H_{2,6}$, $\, H_{2,8}$, $\, H_{2,9}$,   
$\, H_{3,3}$,  $\, H_{3,4}$,  $\, H_{3,6}$, $\, H_{3,8}$,  $\, H_{3,9}$,  ...
Their exterior squares, which are of order six, do not have 
rational solutions\footnote[1]{They cannot be 
homomorphic to Calabi-Yau ODEs.}. 

The order-four 
operators $\, H_{3,3}$, $\, H_{3,4}$, 
are all MUM operators\footnote[5]{Note that the 
Hadamard product of two MUM ODEs is not necessarily a MUM ODE:
the order-six operator $\, H_{3,7}$ is {\em not} MUM.},  
but, similarly to the situation
encountered with  $\, H_{2,2}$, their nome, mirror map and Yukawa 
couplings are {\em not globally bounded}.

\vskip .1cm 

The following operators are of order six:
$\, H_{2,5}$,  $\, H_{2,7}$,    
$\, H_{3,5}$,  $\, H_{3,7}$,  $\, H_{4,5}$, $\, H_{4,7}$,  $\, H_{5,5}$, 
 $\, H_{5,6}$,  $\, H_{5,8}$,  $\, H_{5,9}$, 
 $\, H_{6,7}$, $\, H_{7,7}$, $\, H_{7,8}$,$\, H_{7,9}$, ...
Their  exterior square, which are of order fifteen, do not have 
rational solutions (and cannot be homomorphic to
 higher order Calabi-Yau linear ODEs). 

\vskip .1cm 

Remarkably the following ten order-four operators 
$\, H_{4,4}$,  $\, H_{4,6}$,   $\, H_{4,8}$,  $\, H_{4,9}$, $\, H_{6,6}$, 
$\, H_{6,8}$, $\, H_{6,9}$, $\, H_{8,8}$, $\, H_{8,9}$, 
$\, H_{9,9}$ (with a star in the previous table)
are all MUM, and {\em are such that their exterior 
squares are of order five}\footnote[3]{They
are conjugated to their (formal) adjoint by a function.}:
they are  {\em Calabi-Yau ODEs}. 
Actually the nome, mirror map and Yukawa coupling series 
are {\em series with integer coefficients for all these 
order-four Calabi-Yau operators}. Their Yukawa coupling
and their adjoint Yukawa coupling identify. 
The Yukawa coupling series of these Calabi-Yau operators 
are respectively, for $\, H_{4,4}$
\begin{eqnarray}
\label{YukuM44}
\hspace{-0.9in}&&K(q) \, \,= \,\, \,K^{\star}(q) 
\,\, \,  \,= \,\, \, \,\,\,  \,  \,1 \,\,\, \, \,  + \, 32 \cdot \,q\,  \, 
 +  4896 \cdot \, q^2 \,\,  +702464 \cdot \, q^3 
\,\, \, + 102820640 \cdot \, q^4
  \,\,  \nonumber \\
\hspace{-0.9in}&&\qquad \,\,  + 15296748032 \cdot \, q^5
\, + \,  2302235670528\cdot \, q^6\, \,  
 \,\,  \,    \, + \,\, \,\cdots  
\end{eqnarray}
which is \#3  in Almkvist et al. large
 tables of Calabi-Yau ODEs~\cite{TablesCalabi}, 
and is the well-known one for
 $\, _4F_3([1/2,1/2,1/2,1/2], \, [1,1,1]; \, 256 \,x)$,
and for $\, H_{4,6}$ 
\begin{eqnarray}
\label{YukuM46}
\hspace{-0.9in}&&K(q) \, \,\, \,= \,\, \,K^{\star}(q) \, \,= \,\,\,
 \, \, \, \,  \,1 \,\, \,\, \,  + \, 20 \cdot \,q\,  
 +  36 \cdot \, q^2 \, +15176 \cdot \, q^3 \,\,  + \, 486564 \cdot \, q^4 \,\,  
\,\,    \nonumber \\
\hspace{-0.9in}&&\qquad \quad  \quad + \,21684020 \cdot \, q^5 \,\, 
+ \,1209684456 \cdot \, q^6 \,\,
\,\, \, \, \,    + \,\, \, \cdots 
\end{eqnarray}
which is \#137 in tables~\cite{TablesCalabi}.

We give, in~\cite{Big}, the expansion of the Yukawa coupling 
for a set of other $\, H_{m,n}$ such that 
their exterior squares are order {\em five} (not 
six as one could expect for a generic irreducible 
order-four operator), that {\em actually are Calabi-Yau} operators. Actually
operator $\, H_{4,8}$  is \#36 
in Almkvist et al. large tables of 
Calabi-Yau ODEs~\cite{TablesCalabi}. Operators $\, H_{4,9}$, $\, H_{6,6}$,
 $\, H_{6,8}$  and $\, H_{6,9}$   are respectively~\cite{TablesCalabi}
 \#133,  \#144, \#176
and \#178. Furthermore, operators 
$\, H_{8,8}$, $\, H_{8,9}$ and  $\, H_{9,9}$ are 
respectively~\cite{TablesCalabi} \#107, \#163 
and \#165. 

It will be shown, in a forthcoming publication, that
the occurrence of an order five for the exterior power
 (the ``Calabi-Yau condition'') means 
that these operators are necessarily {\em conjugated} 
(by an algebraic function)
to their adjoints. Thus, the ``adjoint Yukawa coupling'' $\, K^{\star}(q)$
is necessarily equal to the  Yukawa coupling 
 $\, K(q)$ for these operators.

On the other hand, {\em the ten linear differential operators denoted by a star $\, *$ 
in the previous table all share the same property}: they have, 
as a solution, the Hadamard product of two HeunG functions 
solutions  of the form 
$\, HeunG(a, \, q, \, 1, \, 1, \,  1, \, 1; \, x)$.
Note, however, that this  HeunG-viewpoint of 
the most interesting $\, H_{m,n}$'s
does not really help. Even inside this restricted set of  HeunG functions 
solutions  of the form $\, HeunG(a, \, q, \, 1, \, 1, \,  1, \, 1; \, x)$
it is hard to find exhaustively the values of the parameter
$\, a$ and of the accessory parameter $\, q$, such the 
series $\, HeunG(a, \, q, \, 1, \, 1, \,  1, \, 1; \, x)$
is globally bounded,
or, just, such that the order-two operator, having 
$\, HeunG(a, \, q, \, 1, \, 1, \,  1, \, 1; \, x)$ 
as a solution, is globally nilpotent~\cite{bo-bo-ha-ma-we-ze-09}. 

\vskip .1cm 

Many $\,H _{m,n}$ are not MUM, for instance the
 order-eight operator $\, \, H_{12,12}$, or 
the order-six operator $\, H_{3,7}$, are {\em not MUM}. Concerning $\,H_{3,7}$,
and  as far as its
six solutions are concerned, they are structured
 ``like'' the four solutions of an
order-four MUM operator, together with the two solutions 
of another order-two MUM operator, but the  order-six operator $\, H_{3,7}$
is not a direct-sum of an  order-four and order-two operator.
We have two solutions analytical at $\, x\, = \, \, 0$
 (with no logarithmic terms), and
 two solutions involving $\ln(x)$. A linear combination of
these two solutions analytical at $\, x\, = \, \, 0$ is, 
by construction a series with 
{\em integer coefficients} (the Hadamard product of the 
two series with integer coefficients which are the initial
ingredients in this calculation), when the 
other linear combinations 
are {\em not globally bounded}.

\vskip .1cm 

\section{Calabi-Yau Modularity}
\label{hadamardmore}

The previous examples correspond to a ``modularity'' inherited from 
elliptic curves, more precisely Hadamard products of modular forms.
Let us consider, here, two Calabi-Yau examples that do not seem to be
reducible\footnote[2]{The possibility that a solution of an order-four
operator, non-trivially equivalent to these Calabi-Yau operators~\cite{Batyrev}, 
could be written as a $\, _4F_3$ hypergeometric function, {\em up to an involved
algebraic pullback}, is not totally excluded. However it is extremely 
difficult to rule out such highly non-trivial hypergeometric scenario.}
to $\, _4F_3$ hypergeometric functions. 

A first order-four Calabi-Yau operator,  
found by Batyrev and van Straten~\cite{Batyrev},
which is self-adjoint and also corresponds to Hadamard products of simple 
hypergeometric functions (see (\ref{sumup})),
is given in \ref{B1}. All the associated series (solution (\ref{sumup}), nome,
 Yukawa coupling) are series expansion with {\em integer coefficients}. We 
do not have a representation of the solution (\ref{sumup}), as an $\, n$-fold 
integral of the form (\ref{form}). However, since  (\ref{sumup}) 
can be expressed as a  sum 
of products of binomials (see (\ref{nested})), we can conclude, again, 
that (\ref{sumup}) is {\em actually a diagonal of a rational function}. 
 
\subsection{A Batyrev and van Straten Calabi-Yau ODE ~\cite{Batyrev} } 
\label{B2}

A second example  of order-four operator,
 corresponding to  Calabi-Yau 3-folds 
in $\, P_1 \times  P_1 \times  P_1 \times  P_1$, has been 
found by Batyrev and van Straten~\cite{Batyrev}
 (see\footnote[1]{There is a 
small misprint in~\cite{Batyrev} page 34:
 $(2\, \theta \, +\, 1)$ must be replaced 
by $(2\, \theta \, +\, 1)^2$ in the $\, 4\, x$ term.} page 34):
\begin{eqnarray}
\label{Batyrev2}
\hspace{-0.5in}&&B_2\,\,  \, \, = \,\,  \, \, \, \,  \, \,  
\theta^4 \, \, \, \, \,   -4 \, x \cdot 
(5\, \theta^2 \, +5 \,\theta \, +2) \cdot   (2\, \theta \, +\, 1)^2
\,  \nonumber \\
\hspace{-0.5in}&&\quad \quad \quad \quad \quad \quad \quad \quad
 +64 \, x^2 \cdot  
   (2\, \theta \, +\, 3) \cdot   (2\, \theta \, +\,1) \cdot
   (2\, \theta \, +\, 2)^2.   
\end{eqnarray}
It corresponds to the series-solution with coefficients:
\begin{eqnarray}
\label{integB2}
\hspace{-0.5in}&&{2n \choose n} \cdot \sum_{k=0}^{n} \,
 {n\choose k}^2 \cdot {2k \choose k} \cdot {2n-2k \choose n-k}.
\end{eqnarray}
Its Wronskian $\, W_4$ is a rational function such that:
\begin{eqnarray}
\label{wronskB2}
\hspace{-0.9in} \qquad \qquad \qquad 
x^3 \cdot \, W_4^{1/2} \, = \, \,
 {{1} \over { (1 \, -64 \, x) \, (1\, -16\, x)}}.
\end{eqnarray}
This operator is also a Calabi-Yau operator: it is MUM, and it is such
 that its exterior square is {\em order five}.  This order five property 
is a consequence of  $\, B_2$ being conjugated to its adjoint:
$\, B_2 \cdot \, x\, = \, \, x \cdot \, adjoint(B_2)$.

The series-solution of (\ref{Batyrev2}) 
 can be written as an Hadamard product 
\begin{eqnarray}
\label{serB2}
\hspace{-0.9in}&&{\cal S}  \,  \,   \, = \,\, \,\, \, 
 (1\, -4\, x)^{-1/2} 
\, \star \, HeunG(4, 1/2, 1/2, 1/2, 1, 1/2; \,  16\, x)^2 \\
\hspace{-0.9in}&&\, \,  \,   \, = \,\, \,\, \,
 1\,+8\,x\,+168\,x^2\,
+5120\,x^3\,+190120\,x^4\,+7939008\,x^5\,+357713664\,x^6 
\, \, \,  + \, \cdots, 
\nonumber 
\end{eqnarray}
the modular form character of 
$\, HeunG(4, 1/2, 1/2, 1/2, 1, 1/2; \,  16\, x)$
being illustrated with identities (A.4) in
 Appendix A of~\cite{Big}.
Its nome reads:
\begin{eqnarray}
\label{nomeB2}
\hspace{-0.9in}&&q \, \, = \, \, \,  \, \,
x \,\, \,  \, +20\,x^2 \,+578\,x^3 \,+20504\,x^4 \,+826239\,x^5 \,
+36224028\,x^6 \,+1684499774\,x^7 \,
\nonumber  \\
\hspace{-0.9in}&& \quad +81788693064\,x^8 \,
 \,+4104050140803\,x^9 \,+211343780948764\,x^{10}
 \, \,\, \,  \,+\,\, \,\cdots 
\end{eqnarray}
The {\em mirror map} of (\ref{Batyrev2}) reads:
\begin{eqnarray}
\label{mirrorB2}
\hspace{-0.9in}&&x(q) \, \, = \, \, \,  \, \, \,\,
q \, \, \, \, \,  -20\,q^2\, \,  +222\,q^3\, \,
  -2704\,q^4\, \,  +21293\,q^5\,  \, 
-307224\,q^6\, \,  +80402\,q^7\,   \nonumber  \\
\hspace{-0.9in}&& \qquad  \qquad  \, \, \,
 -67101504\,q^8\, \,  -1187407098\,q^9\, \, 
 -37993761412\,q^{10}\,\, 
 \, \, \,\,   + \, \, \cdots 
\end{eqnarray}

The Yukawa coupling of (\ref{Batyrev2}) reads: 
\begin{eqnarray}
\hspace{-0.9in}&&K(q) \, \, = \, \, \, K^{\star}(q) 
\, \,\, = \, \, \, \, \,  \, \, \,
1\, \,\, \,  +4\, q\, \,  +164\, q^2\, \,  +5800\, q^3\, \,  +196772\, q^4\, \, 
 +6564004\, q^5 \, 
\nonumber  \\
\hspace{-0.9in}&&\qquad \quad  \quad \quad +222025448\, q^6 \, \,  
 +7574684408\, q^7 \, +259866960036\, q^8 
\,\,\,  \,  \,   + \, \, \, \cdots 
\end{eqnarray}
The equality of the Yukawa coupling with the
``adjoint'' Yukawa coupling, $\,K(q) \, \, = \, \, \, K^{\star}(q)$, is a straight
consequence of relation  $\,\, B_2 \cdot \, x\, = \, \, x \cdot \, adjoint(B_2)$.

\vskip .1cm 

Recalling Batyrev and van Straten~\cite{Batyrev},
 and following Morrison~\cite{Morrison},
do note that one can also write the Yukawa coupling as:
\begin{eqnarray}
\label{otherYuk}
\hspace{-0.65in}K(q) \, \, \, \, = \, \, \, \, \, \,
  {{ x(q)^3 \, \cdot W_4^{1/2}} \over { y_0^2}} \cdot
 \, \Bigl({{q} \over {x(q)}} \cdot {{d x(q)} \over {dq}}\Bigr)^3
\, \, = \, \, \, \, \,
  {{ W_4^{1/2}} \over { y_0^2}} \cdot
 \, \Bigl( q \cdot {{d x(q)} \over {dq}}\Bigr)^3,
\end{eqnarray}
where $\, W_4$ is the Wronskian (\ref{wronskB2}).
{}From this alternative expression 
for the Yukawa coupling, {\em valid when the operator is 
conjugated to its adjoint} (see (\ref{condmagic})), it is obvious that if 
the analytic series $\, y_0(x)$, {\em as well as} the nome (\ref{nomeB2}) 
are series with {\em integer coefficients}, then, 
the mirror map (\ref{mirrorB2}) 
is also a series with integer coefficients, and, therefore,
  $\, y_0$ seen as a function
of the nome $\, q$, as well as $\,x^3 \, W_4^{1/2}$ (since it
is a  {\em rational function}) are also  series with integer coefficients.
{\em Consequently, the Yukawa coupling  is a series with integer coefficients} 
(as a series in $\, q$ or in $\, x$).

More generally, if one assumes that a linear 
differential operator has a globally bounded 
solution-series, one knows that this operator is a $\, G$-operator, 
necessarily globally nilpotent, and, consequently, its Wronskian,
or the square root of the Wronskian (see $\, W_4^{1/2}$ in (\ref{otherYuk})),
will be a $\, N$-th root of a rational function, and, thus,
{\em will correspond to a globally bounded series}. 
{\em Thus, the globally bounded 
character of the analytic series $\, y_0(x)$
together with the nome, yields the globally bounded character of 
the mirror map, Yukawa coupling, that we associate 
with the modularity}\footnote[1]{Similar results can be found in Delaygue's 
thesis~\cite{Delaygue}, in a framework where the coefficients of
hypergeometric series are {\em ratio of factorials} 
(see \ref{ratiooffac}).}. In contrast the globally bounded character of the analytic 
series $\, y_0(x)$, {\em together with} the globally bounded character of 
the Yukawa coupling (seen for instance as a series in $\, x$) 
{\em does not imply} that the nome, or the mirror map, are globally bounded
as can be seen on example (\ref{saoud}) (see (\ref{Ksaoudx}) and (\ref{Ksaoudq})).

\vskip .1cm  

\subsection{An operator non trivially homomorphic to $\, B_2$}
\label{operatornontriv}

Let us, now, consider the order-four operator 
\begin{eqnarray}
\label{calB2}
\hspace{-0.6in}&&{\cal B}_2 \, \, = \, \, \,\, \,
 256\, x^2 \cdot \, \theta^2\, (2\, \theta+3)\, (2\, \theta+1)\,
 \nonumber  \\
\hspace{-0.6in}&&  \quad  \quad  \qquad \qquad  
-4\, x \cdot \, (2\, \theta+1)\, (2\, \theta-1)\, (5\, \theta^2-5\, \theta+2)
\, \, \,  + \, (\theta-1)^4. 
\end{eqnarray}
This operator is non-trivially\footnote[2]{The
 intertwiners between $\,{\cal B}_2 $
 and $\,B_2 $ are operators not simple functions.} homomorphic to
 the Calabi-Yau operator (\ref{Batyrev2}):
\begin{eqnarray}
{\cal B}_2 \cdot \, x \cdot (2\, \theta\, +1)  \, 
\, \,\, = \,\, \, \,\, \, x \cdot (2\, \theta\, +1) \cdot \, B_2. 
\end{eqnarray}
As a consequence of the previous intertwining relation,
one immediately finds that the series-solution analytic at $\, x\, = \, 0$
of this new MUM operator (\ref{calB2}) is nothing but the action of the
order-one operator $\,\, x \cdot (2\, \theta\, +1)\, $
 on the series (\ref{serB2}), 
and reads:
\begin{eqnarray}
\hspace{-0.9in}&&x \cdot (2\, \theta\, +1)[{\cal S}]
 \, \, \, = \,\, \, \, \, \,  \, 
x \,\, \,  \,  +24\, x^2 \, \,  +840\, x^3 \, \,  +35840\, x^4 \, \, 
 +1711080\, x^5 \, \,  +87329088\, x^6 \, 
\nonumber \\
\hspace{-0.9in}&& \qquad \qquad \quad   +4650277632\, x^7 \, \,  
+254905896960\, x^8 \,\, \,  \,\, + \,\,\, \, \cdots 
\end{eqnarray}
It is obviously also a series with {\em integer coefficients} (the action of 
 $\, x \cdot (2\, \theta\, +1) $ on the series with integer coefficients
is straightforwardly a series with integer coefficients). More generally, the
globally bounded series remain globally bounded series
 by non trivial operator equivalence,
namely homomorphisms between operators (generically the intertwiner 
operators are not simple functions).

The exterior square of the order-four operator 
(\ref{calB2}) is an order-six operator which is, in fact, 
the direct sum of an order-five operator $\,{\cal E}_5$
 and an order-one operator.

Operator $\,{\cal B}_2$  is non-trivially homomorphic to its adjoint:
\begin{eqnarray}
\hspace{-0.7in}{\cal B}_2 \cdot \, x^3 \cdot
 (2\, \theta\, +3) \cdot (2\, \theta\, +5)  \, 
\, \, = \, \, \,\,\, \, x^3 \cdot (2\, \theta\, +3)
 \cdot (2\, \theta\, +5)  \cdot \, adjoint({\cal B}_2). 
\end{eqnarray}
The Yukawa coupling of this order-four operator (\ref{calB2}),  
non-trivially homomorphic to (\ref{Batyrev2}), reads: 
\begin{eqnarray}
\label{YucalB2}
\hspace{-0.9in}&&\quad K(q) \, \, \, \, \, = \, \, \, \, \, \, \,
 1\,\, \, \, \,-4\, q\, \, \, -140\, q^2\, \, \, -4040\, q^3\, \, \,
 - 64436  \, \,{{ q^4} \over {3}} \, \,
\, + 1889332 \,  \, {{q^5 } \over {3 }}
\nonumber  \\
\hspace{-0.9in}&&\quad \quad \quad \quad \quad  
+ 88331368 \, \, {{q^6} \over {5 }}\, \, + 1652707624 \,\, {{q^7} \over { 9}}\, \, 
- 69295027684 \, \, {{q^8} \over {63 }} \, 
\,\,\, \,   + \, \, \, \cdots 
\end{eqnarray}
The Yukawa coupling series (\ref{YucalB2}) is {\em not globally bounded}.

The ``adjoint Yukawa coupling'' of this order-four operator (\ref{calB2})
reads:
\begin{eqnarray}
\label{adjYucalB2}
\hspace{-0.9in}&&K^{\star}(q) \, \, \, \, = \, \, \, \, \, \, \, \,
1\,\, \, \, \, +12\, q\, \, \, +564\, q^2\, \, \, +20440\, q^3\, \, \, 
+865732\, q^4\, \, \, +37162444\, q^5
\nonumber  \\
\hspace{-0.9in}&& \quad  \, \, \, \, \, \,  
 +8255346664 \, \, {{q^6} \over {5}} \, \, 
 + 1121762648248 \, \, {{q^7} \over {15}}\, \, 
+72336859374772 \, \, {{q^8} \over {21}} \,  \,  \,  \,  +  \,  \,     \, \cdots 
\end{eqnarray}
Again, the  adjoint Yukawa coupling series (\ref{adjYucalB2}) 
is {\em not globally bounded}.

On this example one sees that the Yukawa coupling of two 
non-trivially homomorphic operators
are {\em not necessarily equal}. The Yukawa couplings of two  homomorphic 
operators are equal  {\em when} the two operators are
 {\em conjugated by a function}
(trivial homomorphism). The modularity property is {\em not preserved}
 by (non-trivial) operator equivalence: it may depend on a condition 
that the exterior square of the order-four operators is of order {\em five}.
The Calabi-Yau property is not preserved by operator equivalence. 

\vskip .1cm

{\bf To sum-up:} All these examples show that the 
{\em integrality} (globally bounded series) 
is {\em far from identifying with modularity}.

\section{Conclusion}
\label{concl}

Seeking for the linear differential operators for the  $\, \chi^{(n)}$'s,
we discovered, some years ago, that they were Fuchsian
 operators~\cite{ze-bo-ha-ma-04,ze-bo-ha-ma-05c}, and,
in fact, ``special'' Fuchsian operators, namely 
Fuchsian operators with rational exponents for all their singularities,
and with Wronskians that are $\, N$-th roots of rational functions.
Then we discovered that they were $\, G$-operators
(or equivalently globally nilpotent~\cite{bo-bo-ha-ma-we-ze-09}), and 
more recently, we accumulated 
results~\cite{CalabiYauIsing} indicating that 
they are ``special'' $\, G$-operators.
 There are, in fact, {\em two quite different 
kinds of ``special features''} of these $\, G$-operators. On one side,
we have the fact that one of their solutions is not only $\, G$-series, 
but is a {\em globally bounded} series. This special character 
has been addressed in this very paper,
 and we have seen that, in fact, this ``{\em integrality} 
property~\cite{Kratten}'' is a consequence of
 {\em quite general mathematical assumptions}
 often satisfied in physics (the  integrand is not only algebraic but
has an expansion at the origin of the form\footnote[1]{Puiseux 
series are excluded.} (\ref{form2}). However, we have also 
seen another special property of these $\, G$-operators, namely the fact 
that they seem to be quite systematically
{\em homomorphic to their adjoints}~\cite{CalabiYauIsing}. We will 
show, in a forthcoming publication, 
that this last property amounts, on the associated linear differential systems,
to having {\em special differential Galois groups}, and  
that their exterior or symmetric squares, have {\em rational solutions}. This 
last property is a  property of a more ``physical'' nature than the previous one,
related to an underlying {\em Hamiltonian structure}~\cite{Manin},
 or as this is the case,
for instance in the Ising model,
related to the underlying isomonodromic structure in the problem, which
yields the occurrence of some underlying Hamiltonian structure~\cite{Manin}. 
In general the {\em integrality} of $\, G$-operators 
{\em does not} imply the operator to be
homomorphic to its adjoint, and conversely being homomorphic to its adjoint
{\em does not} imply\footnote[2]{See Appendix M and  Appendix O in~\cite{Big}
which give an example of a (hypergeometric) family 
of order-four operators satisfying the Calabi-Yau condition that their
exterior square is of order five, and, even, a  family 
of self-adjoint order-four operators, the 
corresponding hypergeometric solution-series
being {\em not globally bounded}.}
 integrality (and even does not
 imply\footnote[1]{For 
instance the operator $\,\, D_x^n \, -x\, D_x\, -1/2\,$
 (see page 74 of~\cite{Katz})
 with an irregular singularity is self-adjoint.}
 the operator to be Fuchsian). Interestingly,
 the $\,\chi^{(n)}$'s, as well as many important 
problems of theoretical physics, correspond to $\, G$-operators that present 
these two complementary ``special characters'' 
(integrality and, up to homomorphisms,
self-adjointness), and, quite often, this is seen in the framework of the
emergence of ``modularity''. 

Nomes, mirror maps, and Yukawa couplings are {\em not D-finite} functions:
they are solutions of quite involved {\em non-linear} (higher order
Schwarzian) ODEs (see for instance Appendix D in~\cite{bo-bo-ha-ma-we-ze-09}).
Therefore, the question of the series integrality of the nomes, mirror maps,
Yukawa couplings, and other pullback-invariants (see \ref{Yukawaratio})
requires to address the very difficult question of series-integrality for
(involved) {\em non-linear} ODEs. Note, however, as seen in
Section~\ref{B2}, in particular in~\eqref{otherYuk}, that the integrality of the
series $\,y_0(x)$ and of the nome $\, q(x)$ are
 {\em sufficient to ensure}, {\em provided the operator is conjugated to its
adjoint} (see (\ref{condmagic})), 
the integrality of the other quantities such as the
Yukawa coupling, mirror maps. However the integrality of the nome remains 
an involved problem. These questions will certainly remain open for some time. 

In contrast, and more modestly, we have shown that
 a {\em very large sets of
problems in mathematical physics} (see sections (\ref{gener}), (\ref{plan})
 and (\ref{subtheorem})) 
{\em actually corresponds to diagonals of rational
functions}. In particular,  we have been able to show
that the $\, \chi^{(n)}$'s $\, n$-fold integrals of the
susceptibility of the two-dimensional Ising model
 are actually {\em diagonals of rational
functions for any value of the integer} $\, n$, thus proving that the
 $\,\chi^{(n)}$'s are 
{\em globally bounded for any value of the integer} $\, n$.
As can be seen in the ``ingredients'' of our simple 
demonstration (see (\ref{gener})),
no elliptic curves, and their modular forms~\cite{Kean},
 no Calabi-Yau~\cite{Huse}, 
or Frobenius manifolds~\cite{Manin}, 
or Shimura curves, or arithmetic lattice
 assumption~\cite{Bouw,Dettweiler} is required to prove the result. 
We just need to have an $\, n$-fold integral
 such that its integrand is {\em not only
algebraic}, but has an expansion at the origin of the form (\ref{form2}).

The integrality of all the $\, \chi^{(n)}$'s,
consequence of the remarkable result that the all $\, \chi^{(n)}$'s
are diagonal of rational functions, raises the question of the 
{\em modularity} of the $\, \chi^{(n)}$'s. Now, the full susceptibility 
can, formally, be seen as the  diagonal of an {\em infinite sum of
rational functions}. This also raises the question of defining,
 and addressing, modularity for non-holonomic 
functions\footnote[3]{Along this line, 
recall Chazy's equations~\cite{Chazy}
 and their (circle) natural boundaries, and, especially, Harnad and McKay 
paper~\cite{Harnad} on {\em modular} solutions to equations of generalized 
Halphen type.} like the 
full susceptibility.

\vskip .1cm 

\vskip .1cm 

{\bf Acknowledgments} 
We would like to thank A. Enge and F. Morain for interesting and detailed 
discussions on Fricke and Atkin-Lehner involutions.
S. B. would like to thank the LPTMC and the CNRS for kind support. 
A. B. was supported in part by the Microsoft Research--Inria Joint Centre.
As far as physicists authors 
are concerned, this work has been performed without
 any support of the ANR, the ERC or the MAE. 

\vskip .1cm 

\appendix

\section{Modular forms and series integrality}
\label{modularapp}

{\bf First example}:
The generating function of the integers 
\begin{eqnarray}
\label{hadaprod}
\hspace{-0.6in}&&\quad \sum_{k=0}^{n} \, {n\choose k}^2 \cdot {2k \choose k}
 \cdot {2n-2k \choose n-k}  \\
\hspace{-0.6in}&&\quad \quad \quad \quad  \quad \, \, = \, \, \, \, \,  {2n \choose n} \cdot
\,  _2F_1\Bigl([{{1} \over {2}}, \, -n,\,  -n,\,  -n],
\,  \,[1,\,  1,\,  -\, {{2\, n \, -1} \over {2}}]; \, 1\Bigr),  
\nonumber 
\end{eqnarray}
is nothing else but the expansion of the square of a HeunG function
\begin{eqnarray}
\label{this}
\hspace{-0.7in}&& HeunG\Bigl(4,\, {{1} \over {2}},\, {{1} \over {2}},
 \, {{1} \over {2}},\, 1,\, {{1} \over {2}}; 
\, 16 \cdot x\Bigr)
\,  \, = \, \, \,  \, \, \,\, \, \,
1\, \,\,  \,\,\,+2\,x\,\, \, \,+12\,x^2\,\,+104\,x^3\,
 \nonumber \\
\hspace{-0.7in}&&\qquad \quad \quad  \, \, +1078\,x^4\,\,+12348\,x^5\,\,+150528\,x^6\,\,
+1914432\,x^7\,\,\,\, \, + \, \, \cdots  
\end{eqnarray}
solution of the  order-two operator
\begin{eqnarray}
\label{GoodHeunDiam}
\hspace{-0.4in}H_{diam} \,\, = \, \, \,\,\,\, \, \,
 \theta^2 \, \, \, \, \, -2 \cdot x \cdot (10 \, \theta^2\, +5\, \theta\, +1)
 \, \,\, \,
 +\,  16 \, x^2 \cdot  (2\, \theta\, +1)^2.
\end{eqnarray}
which corresponds to the {\em diamond lattice}~\cite{GoodGuttmann}. 
This HeunG function (\ref{this}) is {\em actually 
a modular form} which can be written in two different ways:
\begin{eqnarray}
\label{twopull}
\hspace{-0.4in}&&HeunG(4, {{1} \over {2}}, {{1} \over {2}}, 
{{1} \over {2}}, 1, {{1} \over {2}}; \,  16\, x)
 \,  \, \,\, 
\nonumber \\
\hspace{-0.4in}&&\qquad 
\,  \, = \, \,\,  \, \, (1\, -4 \, x)^{-1/2} \, \cdot \,
 _2F_1\Bigl([{{1} \over {6}},\, {{1} \over {3}}], \, [1]; \,
 {{108 \, x^2} \over {(1\, -4 \, x)^3 }} \Bigr)  \\
\hspace{-0.4in}&&\qquad 
\,  \, = \, \, \, \, \, (1\, -16 \, x)^{-1/2} \, \cdot \,
 _2F_1\Bigl([{{1} \over {6}},\, {{1} \over {3}}], \, [1]; \,
 -\, {{108 \, x } \over {(1\, -16 \, x)^3 }} \Bigr). 
\nonumber
\end{eqnarray}

These two pullbacks are {\em related by an ``Atkin'' involution}
 $\, x \, \leftrightarrow \, 1/64/x$. 
The associated modular curve, relating these
 two pullbacks (\ref{twopull}) 
yielding {\em  the modular curve}:
\begin{eqnarray}
\label{encore2}
\hspace{-0.5in}&&4\cdot \, y^3 \,z^3 \,\,\,\, -12 \,y^2\, z^2\, \cdot (y+z)\,\,\,\,
+3  \,y \, z \cdot(4\, y^2\,+ \, 4\, z^2 \, -127 \,y \, z)\,\,
\nonumber \\
\hspace{-0.5in}&&\quad \quad  \quad  \quad 
 -4 \cdot (y \, +z) \cdot (y^2\, + z^2 \, +83 \,y \,z)
\,\,\,\, +432  \,y \, z
\,\,\,\, \,\,\, = \,\,\,\,\,  \,\, 0, 
\end{eqnarray}
which is $(y, \, z)$-symmetric and is 
{\em exactly the rational modular
 curve} in eq. (27) already found
for the order-three operator $\, F_3$ in~\cite{CalabiYauIsing}
for the five-particle contribution $\, \tilde{\chi}^{(5)}$ 
of the magnetic susceptibility
of the Ising model. 
\vskip .3cm 
This result in~\cite{Prell,GoodGuttmann} 
can be rephrased as follows. 
One introduces
the order-three operator which has the following $\, _3F_2$
solution
\begin{eqnarray}
{{1} \over {(4\, -\, x^2)^3}} \, \cdot \, \, \, 
_3F_2\Bigl([{{1 } \over {3}}, \, {{1 } \over {2}}, \,{{2 } \over {3}}],\, 
[1, \, 1], \, \, {{27 \, x^4 } \over { (4\, -\, x^2)^3}}   \Bigr), 
\end{eqnarray}
associated with the {\em Green function of the diamond lattice}.
Along a {\em modular form line} lets us note that
this hypergeometric function actually has {\em two} pullbacks:
\begin{eqnarray}
\hspace{-0.5in}&&_3F_2\Bigl([{{1} \over {3}}, \,
 {{1} \over {2}}, \,{{2 } \over {3}}],\, 
[1, \, 1], \, \, {{27 \, x^4 } \over { (4\, -\, x^2)^3}}\Bigr) 
\,\,\,\,   \\
\hspace{-0.5in}&& \quad \quad \quad \, = \, \, \,  \,  \,\, 
 {{x ^2 \, -4  } \over {4  \cdot (x^2-1) }} \cdot \, \, 
_3F_2\Bigl([{{1 } \over {3}}, \, \, {{1 } \over {2}}, \,{{2 } \over {3}}],\, 
[1, \, 1],\,\, {{27 \, x^2 } \over {4 \cdot \, (x^2 -1)^3}}\Bigr). 
\nonumber  
\end{eqnarray}
These two pullbacks related by the ``Atkin'' involution 
$ \, \,  x \, \rightarrow \, 2/x$:
\begin{eqnarray}
\label{pull}
\hspace{-0.3in}u(x)   \, = \, \, {{27 \, x^4} \over { (4\, -\, x^2)^3}}, 
\qquad \quad 
v(x) \,\, = \, \,\, \,  u({{2} \over {x}}) \,\, \,  = \,  \, \, \,
{{27 \, x^2} \over {4 \cdot \, (x^2 -1)^3}},
\end{eqnarray}
corresponding, again, to the modular curve (\ref{encore2}).

\vskip .1cm 

{\bf Second example}. The HeunG function 
\begin{eqnarray}
\label{HeunGminus3}
\hspace{-0.6in}&&HeunG(-3, 0,1/2,1,1,1/2; \, 12\cdot x)
\, \, \, \, \, \,  \\
\hspace{-0.6in}&& \quad \, \,\, = \, \, \, \, (1\, +4 \, x)^{-1/4} \cdot \, 
HeunG\Bigl(4, \, {{1} \over {2}}, \, {{1} \over {2}}, \, 
{{1} \over {2}}, \, 1, \, {{1} \over {2}}, \, {{16 \, x} \over {1\, +4 \, x}}\Bigr)
\, \,\,  
 \nonumber \\
\hspace{-0.6in}&& \quad \quad \quad  \, \,\, = \, \, \, \,  \,\,\,\,
1\,\,\, \,\, +6\, x^2\,\, +24\,x^3\, \,  +252\,x^4\, \, +2016\,x^5  \, \,
+19320\,x^6\, \,  +183456\,x^7\,
 \nonumber \\
\hspace{-0.6in}&& \quad\qquad \qquad \quad  \quad 
  \,+1823094\,x^8 \,\, \, +18406752\,x^9 \,\, \, +189532980\,x^{10}
\, \, \,\,\, + \,\,\cdots 
\nonumber 
\end{eqnarray}
is solution of the order-two operator
\begin{eqnarray}
\label{Heunfcc}
\hspace{-0.8in}H_{fcc}\,\, \, = \,\,\, \, \, \, \, 
\theta^2 \,\, \,\,-2 \, x \cdot \theta \cdot (4\, \theta \, +\, 1)
 \, \, \,  \, 
-24 \cdot x^2 \cdot  (2\, \theta \, +\, 1) \cdot  (\theta \, +\, 1),
\end{eqnarray}
The  square of (\ref{HeunGminus3})
is actually the solution of an order-three operator
 (see equation (19) in~\cite{GoodGuttmann})
emerging for lattice Green functions of the 
face-centred cubic (fcc) lattice
which is thus the symmetric square of (\ref{Heunfcc}).
This  hypergeometric function with a polynomial
pull-back can also be written:
\begin{eqnarray}
\label{identitybis}
\hspace{-0.5in}&&HeunG(-3,\,  0,\, 1/2,\, 1,\, 1,\, 1/2; \, 12 \cdot x)
\, \,\, \, \, 
  \nonumber \\
\hspace{-0.5in}&&\qquad \, \, = \, \, \, \,  \, 
 _2F_1\Bigl([{{1} \over {6}}, \, {{1} \over {3}}],[1];\, 
 108\cdot x^2 \cdot(1+4\, x)\Bigr) \\
\hspace{-0.5in}&& \qquad \, \, = \, \, \,  \, 
(1-12\, x)^{-1/2} \, \cdot \,
 _2F_1\Bigl([{{1} \over {6}}, \, {{1} \over {3}}],[1];\, 
 -\, {{ 108 \cdot x \cdot(1+4\, x)^2} \over {(1-12\, x)^3}} \Bigr),
 \nonumber 
\end{eqnarray}
where the involution 
$\, x  \, \leftrightarrow \, -1/4\cdot (1+4\, x)/(1-12\, x)\, \, $
takes place.
The modular curve relating these two pullbacks reads {\em exactly 
the rational curve} (\ref{encore2}) already 
obtained in~\cite{CalabiYauIsing}.

\vskip .1cm 

{\bf Third example}. The HeunG function 
$\, HeunG(1/9, 1/12, 1/4, 3/4, 1, 1/2; \,  4\,x)$
is solution of the order-two operator corresponding to the 
simple cubic lattice Green function 
\begin{eqnarray}
\label{364bis}
\hspace{-0.6in}&&H_{sc} \, \, \, = \, \, \, \, \, \, \,\,
 \theta^2 \, \,  \, \, \,  \, -x \cdot (40\, \theta^2+20\, \theta\, +3)
\, \,  \,  \,
 +9 \cdot x^2  \cdot (4\, \theta\, +3) \cdot (4\, \theta\, +1). 
\nonumber 
\end{eqnarray}
The square of this HeunG function is a series with
{\em integer} coefficients which identifies with the
 Hadamard product of  $\, (1\, -4\, x)^{-1/2}$
with a modular form :
\begin{eqnarray}
\hspace{-0.9in}&& HeunG(1/9, 1/12, 1/4, 3/4, 1, 1/2; \,  4\,x)^2 
\, \,\,    \\
\hspace{-0.9in}&& \qquad \quad = \, \, \, \,\, \,
 (1-4\, x)^{-1/2}\, \star \, HeunG(1/9, 1/3, 1, 1, 1, 1; \, x)  
\nonumber \\
\hspace{-0.9in}&& \qquad \quad= \, \, \, \, \, \,\,
1\,  \,\,\, \,+6\, x\, \, + 90\, x^2 \,  \,+ 1860\, x^3 \, \,
 +44730\, x^4\, \,
+ 1172556\, x^5\, \,+ 32496156\, x^6\, \nonumber \\
\hspace{-0.9in}&& \qquad \quad \quad \quad \qquad  + 936369720\, x^7 \, \,
 + 27770358330\, x^8\, \, +842090474940\, x^9\,\, \, \,\,
\, + \, \, \cdots 
\nonumber
\end{eqnarray}
The HeunG function 
$\, HeunG(1/9, 1/12, 1/4, 3/4, 1, 1/2; \,  4\,x)$ is globally bounded:
 the series of 
$\, HeunG(1/9, 1/12, 1/4, 3/4, 1, 1/2; \,  8\,x)$ 
 is a series with integer coefficients. 
One can also write this HeunG function
in terms of a $\, _2F_1([1/6,1/3],\, [1],x)$ hypergeometric function 
up to a simple algebraic pullback (with a square root), or in terms 
 of a $\, _2F_1([1/8,3/8],\, [1],x)$ hypergeometric function:
\begin{eqnarray}
\label{even}
\hspace{-0.9in}&&HeunG(1/9, 1/12, 1/4, 3/4, 1, 1/2; \,  4\,x) 
\, \,  =  \, \,  \, \, \, C_2^{1/4} \, \cdot  \,
 _2F_1\Bigl([1/8,\,  3/8],\,  [1]; \, P_2   \Bigr), \quad \hbox{with:} 
\nonumber 
\end{eqnarray}
\begin{eqnarray}
\hspace{-0.9in}&&C_2\, \,  =  \, \, \,
 {{1} \over {9 \cdot \, (1\, +12\, x)^2}} \cdot
 \Bigl(5\, -36\, x \,+4\cdot (1-36\,x)^{1/2}\Bigr), \quad\quad
P_2\, \,  =  \, \, \, {{ 128 \cdot x} \over { (1\, +12 \, x)^4}} \cdot \, p_2, 
 \nonumber \\
\hspace{-0.9in}&&p_2\, \,  =  \, \, \, \, 
(1\, -42\, x \, +352\, x^2\, -288\, x^3) \, \, \,
 +\, (1\, -4\, x) \cdot (1\, -20\, x) \cdot \, (1\, -36 \, x)^{1/2}.
\nonumber
\end{eqnarray}
Do note that taking the Galois conjugate (changing $(1-36\,x)^{1/2}$ into 
$-(1-36\,x)^{1/2}$) gives the series expansion
of  $\,\,\, 3^{-1/2} \cdot \, HeunG(1/9, 1/12, 1/4, 3/4, 1, 1/2; \,  4\,x)$. 
This shows that there exists an identity for $\, _2F_1([1/8,3/8],\, [1],x)$ 
with {\em two different pullbacks}, namely the previous $\, P_2$  
and its Galois conjugate, these two pullbacks being related by a
(symmetric genus zero) modular curve:
\begin{eqnarray}
\label{A13}
\hspace{-0.9in}&&5308416 \cdot  y^4\, z^4\, \,  \,
+442368 \cdot y^3\, z^3 \cdot (y+z)\,\,  \,
+512\, y^2 \,z^2\cdot (27\,y^2\, +27\,z^2\, -27374\,x\, y)\,
\nonumber \\
\hspace{-0.9in}&&\quad \, \, \, \,   +192\, y\, z \cdot (y\,+z) 
\cdot (y^2\,+z^2\,+10718 \,y\, z)
\, \, \,\, +y^4\, +z^4\,\,  +3622662\, y^2\, z^2
\nonumber \\
\hspace{-0.9in}&&\, \, \, \, \,   \quad  \, 
-19332 \cdot y\, z \cdot \, (y^2\, +  \,z^2) \, \,
 +79872 \cdot y \, z \cdot (y\, +z)\, 
 -65536  \cdot y \, z\, \,\, \, = \,\,\, \, \,\, 0. 
\end{eqnarray}

\vskip .1cm 

{\bf Revisiting the examples}. In a recent paper~\cite{spanning} corresponding to 
spanning tree generating functions and Mahler measures, a result from Rogers 
(equation (36) in~\cite{spanning}) is given where the two following $\,\, _5F_4\,$
 hypergeometric functions take place:
\begin{eqnarray}
\hspace{-0.3in}&&_5F_4\Bigl([{{5} \over {4}}, \,
 {{3} \over {2}}, \, {{ 7} \over {4}}, \, 1, \, 1], \, 
[2, \, 2, \, 2, \, 2], \, \, {{ 256 \, x^3} \over {9 \cdot \, (x+3)^4 }} \Bigr), 
\nonumber \\
\hspace{-0.3in}&&
_5F_4\Bigl([{{5} \over {4}}, \, {{3} \over {2}}, 
\, {{ 7} \over {4}}, \, 1, \, 1], \, 
[2, \, 2, \, 2, \, 2], \, \, {{ 256 \, x} \over {
9 \cdot \, (1\, + \, 3\, x)^4 }} \Bigr). 
\end{eqnarray}
The corresponding order-five linear differential operators (annihilating
these two $\, _5F_4$
 hypergeometric functions) are actually homomorphic (the intertwiners 
being order-four operators).
The relation between these two pullbacks $\, y \, = \, \, 256 \, x^3/9/(x+3)^4$
and $\, z \, = \, \, 256 \, x/9/(1\, +3\, x)^4$, remarkably {\em gives, again,
 the previous} $(y, \, z)$-{\em symmetric modular curve} (\ref{A13}). 

The  order-five linear differential operator, corresponding to the first  $\, _5F_4$
 hypergeometric function, factorizes in an order-one operator,
an order-three operator and an order-one operator, the order-three
operator being, in fact, exactly the symmetric square of an order-two 
operator:
\begin{eqnarray}
\hspace{-0.1in} L_1 \,  \,  \cdot\,  \,  Sym^2(W_2)  \,  \,  \cdot \, \,   
{\frac {{x}^{4}}{ (x\, -9)  \, (x\, +3)^{4}}} \, \,   \cdot \, \,  R_1,
\nonumber 
\end{eqnarray}
where the order-one operators read respectively
\begin{eqnarray}
\hspace{-0.95in}&& L_1  = \,  D_x \, - \, {{d} \over {dx}} \ln\Bigl(  {{x-9} \over {
(9\,x^2\,+14\,x\,+9) \cdot \, (x+3)^4 }}  \Bigr), \quad \, \, \, \,
R_1 \, = \,  \, D_x \, \, - \, {{d} \over {dx}} \ln\Bigl( {{ (x+3)^4 } \over { x^3}}\Bigr), 
\nonumber
\end{eqnarray}
and where the order-two operator $\, W_2$ reads:
\begin{eqnarray}
\hspace{-0.9in}W_2 \,\, = \, \,\,\, \, D_x^2 \,  \, \, \,
 +3\,{\frac { ( 6 \cdot \,{x}^{2} \, +7\,x \, +3)
 }{ ( 9\,{x}^{2}\, +14\,x\,+9)\cdot \,  x}}
\cdot \, D_x 
 \,\, \, + \,  {{3} \over {4}} \cdot \,
{\frac {3\,x \, +2}{ (9\,{x}^{2}\,+14\,x\,+9)\cdot \,  x}}. 
\end{eqnarray}

We have a similar result for the  order-five linear differential operator 
corresponding to the second  $\, _5F_4$ hypergeometric function.

Another solution of this order-five linear differential operator reads:
\begin{eqnarray}
\label{3F2pull}
\hspace{-0.8in}\,\,\, {{(x\, +3)^4} \over {x^3}} \cdot \, 
\int \, {{x\, -9 } \over {( x+3) \cdot \, x }} 
 \, \cdot  \,  \, 
_3F_2\Bigl( [{{1} \over {4}}, \, {{1} \over {2}}, \, {{ 3} \over {4}}], \, 
[1, \, 1], \, \, {{ 256 \, x^3} \over {9 \cdot \, (x+3)^4 }}  \Bigr)
     \, \cdot \, dx. 
\end{eqnarray}
The expansion of the $\, _3F_2$ hypergeometric function 
in (\ref{3F2pull}) is globally bounded
(change $\,\, x \, \rightarrow \, 9 \, x\,$ to get a series with integer coefficients). 

Recalling the two previous pullbacks we have, in fact, the following identity:
\begin{eqnarray}
\label{followident}
\hspace{-0.5in}&& 3 \cdot \, (1 \, + \, 3 \, x) \cdot \, \,
 _3F_2\Bigl( [{{1} \over {4}}, \, {{1} \over {2}}, \, {{ 3} \over {4}}], \, 
[1, \, 1], \, \, {{ 256 \, x^3} \over {9 \cdot \, (x+3)^4 }}  \Bigr)
\nonumber \\
\hspace{-0.5in}&& \qquad \quad  \,  \, = \,\, \,  \,  \, \,
 (x\, +\, 3) \cdot \, \,  _3F_2\Bigl( [{{1} \over {4}},
 \, {{1} \over {2}}, \, {{ 3} \over {4}}], \, 
[1, \, 1], \, \, {{ 256 \, x} \over {9 \cdot \, (1\, + \, 3\, x)^4 }}  \Bigr).
\end{eqnarray}
However this $\, _3F_2$ hypergeometric function is nothing but the
square of a $\, _2F_1$ hypergeometric function
\begin{eqnarray}
\hspace{-0.3in} _3F_2\Bigl( [{{1} \over {4}}, \, {{1} \over {2}}, \, {{ 3} \over {4}}], \, 
[1, \, 1], \, \, x \Bigr)
 \,  \,\, \,  = \, \, \, \,  \,  \,
 _2F_1\Bigl( [{{1} \over {8}}, \,  {{ 3} \over {8}}], \, 
[1], \, \, x  \Bigr)^2.
\end{eqnarray}
Thus, the previous identity (\ref{followident}) is nothing but the
 identity on a $\, _2F_1$ hypergeometric function with {\em two different
pullbacks}:
\begin{eqnarray}
\label{newident}
\hspace{-0.6in}&& (1+3\,x)^{1/2} \cdot \, \,
  _2F_1\Bigl( [{{1} \over {8}}, \,  {{ 3} \over {8}}], \, 
[1], \, \,  {{ 256 \, x^3} \over {9 \cdot \, (x+3)^4 }} 
 \Bigr) \nonumber \\
\hspace{-0.6in}&& \qquad \quad  \quad \,  \, = \, \,  \, \,   \,   \, 
\Bigl(1 \, +{{x} \over {3}}\Bigr)^{1/2} \, \cdot \, 
 _2F_1\Bigl( [{{1} \over {8}}, \,  {{ 3} \over {8}}], \, 
[1], \, \,  {{ 256 \, x} \over {9 \cdot \, (1\, + \, 3\, x)^4 }}   \Bigr).
\end{eqnarray}
The expansion of (\ref{newident}) is globally bounded. One gets a series 
with {\em positive integer}  coefficients using the simple rescaling
 $\, x \, \rightarrow \, 36 \cdot \, x$.
Note that the two pullbacks can be exchanged by the simple ``Atkin'' involution
 $\, x \, \leftrightarrow \, \, 1/x$, being related by the modular curve 
occurring for the simple cubic lattice, namely (\ref{A13}).

We have a similar result for the  other $\,\, _5F_4\,$
 hypergeometric functions popping out in~\cite{spanning}.

 For instance, for 
the diamond lattice one gets an expression (see eq. (50)
 in~\cite{spanning}) where the two following $\, _5F_4$
 hypergeometric functions take place\footnote[2]{Note a small misprint 
in eq. (50) of~\cite{spanning}: one should read $\, -27 z^2/4/(1-z^2)^3$ 
instead of $\, -27 z^4/4/(1-z^2)^3$.}:
\begin{eqnarray}
\label{diampull}
\hspace{-0.3in}&&_5F_4\Bigl([{{5} \over {3}}, \,
 {{3} \over {2}}, \, {{ 4} \over {3}}, \, 1, \, 1], \, 
[2, \, 2, \, 2, \, 2], \, \, {{ - 27 \, x^2 } \over { 4 \cdot \, (1\, - \, x^2)^3 }} \Bigr), 
\nonumber \\
\hspace{-0.3in}&&
_5F_4\Bigl([{{5} \over {3}}, \, {{3} \over {2}}, 
\, {{ 4} \over {3}}, \, 1, \, 1], \, 
[2, \, 2, \, 2, \, 2], \, \,  {{  27 \, x^4 } \over {  (4\, - \, x^2)^3 }}  \Bigr). 
\end{eqnarray}
These two pullbacks can be exchanged by the simple ``Atkin'' involution
 $\, x \, \leftrightarrow \, \, 2/x$. These two pullbacks have been seen to be
related by the (genus-zero) $(y, \, z)$-symmetric modular curve (\ref{encore2}):
\begin{eqnarray}
\label{encore3}
\hspace{-0.6in}&&4\,{y}^{3} \, {z}^{3} \, \, \,   -12\,{y}^{2} \, {z}^{2} \cdot \, (y \, +z)\,
  \,  \, \,  +3\,y\, z \left( 4\,{y}^{2}+4\,{z}^{2}-127\,y \, z \right)\,
 \nonumber \\ 
\hspace{-0.6in}&&\qquad \qquad \quad   -4 \cdot \, (y\, +z) \cdot \, 
 (y^2 \, + \, {z}^{2}+83\, y\, z) \, \, \, \,   +432\, y\, z
 \, \,\,\,\,   = \,\, \,\,\,  \, 0. 
\end{eqnarray}

Similarly to (\ref{followident}) we have an identity between two
$\, _3F_2$ hypergeometric functions 
(namely $\, _3F_2([2/3,\,1/2, \, 1/3],[1, \, 1],z)$) with 
the two pullbacks (\ref{diampull}), and these
$\, _3F_2$ hypergeometric functions being the square of 
$\, _2F_1$ hypergeometric functions, one finds that the 
``deus ex machina'' is the 
identity similar to (\ref{newident}):
 \begin{eqnarray}
\label{newidentdiam}
\hspace{-0.6in}&& (1 \,- x^2)^{1/2} \cdot \, \,
  _2F_1\Bigl( [{{1} \over {3}}, \,  {{ 1} \over {6}}], \, 
[1], \, \,   {{  27 \, x^4 } \over {  (4\, - \, x^2)^3 }} 
 \Bigr) \nonumber \\
\hspace{-0.6in}&& \qquad \quad  \quad \,  \, = \, \,  \, \,   \, \,
(1 \, -{{x^2} \over {4}})^{1/2} \cdot \, 
 _2F_1\Bigl( [{{1} \over {3}}, \,  {{ 1} \over {6}}], \, 
[1], \, \,  {{ - 27 \, x^2 } \over { 4 \cdot \, (1\, - \, x^2)^3 }}   \Bigr).
\end{eqnarray}
The series expansion of (\ref{newidentdiam}) 
is globally bounded. Rescaling the
$\, x$ variable as $\, x \,\rightarrow  \, 4 \, x$, the series expansion 
becomes a series with {\em positive integer} coefficients 
(up to the first constant term).

 For the face-centred cubic lattice one gets an expression (see eq. (52)
 in~\cite{spanning}) where the two following $\, _5F_4$
 hypergeometric functions take place\footnote[1]{There is one more misprint
in~\cite{spanning}:
the pullback $\, -x \, (x+3)/(x-1)^3$ must be changed into
 $\, x \, (x+3)/(x-1)^3$.}:
\begin{eqnarray}
\label{fcc}
\hspace{-0.3in}&&_5F_4\Bigl([{{5} \over {3}}, \,
 {{3} \over {2}}, \, {{ 4} \over {3}}, \, 1, \, 1], \, 
[2, \, 2, \, 2, \, 2], \, \, {{  x \cdot \, (x\,+3)^2} \over { (x\, -1)^3 }} \Bigr), 
\nonumber \\
\hspace{-0.3in}&&
_5F_4\Bigl([{{5} \over {3}}, \, {{3} \over {2}}, 
\, {{ 4} \over {3}}, \, 1, \, 1], \, 
[2, \, 2, \, 2, \, 2], \, \, {{  x^2 \cdot \, (x\,+3)} \over { 4 }}  \Bigr). 
\end{eqnarray}
This example is nothing but the previous diamond lattice example (\ref{diampull}) 
with the change of variable $\, x \, \rightarrow \, -3\, x^2/(x^2-4)$
in (\ref{fcc}). Therefore, the two pullbacks in (\ref{fcc}) are, again,
related by the modular curve (\ref{encore2}). The two pullbacks in (\ref{fcc}) can 
actually be seen directly in the following identity
 (equivalent to (\ref{newidentdiam})): 
\begin{eqnarray}
\hspace{-0.9in}&&\quad  _2F_1\Bigl( [{{1} \over {3}}, \,  {{ 1} \over {6}}], \, 
[1], \, \,    {{  x \cdot \, (x\,+3)^2} \over { (x\, -1)^3 }}   \Bigr)
\,  \,\, = \, \,  \, \,   \, \, 
 (1 \,- x^2)^{1/2} \cdot \, \,
  _2F_1\Bigl( [{{1} \over {3}}, \,  {{ 1} \over {6}}], \, 
[1], \, \,   {{  x^2 \cdot \, (x\,+3)} \over { 4 }} 
 \Bigr).
\nonumber 
\end{eqnarray}

Finally, the equation (17) of~\cite{spanning} on Mahler measures, 
the two following $\, _4F_3$
 hypergeometric functions take place:
\begin{eqnarray}
\label{lastpull}
\hspace{-0.3in}&&_4F_3\Bigl([{{5} \over {3}}, \,
 {{4} \over {3}},  \, 1, \, 1], \, 
[2, \, 2, \, 2], \, \, {{ 27 \, x} \over { (x \, -2)^3 }} \Bigr), 
\nonumber \\
\hspace{-0.3in}&&
_4F_3\Bigl([{{5} \over {3}}, \,
 {{4} \over {3}},  \, 1, \, 1], \, 
[2, \, 2, \, 2], \, \, {{ 27 \, x^2} \over { (x \, +4)^3 }} \Bigr). 
\end{eqnarray}
These two previous pullbacks can be exchanged by an ``Atkin'' involution
$\, x \, \leftrightarrow \, -8/x$ and are related by the (genus-zero)
$(y, \, z)$-symmetric modular curve: 
\begin{eqnarray}
\hspace{-0.6in}&&8\,{y}^{3} \, {z}^{3} \,  \,  \, 
-12\,\,{y}^{2}{z}^{2}\cdot \,  (y \, +z) \,  \, \, 
  +3\, y \, z \cdot \, ( 2\,{y}^{2} +2\,{z}^{2} \, +13\,y\, z)
  \nonumber \\
\hspace{-0.6in}&& \quad \quad \quad \quad \quad \quad \,  
\, \,  - \, (y \,+ z)  \cdot \, (y^2 \, +{z}^{2} \, +29\,y \, z) \, \,  \,  +27\,y \, z 
\,  \,  \,  \,\,\, =  \, \,   \,  \, \,  \, 0.  
\end{eqnarray}
The underlying identity on $\, _2F_1$ hypergeometric functions with
the two pullbacks (\ref{lastpull}) read:
\begin{eqnarray}
\label{newidentdiam3}
\hspace{-0.6in}&& -\, 2 \cdot \, (x \,- 2) \cdot \, \,
  _2F_1\Bigl( [{{1} \over {3}}, \,  {{ 2} \over {3}}], \, 
[1], \, \,   {{  27 \, x^2 } \over {  (x \, + \, 4)^3 }} 
 \Bigr) \nonumber \\
\hspace{-0.6in}&& \qquad \quad \quad \quad \,  \, = \, \,  \, \,   \,   \,   \, 
\, (x \, + \, 4)\cdot \, 
 _2F_1\Bigl( [{{1} \over {3}}, \,  {{ 2} \over {3}}], \, 
[1], \, \,  {{  27 \, x } \over { (x \, - \, 2)^3 }}   \Bigr).
\end{eqnarray}
The series expansion of (\ref{newidentdiam3}) is globally bounded. Rescaling the
$\, x$ variable as $\, x \,\rightarrow  \, -8 \, x$, the series expansion 
becomes a series with {\em positive integer} coefficients.

\vskip .1cm 

\section{Seeking for the ``minimal''  rational function}
\label{minimalrational}

 The  effective calculations of section (\ref{theorem}) guarantee to obtain
 an {\em explicit expression} for 
the rational function associated with (\ref{gooddiag}), however the rational function is 
far from being unique.
 Recalling the well-known Ap\'ery series $\, \mathcal{A}(x)$, 
and its rewriting due to Strehl and Schmidt~\cite{Strehl,Schmidt,Zudilin},
\begin{eqnarray}
\label{Aper}
\hspace{-0.9in}&&\mathcal{A}(x) \,\, =  \, \, \sum_{n=0}^{\infty} \, \sum_{k=0}^{n}
 \, \, {n\choose k}^2 \,  {n+\, k \choose k}^2 \cdot x^n 
\, \, \,\, = \,\,  \,\,  \,  \,
\sum_{n=0}^{\infty} \, \sum_{k=0}^{n} \, \sum_{j=0}^{k}
 \, \, {n\choose k} \,  {n+\, k \choose k} {k\choose j}^3 \,  \cdot x^n 
\nonumber  \\
\hspace{-0.9in}&& \quad  \quad \quad  \quad  \qquad 
\, \, = \, \, \, \,  \,\,  \,\,
1 \, \, \,\, \,  + 5\, x \, \,  \, + \, 73 \, x^2 \, \,  \, +1445 \, x^3 
\, \,  \, + \, 33001 \, x^4 
\,\,  \,\,  \, \, + \, \cdots,  
\end{eqnarray}
$\mathcal{A}(x)$ is known to be 
the diagonal of the rational function in five variables 
$\, 1/R_1/R_2$
where $\, R_1, \, R_2$ read~\cite{Christol84}: 
\begin{eqnarray}
\hspace{-0.9in}\,\, R_1 \, \, = \, \, \, \,  \,
 1 \, - \, z_0, \quad \quad \quad \, \,\, \,
R_2 \, \, = \, \, \, \,  \,
 (1 \, - \, z_1)(1 \, - \, z_2) (1 \, - \, z_3)(1 \, - \, z_4) \, \, 
- \, z_0 z_1 z_2, \nonumber
\end{eqnarray}
as well as the diagonal of the rational function in five variables 
$\, 1/Q_1/Q_2$
where $\, Q_1, \, Q_2$ read~\cite{Christol85,Christol}:
\begin{eqnarray}
\hspace{-0.9in}\,\, \,\, Q_1 \, \, = \, \, \, \,  \,
 1 \, - \, z_1\,  z_2\,  z_3\, z_4, \quad \quad \, \,\, \, 
Q_2 \, \, = \, \, \, \,  \, (1 \, - \, z_3)(1 \, - \, z_4) \, \, \,  
- \, z_0 \cdot \, (1 \, + \, z_1)(1 \, + \, z_2), \nonumber
\end{eqnarray}
and {\em also}  the diagonal of the rational function 
in six variables $\, 1/P_1/P_2/P_3$
where $\, P_1, \, P_2, \, P_3$ read~\cite{Christol84}:
\begin{eqnarray}
\hspace{-0.9in} P_1  \, = \, 
 1\,\, -z_0 \, z_1 ,\quad \, \,\,  P_2  \, = \, 
1\,\, -z_2 \, - z_3 \, - z_0 \, z_2 \, z_3,  \quad \, \,\,  P_3  \, = \, 
1\,\, -z_4 \, - z_5 \, - z_1 \, z_4 \, z_5. \nonumber
\end{eqnarray}

A yet different diagonal representation for the Ap\'ery series, due 
to Delaygue\footnote[1]{Personal communication.}, is provided by 
the diagonal of the rational function in eight variables:
\begin{eqnarray}
\hspace{-0.95in}{{1} \over {(1\; \, - \; z_4 z_5 z_6 z_7)
 \cdot \,  (1\; - \; z_0 \cdot \,(1+z_4))
 \cdot(1 \, -z_1 \cdot \,(1+z_5)) \cdot (1-z_2-z_6) \cdot (1-z_3-z_7)}}.
\nonumber 
\end{eqnarray}

Calculations similar to  (\ref{112bis}) on these new binomial
expressions 
provide two new rational functions such that (\ref{Aper}) 
can be written as the diagonal of one of these two rational functions.
One is a rational function of five variables, of the form $\, 1/Q^{(5)}_1/Q^{(5)}_2$ 
\begin{eqnarray}
\hspace{-0.6in}&&Q^{(5)}_1 \, \,\, = \, \,  \, \, \,\,
 1\,\,\, \, -z_0 \, z_1 \, z_2 \, z_3 \, z_4  \, \cdot \,
(1+z_1)\, (1+z_2)\, (1+z_3)\, (1+z_4),
\nonumber  \\
\hspace{-0.6in}&&Q^{(5)}_2 \, \,\, = \, \,  \, \, \,\,
1\,\,\, \, -z_0\,  \cdot \,(1+z_1)\, (1+z_2)\, (1+z_3)^2\, (1+z_4)^2,
\end{eqnarray}
and the other one, is a rational function of six variables, 
 of the form $\, 1/Q^{(6)}_1/Q^{(6)}_2/Q^{(6)}_3$ 
\begin{eqnarray}
\hspace{-0.6in}&&Q^{(6)}_1 \, \, = \, \,  \,  \,  \,\,
 1\,\, \, \,  -z_0 \, z_3 \, z_4 \, z_5 \, \cdot
 \,(1+z_1)\, (1+z_2)^2\, (1+z_3)\, (1+z_4)\, (1+z_5),
\nonumber  \\
\hspace{-0.6in}&&Q^{(6)}_2 \, \, = \, \,  \, \, \, \,
1\,\, \,  \, -z_0\,z_1\, z_2 \, z_3 \, z_4 \, z_5 \cdot \,(1+z_1)\, (1+z_2),
\nonumber  \\
\hspace{-0.6in}&&Q^{(6)}_3 \, \, = \, \,  \, \,  \,\,
1\,\, \,  \, -z_0 \,\cdot  \,(1+z_1)\, (1+z_2)^2\, (1+z_3)\, (1+z_4)\, (1+z_5).
\end{eqnarray}

\section{Hypergeometric series with coefficients ratio of factorials}
\label{ratiooffac}

As a consequence of the classification by Beukers and Heckman~\cite{BeHe89}
 of all algebraic $_{n}F_{n-1}$'s, 
the $\, \, _8F_7$ hypergeometric series
\begin{eqnarray}
\hspace{-0.75in}_8F_7\biggl(
\left[\frac{1}{30}, \, \frac{7}{30},\,\frac{11}{30},\,
\frac{13}{30},\,\frac{17}{30},\,\frac{19}{30},\,\frac{23}{30},\,
\frac{29}{30}\right],\,
\left[\frac{1}{5},\,\frac{2}{5},\,\frac{3}{5},\,\frac{4}{5},\,
\frac{1}{2},\,\frac{2}{3},\,\frac{1}{3}\right],
\,\, \, 
2^{14} \, \, 3^9 \, \, 5^5   \, \, x \biggr),  
\nonumber 
\end{eqnarray}
has {\em integer coefficients}, and is an {\em algebraic function}. 
The Galois group belonging to this function is the Weyl group $\, W(E_8)$
which has 696729600 elements~\cite{Varilly}.
It is an {\em algebraic series} of degree 483840.
More precisely, it was noticed by Rodriguez-Villegas~\cite{Villegas}
 that the previous power series reads: 
\begin{eqnarray}
\sum_{n =0}^{\infty} \,\,
 {{(30 \, n)! \, \, n!} \over {(15 \,n)! \,  \,
 (10 \,n)! \, \, (6 \,n)!}} \cdot \, x^n,
\end{eqnarray}
which is closely related to the series introduced by Chebyshev 
in his work~\cite{Chebyshev} on the
 distribution of prime numbers to establish 
the estimate~\cite{Big} 
on the prime counting function $\, \pi(x)$.

\vskip .1cm  

Considering hypergeometric series such that their coefficients are ratio
of factorials, reference~\cite{Villegas} 
 gives the conditions of these factorials for the hypergeometric series to be 
algebraic (all the coefficients are thus integers). A simple example is, 
for instance the algebraic function:
\begin{eqnarray}
\hspace{-0.4in}\quad _3F_2\Bigl([{{1} \over {4}}, \,{{1} \over {2}}, \, {{3} \over {4}}], \, 
[{{1} \over {3}}, \,{{2} \over {3}}]; \, \,  {{2^8} \over{3^3}} \cdot \, x\Bigr) 
\, \, \, \, = \,\,\, \, \,  \,  \, \sum_{n=0}^{\infty} \, \, {4 \, n \choose n} \cdot \, x^n. 
\end{eqnarray}

\vskip .1cm

Along this line it is worth recalling Delaygue's Thesis~\cite{Delaygue} 
(see also Bober~\cite{Bober})
which gives some results\footnote[8]{Necessary and sufficient conditions for
the integrality of the mirror maps series.} 
for series expansions\footnote[5]{These series are not algebraic functions.}
 such that their coefficients are
{\em ratio of factorials}, namely $\, _2F_1([1/3, \, 2/3], [1]; \, 27 \, x) \,$,
and $\, _4F_3([1/2, 1/2,1/2, 1/2], \, [1,1,1]; \, 256\, x)$, 
giving respectively the series 
\begin{eqnarray}
\hspace{-0.7in}&& \quad \quad \quad 
 \sum_{n=0}^{\infty} \, {{(3n)!} \over {(n!)^3}}   \cdot \, x^n, \quad \, \, 
\quad\quad 
 \sum_{n=0}^{\infty} \, {{((2n)!)^4 } \over { (n!)^8}} \cdot \, x^n,
 \qquad \quad \quad \hbox{and:}
\nonumber 
\end{eqnarray}
\begin{eqnarray}
\hspace{-0.9in}&&\quad _4F_3 \Bigl([{{1} \over {2}}, {{1} \over {2}}, 
{{1} \over {6}}, {{5} \over {6}}],[1,1,1]; \,\,\, 2^8 \, \, 3^3 \cdot \,  x\Bigr)
\, \,\,  \,= \, \, \, \, \,  \, \,
\sum_{n=0}^{\infty} \, {{(6n)! \, \, (2n)! } \over {(3n)! \, (n!)^{5}}} \cdot \, x^n.
\end{eqnarray}
These ratio of factorials are all {\em integer numbers}.

\vskip .1cm 

\section{Proof of integrality of series (\ref{contre1})}
\label{proof}

Let us sketch the proof of the integrality of series (\ref{contre1}),
namely, the integrality of coefficients (\ref{far1}). 
For each power of the integer number $\, q \, = \, \, p^n$ 
a term like $\, 4\, + 9 \, n$ is periodically divisible (period $p$) 
by $\, q$. In order to have the ratio (\ref{ratio1}) be an integer, 
one needs the numerator to be divisible by this factor $\, q$
{\em before} the denominator. The case $\, p\, = \, 3$ is an easy one.
The other prime $\, p$ do not divide $\, 9$. One needs to find 
the first case of divisibility, namely the first integer $\, n$ such that 
$\,4\,+9\,n\, =\,\, k\, q$ (this corresponds to the smallest $\, k$). 
If $\, d \, q \, = \, \, 1$, $mod.\,  9$ then 
$\, k \, = \, \, 4 \, d$, $mod.\,  9$.
In other words, the smallest $\, k$ is the rest of  $4 \, d$, $mod.\, 9$. 
Consequently, we have replaced the calculations, for every integer $\, q$,
by a {\em finite set of} calculations for 
$\, d\, = \, \, 1, \, 2, \, 4, \, 5, \, 7, \, 8$.
Let us use this approach for the ratio (\ref{ratio1}). 

\vskip .1cm
{\bf Remark:} The terms $\, n\, +1$ are always the last to be divisible by $\, q$.
Hence, one can forget its factors. However, one needs as many factors 
at the numerator than at the denominator.
For the other terms, the following table of the rest of  $\, d \cdot a$
gives the complete proof:
\begin{eqnarray}
&&. \quad 	1\quad 	2\quad 	4\quad 	5\quad 	7\quad 	8 \nonumber  \\
&&1\quad 	1\quad 	2\quad 	4\quad 	5\quad 	7\quad 	8 \nonumber \\
&&4\quad 	4\quad 	8\quad 	7\quad 	2\quad 	1\quad 	5\nonumber \\
&&5\quad 	5\quad 	1\quad 	2\quad 	7\quad 	8\quad 	4\nonumber \\
&&3\quad 	3\quad 	6\quad 	3\quad 	6\quad 	3\quad 	6\nonumber \\
\end{eqnarray}
One finds out that this is always a factor of the numerator, before the occurrence
of a factor at the denominator.

\vskip .1cm 

\section{Yukawa couplings}
\label{Yukawaratio}

\subsection{Yukawa couplings as ratio of determinants}
\label{Yukawaratio1}

Consider an order-four MUM linear differential operator.
Let us introduce the determinantal variables $\, W_m \, = \, \, \det(M_m)$ 
which are the determinants\footnote[2]{For an order-four 
operator the Wronskian is $\, W_4$.}
of  the following $\, m \times m$ matrices $\, M_m$,
 $\, m\,= \,  \,  1, \, \ldots, \, 4$, 
with entries expressed in terms of derivatives of the four solutions
$y_0(x)$, $y_1(x)$, $y_2(x)$ and $y_3(x)$ 
of the MUM linear differential operator 
(see Section (\ref{planb}) for the definitions). 
One takes
$W_1 (x) \,\, = \, \, \,   y_0 (x)\, \, $ and:
\begin{eqnarray}
\label{filtration}
\hspace{-0.9in}&&M_2 \, = \, \,  
 \left[ \begin {array}{cc} 
y_0  &y_1
 \\ \noalign{\medskip} y_0'  &y_1' 
 \end {array}
 \right], \quad 
M_3 \, = \, \,  
\left[ \begin {array}{ccc} y_0  &y_1
  &y_2  \\ 
\noalign{\medskip} y_0'  &y_1' &y_2' \\ 
\noalign{\medskip}y_0'' &y_1''  &y_2'' 
 \end {array} \right],   
 \quad M_4 \, = \, \,  
 \left[ \begin {array}{cccc} 
y_0  &y_1  &y_2  &y_3  \\ 
\noalign{\medskip}y_0'  &y_1'  &y_2' &y_3' \\ 
\noalign{\medskip}y_0''  &y_1''  &y_2'' 
 &y_3''  \\ 
\noalign{\medskip}y_0'''  & y_1'''  &y_2'''  &y_3'''
  \end {array} \right],
  \nonumber \\
\hspace{-0.9in}&&\, \,  \hbox{where:} \qquad  \quad  
y_i' \, = \, \, {\frac {d}{dx}}{y_i}, 
\quad \quad \quad 
y_i'' \, = \, \, {\frac {d^2}{dx^2}}{y_i}, 
\quad \quad  \quad 
y_i''' \, = \, \, {\frac {d^3}{dx^3}}{y_i}. 
\end{eqnarray}

Since $q$ is equal to $\,  \, q \, = \, \, \exp(y_1/y_0)$, and its derivative verifies
\begin{eqnarray}
\label{derivq}
 q \cdot {{d} \over {dq }} 
 \, \, \, = \, \, \,  \, \, 
{{W_1^2} \over {W_2}} \cdot  {{d} \over {dx }} \, \, \, = \, \, \,  \, \, 
{{y_0^2} \over {W_2}} \cdot  {{d} \over {dx }},
\end{eqnarray}
we have that 
\begin{eqnarray}
\hspace{-0.5in}\Bigl(q \cdot {{d} \over {dq }}  \Bigr)^2
\, \,  \, \, = \, \, \, \,  \, \,  \, 
{{y_0^4} \over {W_2^2}}\,  \cdot \,  {{d^2} \over {dx^2 }} \,\,  
\, + \, \, 2 \, {{y_0^3} \over {W_2^2}}\,  
 {{d y_0} \over {dx }} \cdot \,  {{d} \over {dx }} \, \, 
\,  - {{y_0^4} \over {W_2^3}}\,  {{d W_2} \over {dx }}
  \cdot \,  {{d} \over {dx }}.
\end{eqnarray}
We deduce, after some simple algebra,
 an alternative definition for the 
{\em Yukawa coupling}: 
\begin{eqnarray}
\label{Yukawa}
\hspace{-0.2in}K(q) \,\, = \, \, \, \,
 \Bigl( q \cdot {{d} \over {dq }} \Bigr)^2
 \Bigl(  {{y_2} \over {y_0}}\Bigr)
 \, \,\, = \, \,\,\, \,\, 
 {{W_1^3 \cdot W_3 } \over {W_2^3 }}
\, \,\, = \, \,\,\, \,\, 
 {{y_0^3 \cdot W_3 } \over {W_2^3 }}. 
\end{eqnarray}
to be compared with the other previous 
alternative expression previously given (\ref{otherYuk}) 
for the {\em Yukawa coupling} 
\begin{eqnarray}
\label{otherYuk2}
\hspace{-0.65in}K(q) \, \, \, = \, \, \, \,
  {{ x(q)^3 \, \cdot W_4^{1/2}} \over { y_0^2}} \cdot
 \, \Bigl({{q} \over {x(q)}} \cdot {{d x(q)} \over {dq}}\Bigr)^3
\, \, = \, \, \, \, 
  {{ W_4^{1/2}} \over { y_0^2}} \cdot
 \, \Bigl( q \cdot {{d x(q)} \over {dq}}\Bigr)^3.
\end{eqnarray}

In fact from (\ref{derivq}) we deduce 
\begin{eqnarray}
 \Bigl(q \cdot {{d x(q)} \over {dq }} \Bigr)^3
 \, \, \, = \, \, \,  \, \, 
{{W_1^6} \over {W_2^3}}  \, \, \, = \, \, \,  \, \, 
{{y_0^6} \over {W_2^3}},
\end{eqnarray}
and, so, (\ref{otherYuk2}) is compatible with (\ref{Yukawa})
{\em if the following identity is verified}:
\begin{eqnarray}
\label{condmagic}
  W_3^2  
\, \, \, = \, \, \, \,\,   W_4 \cdot \, y_0^2 
\,\,  \, = \, \, \, \,\,   W_4 \cdot \, W_1^2. 
\end{eqnarray}
This identity is in fact specific of {\em order-four operators 
conjugated to their adjoints} (see below (\ref{cond})). 
Therefore we prefer to use definition (\ref{Yukawa}) 
for the Yukawa coupling, instead of the more restricted 
definition (\ref{otherYuk2}).

\vskip .1cm 

Let us assume that the pullback $\, p(x)$ 
has a series expansion of the form
\begin{eqnarray}
\label{formpull}
p(x) \, \,  \, = \, \, \, \,\,   \lambda \cdot \, x^r \cdot \, A(x), 
\end{eqnarray}
where the exponent $\, r$ is an integer, where $\,\lambda $
is a constant, and where $\, A(x)$ is a function analytic at
 $\, x \, = \, \, 0$ with the series expansion:
\begin{eqnarray}
\hspace{-0.1in}A(x) \, \, = \, \, \, \,\, \, \,  
 1\,\, \,\,   \, + \alpha_1 \cdot \, x\,  \,  \,
+ \alpha_2 \cdot \, x^2\, \, \,  \, +\, \cdots 
\nonumber 
\end{eqnarray}
The determinantal variables $\, W_m$ transform 
very nicely under pullbacks $\, p(x)$ of the
form (\ref{formpull}):
\begin{eqnarray}
\label{covar}
\hspace{-0.98in}&&(W_1(x), \, W_2(x), \, W_3(x), \, W_4(x)) 
 \, \,\, \, \quad  \longrightarrow  \\
\hspace{-0.98in}&&\qquad 
\Bigl(W_1(p(x)),\, \,{{p'} \over {r}}  \cdot \,W_2(p(x)), \,\, 
 {{p'^3} \over {r^3}}  \cdot \, W_3(p(x)), \,\, 
 {{p'^6} \over {r^6}} \cdot \, W_4(p(x))\Bigr),
 \quad \, \, \, \,    
p' \, = \, \, \, {{d p(x)} \over {d x}}. \nonumber 
\end{eqnarray}

One can show that the nome (\ref{nome}) of an order-$\, N$
operator transforms under a pullback $\, p(x)$:
\begin{eqnarray}
\label{previous}
\hspace{-0.2in}q(x) \, \, \longrightarrow  \, \, Q(x)
 \qquad \, \,   \hbox{with:} \qquad \quad  \,  \, 
 \lambda \cdot \, Q(x)^r \, \, = \, \, \, q(p(x)). 
\end{eqnarray}

{}From the covariance property (\ref{covar}), and 
from the previous transformation
$\, q \, \rightarrow \, \lambda \cdot \,q^r$ for the nome,   
one easily gets the transformation of the Yukawa coupling seen
as a function of the nome
$\,\, K(q) \, \, \rightarrow \, \, K(\lambda \cdot \, q^r)$:
\begin{eqnarray}
\label{Yukawachange}
\hspace{-0.9in}&&K(q(x)) \,
  \, \,\, = \, \,\,\, \,\, 
 {{W_1(x)^3 \cdot W_3(x) } \over {W_2(x)^3 }} \\
\hspace{-0.9in}&& \quad \quad \quad \quad \quad  \,\,\,  
 \, \, \longrightarrow   \quad \quad  \, \, 
 {{W_1(p(x))^3 \cdot W_3(p(x)) } \over {W_2(p(x))^3 }}
\,\,\,  \,  = \, \, \,  \, \,K(q(p(x)))
\,\, = \, \, \, K(\lambda \cdot \, Q(x)^r).
 \nonumber 
\end{eqnarray}

For $\, \lambda \, = \, \, 1$ and $\, r \, = \, \, 1$ (i.e. when
 the pullback is a deformation of the 
identity transformation), one recovers the known 
invariance of the Yukawa coupling 
by pullbacks (see Proposition 3 in~\cite{Prop3}).

\vskip .1cm 

One finds {\em another}
pullback invariant ratio, namely:
\begin{eqnarray}
\label{Kstar}
 K^{\star} \, \,\,  = \, \, \, \,\, \,\,  
{{W_1 \cdot W_3^3} \over {W_4 \cdot W_2^3 }},
\end{eqnarray}
which is, in fact, 
{\em nothing but the Yukawa coupling for the adjoint} 
of the original operator.
\vskip .1cm
Another invariance property is worth noting. Let us consider 
two linear differential operators $\, \Omega_1$ and $\, \Omega_2$ 
of order $\, N$ that are equivalent, in 
the sense of the equivalence of linear differential operators. This 
means that there exists  linear differential operators intertwiners $\, I_1$,
 $\, I_2$,  $\, J_1$, $\, J_2$, of order at most $\, N-1$  
such that 
\begin{eqnarray}
\hspace{-0.2in}\Omega_1 \cdot I_1 \,\,\, = \, \,\, \, I_2 \cdot \Omega_2, 
\quad \, \quad\hbox{and:}   \quad \,\, \quad \,
J_1 \cdot \Omega_1 \,\,\, = \, \,\,\,  \Omega_2 \cdot J_2.
\end{eqnarray}
Let us assume that one of these intertwiners is a linear 
differential operator
of order zero (a function), then the previous homomorphism between
 operators amounts to saying that
the two operators are conjugated by a function:
\begin{eqnarray}
\label{conju}
\hspace{-0.1in}\Omega_2 \,\,\,\, = \, \,\,\, \,\, 
  \rho(x)  \cdot \Omega_1 \cdot \rho(x)^{-1},
\end{eqnarray}
which correspond to changing the four solutions
as follows: 
$\, y_i \, \, \rightarrow \, \, \rho(x) \cdot \, y_i$.
In such a case the previous determinant variables transform, again,
very nicely under the ``gauge'' function $\, \rho(x)$:
\begin{eqnarray}
\label{I11}
\hspace{-0.9in}(W_1, \, W_2, \, W_3, \, W_4) 
 \,\,   \, \,\, \rightarrow \, \,\,\,\,    \, 
(\rho(x) \cdot W_1,\, \, \rho(x)^2 \cdot  W_2, \,\,
  \rho(x)^3 \cdot  W_3, \,\,  \rho(x)^4 \cdot  W_4). 
\end{eqnarray}

It is straightforward to see that the Yukawa coupling 
and the ``dual Yukawa'', $\,K^{\star}$, are 
{\em invariant by such a transformation}\footnote[5]{ $\,K$
and $\,K^{\star}$ (and their combinations) are the only monomials 
$\,\, W_1^{n_1} \, W_2^{n_2} \, W_3^{n_3} \, W_4^{n_4}$
to be invariant by (\ref{covar}) and (\ref{I11}). }.
Two  conjugated operators (\ref{conju}) 
automatically have the same Yukawa coupling.

Do note that the Yukawa couplings for two operators, which are 
non trivially homomorphic to each other
 (intertwiners of order one, two, ...), 
are actually different.
{\em The (pullback invariant) Yukawa coupling is not preserved by 
operator equivalence} (see subsection (\ref{operatornontriv})).
\vskip .1cm
\vskip .1cm
{\bf Remark:} The definition of these determinantal variables
$\, W_i$ heavily relies on the MUM structure of the 
operator between the four solutions the
 definition of $\, W_i$'s, in particular the
log-ordering of the solutions. 
It is worth noting that if one permutes the four solution $\, y_i$,
one would get 24 other sets of $(W_1, \,W_2,  \,W_3,  \,W_4)$
 which are actually {\em also nicely covariant by pullbacks}, 
thus yielding a finite set of other ``Yukawa couplings''
 or  adjoint Yukawa coupling $\, K^{\star}$ {\em also invariant by pullbacks}.

In fact these ``Yukawa couplings'' (\ref{Yukawa}),
 and other adjoint Yukawa $\, K^{\star}$ (\ref{Kstar}),  
can even be defined when the linear differential 
operator is not MUM, and they are still
invariant by pullbacks. 

\subsection{Pullback-invariants for higher order ODEs}
\label{invar}

These simple calculations can straightforwardly be generalised to higher order
linear differential equations. We give here 
the invariants for higher order linear differential operators.

Let us give, for the $\, n$-th order linear differential
operator the list of the $\, K_n$  invariants by pullback transformations:
\begin{eqnarray}
\hspace{-0.9in}&&K_3 \, = \, \, {{W_1^3 \cdot W_3} \over { W_2^3}}, \, \quad\, \,  
K_4 \, = \, \, {{W_1^8 \cdot W_4} \over { W_2^6}}, \, \quad \, \, 
K_5 \, = \, \, {{W_1^{15} \cdot W_5} \over { W_2^{10}}}, \,  \quad \, \, 
K_6 \, = \, \, {{W_1^{24} \cdot W_6} \over { W_2^{15}}}, \,\,  \, \cdots 
 \nonumber \\
\hspace{-0.9in}&&K_n \, = \, \, 
{{W_1^{a_n} \cdot W_n} \over { W_2^{b_n}}},\, \, 
 \quad \hbox{with:}  \, \, \quad \quad
 a_n \, = \, \,  n \cdot (n-2), \qquad 
b_n \, = \, \, {{ n \cdot (n-1)} \over { 2}}.
\end{eqnarray}
A $\, n$-th order linear differential operator has  $\, K_n$ as
an invariant by pullback transformation, as well as all the  $\, K_m$ 
with $\, m \, \le n$.
$K_3$ is the Yukawa coupling, and one remarks, for the order-four
operators, that the other pullback invariant
 $\, K^{\star}$ (see (\ref{Kstar})),
 which is actually also the Yukawa coupling of the adjoint
 operator, is nothing but $\, K_3^3/K_4$.

\vskip .3cm 
For  order-four operators  conjugated to their adjoint
 (see (\ref{conju})) (i.e. operators 
homomorphic to their adjoint, 
the intertwiner being
an order zero differential operator, a function), one has the equality 
\begin{eqnarray}
\label{cond}
\hspace{-0.8in}K_4 \, = \, \, K_3^2, \qquad  \hbox{i.e. } 
\qquad K_3 \, = \, \, K^{\star}, \qquad  \hbox{or} \qquad 
W_3^2      \, = \, \, W_1^2 \cdot W_4, 
\end{eqnarray}
to be compared with the equality in Almkvist et al. 
(see Proposition 2 in~\cite{Almkvist})
\begin{eqnarray}
\hspace{-0.1in}&&\quad y_0 \, y_3' \, - \, \, y_3 \, y_0'
\,\,\,\, = \,\, \,\,\,\,\,
y_1 \, y_2' \, - \, \, y_2 \, y_1',
\end{eqnarray}
which is satisfied when the Calabi-Yau condition that
the exterior square is of order five is satisfied.

If a linear differential operator $\, \Omega_4$
 verifies condition (\ref{cond}),
 its conjugate by a function, 
$\,\rho(x) \cdot \, \Omega_4 \cdot \rho(x)^{-1} $,
also verifies condition (\ref{cond}) (their Yukawa couplings
are equal). 

The condition (\ref{cond}) is not satisfied for linear differential 
operators homomorphic to their adjoint with {\em non-trivial} intertwiner
(of order greater than zero). For instance the order-four operator
(\ref{calB2}) does not satisfy condition (\ref{cond}).

\vskip .1cm

\section{More Hadamard products:  a Batyrev and van Straten Calabi-Yau ODE }
\label{B1}

An order-four operator has been found by Batyrev and van Straten~\cite{Batyrev} 
associated with a Calabi-Yau three-fold on $\, P_2 \times P_2$
\begin{eqnarray}
\label{Batyrev1first}
\hspace{-0.5in}&& B_1\,\,  \,  \, = \,\,  \, \, \,  \, \,
\theta^4 \, \, \,\,  -3 \, x \cdot 
(7\, \theta^2 \, +7 \,\theta \, +2) \cdot 
  (3\, \theta \, +\, 1) \cdot   (3\, \theta \, +\,2)
\, \\
\hspace{-0.5in}&&\quad \quad \quad \quad \quad \,  \,  \,  \, 
 -72 \, x^2 \cdot  
   (3\, \theta \, +\, 5) \cdot   (3\, \theta \, +\,4) \cdot
   (3\, \theta \, +\, 2)   \cdot  (3\, \theta \, +\, 1). 
  \nonumber
\end{eqnarray}
This operator is conjugated to its adjoint:
$\, B_1 \cdot \, x\, = \, \, x \cdot \, adjoint(B_1)$.

Operator (\ref{Batyrev1first}) is a {\em Calabi-Yau}
 operator~\cite{Batyrev}: it is MUM, and it is such
 that its exterior square is of {\em order five}.
It has a solution analytical at $\, x\, = \, 0$ which 
is actually the Hadamard product
of the previous selected hypergeometric $\, _2F_1$: 
 \begin{eqnarray}
\label{sumup}
\hspace{-0.7in}\quad  _2F_1\Bigl([{{1} \over {3}}, \,{{2} \over {3}}], [1]; \,
\, 27 \, x\Bigr) \, \star \, 
\Bigl( {{1 } \over {1\, +4 \, x}}\,  \cdot \,
 _2F_1\Bigl([{{1} \over {3}}, \, {{2} \over {3}}], \, [1]; \, 
{{27 \cdot x } \over {(1\, +4 \, x )^3 }}  \Bigr) \Bigr).
\end{eqnarray}
The coefficients of the series expansion of (\ref{sumup}) are integers:
they can actually be
written as a sum of {\em product of binomials}:
 \begin{eqnarray}
\label{nested}
\hspace{-0.7in}&& \quad \quad \quad {{(3n)!} \over {(n!)^3}} \cdot \, \, 
\sum_{k=0}^{n} \,  {n \choose k}^3 \, \,\, \, = \, \, \, \,  \,
{3\, n \choose n} \cdot \,{2\, n \choose n}\,
   \cdot \, \, 
\sum_{k=0}^{n} \,  {n \choose k}^3.
\end{eqnarray}
The series expansions of the nome and the Yukawa coupling of this 
Calabi-Yau operator are series with {\em integer coefficients}, reading
respectively:
\begin{eqnarray}
\hspace{-0.9in}&& q \, \, = \, \, \, x \, +48\,{x}^{2}\, +4626\,{x}^{3}\, 
+549304\,{x}^{4}\, +74589735\,{x}^{5}\, +11014152048\,{x}^{6}
\,\,\, + \, \, \, \cdots 
\nonumber
\end{eqnarray}
\begin{eqnarray}
\hspace{-0.9in}&& K(q) \, \, = \, \, \,1 \, +21\,q\, +3861\,{q}^{2}\, 
+429159\,{q}^{3}\, +57224661\,{q}^{4}\, +7337893896\,{q}^{5}\, 
\, \, \, + \, \, \, \cdots \nonumber
\end{eqnarray}
\begin{eqnarray}
\hspace{-0.9in}&& K(x) \, \, = \, \, \,1 \, +21\,x\,+4869\,{x}^{2}\,
+896961\,{x}^{3}\,+175176657\,{x}^{4}\,
+34770008997\,{x}^{5}\,\, + \, \, \, \cdots \nonumber
\end{eqnarray}

\vskip .5cm

\pagebreak 

\section*{References}

\def\cprime{$'$}
\providecommand{\newblock}{}

\end{document}